\documentclass[letterpaper,11pt]{article}
\pdfoutput=1

\usepackage{jheppub} 

\usepackage[printonlyused]{acronym}

\usepackage{color}

\usepackage{verbatim}
\usepackage{amsmath}
\usepackage{mathabx}

\usepackage{amssymb}
\usepackage[caption=false]{subfig}

\usepackage{multirow}

\newcommand{\be}{\begin{equation}}
\newcommand{\ee}{\end{equation}}
\newcommand{\iso}[2]{\mathcal{I}^\text{#1}_{#2}}

\newcommand{\remd}{\mathcal{R}_n}

\usepackage{comment}

\DeclareRobustCommand{\Sec}[1]{Sec.~\ref{#1}}
\DeclareRobustCommand{\Secs}[2]{Secs.~\ref{#1} and \ref{#2}}
\DeclareRobustCommand{\App}[1]{App.~\ref{#1}}
\DeclareRobustCommand{\Tab}[1]{Table~\ref{#1}}

\DeclareRobustCommand{\Fig}[1]{Fig.~\ref{#1}}
\DeclareRobustCommand{\Figs}[2]{Figs.~\ref{#1} and \ref{#2}}
\DeclareRobustCommand{\Eq}[1]{Eq.~(\ref{#1})}
\DeclareRobustCommand{\Eqs}[2]{Eqs.~(\ref{#1}) and (\ref{#2})}
\DeclareRobustCommand{\Ref}[1]{Ref.~\cite{#1}}

\begin{document}

\count\footins = 1000
\interfootnotelinepenalty=10000
\setlength{\footnotesep}{0.6\baselineskip}

\renewcommand{\arraystretch}{1.3}

\title{A Robust Measure of Event Isotropy at Colliders}

\author[a]{Cari Cesarotti}
\author[b]{and Jesse Thaler}

\affiliation[a]{Department of Physics, Harvard University, Cambridge, MA 02138, USA}
\affiliation[b]{Center for Theoretical Physics, Massachusetts Institute of Technology,\\ Cambridge, MA 02139, USA}

\emailAdd{ccesarotti@g.harvard.edu}
\emailAdd{jthaler@mit.edu}

\preprint{MIT--CTP 5195}

\abstract{
We introduce a new event shape observable---\emph{event isotropy}---that quantifies how close the radiation pattern of a collider event is to a uniform distribution.
This observable is based on a normalized version of the energy mover's distance, which is the minimum ``work'' needed to rearrange one radiation pattern into another of equal energy. 
We investigate the utility of event isotropy both at electron-positron colliders, where events are compared to a perfectly spherical radiation pattern, as well as at proton-proton colliders, where the natural comparison is to either cylindrical or ring-like patterns.
Compared to traditional event shape observables like sphericity and thrust, event isotropy exhibits a larger dynamic range for high-multiplicity events.
This enables event isotropy to not only distinguish between dijet and multijet processes but also separate uniform $N$-body phase space configurations for different values of $N$.
As a key application of this new observable, we study its performance to characterize strongly-coupled new physics scenarios with isotropic collider signatures.
}
\maketitle

\section{Introduction}
\label{sec:intro}

The Large Hadron Collider (LHC) at CERN has enormous potential for discovery, as demonstrated by the first detection of the Higgs boson in 2012 \cite{Aad:2012tfa,Chatrchyan:2012xdj}.
Other new physics signals near the weak scale may have evaded detection by hiding in unexplored kinematic regimes. 
The challenge is then to design robust searches for new physics that are sensitive to several possible classes of theories. 
One promising strategy is to construct observables sensitive to event characteristics that are fundamentally distinct from those arising from the Standard Model (SM).

Event shape observables have long provided useful insights into the structure of the SM and the underlying dynamics of quantum chromodynamics (QCD) \cite{Farhi:1977sg, Barber:1979bj,Althoff:1983ew,Abrams:1989ez,Li:1989sn,Buskulic:1995aw,Adriani:1992gs, Braunschweig:1990yd,Abe:1994mf,Heister:2003aj,Abdallah:2003xz,Achard:2004sv, Abbiendi:2004qz, Dasgupta:2003iq,Dissertori:2008cn}.
Although QCD is complicated, we can generally characterize its behavior at collider scales by its small 't Hooft coupling $\lambda \equiv g_s^2 N_C$. 
In this perturbative regime, the dynamics is dominated by the emission of soft and collinear quarks and gluons.
As the parton shower evolves to lower scales, these quarks and gluons hadronize, such that the final state is populated with many light mesons.
Because color flux tubes can be severed by light quark-antiquark pairs, the hadronized final state retains much of the kinematic information from the initial perturbative showering.
Therefore, SM events at energy scales well above the QCD confinement scale ($\Lambda_\text{QCD} \simeq 200$ MeV) are inherently jetty due to these initial perturbative soft and collinear splittings \cite{Ellis:1991qj}, and this jet-like structure can be quantified via event shape observables. 
By contrast, if new physics has a large 't Hooft coupling or other non-QCD-like features, then it could generate events that are not collimated at collider scales but rather isotropic or quasi-isotropic.
Event shape observables could help search for isotropic new physics signals and separate them from jet-like QCD backgrounds.

In this paper, we introduce \emph{event isotropy} $\iso{}{}$, a new event shape observable designed to identify uniform radiation patterns.
Event isotropy is defined as the dimensionless distance between a collider event $\mathcal{E}$ and a uniform radiation pattern $\mathcal{U}$ of the same energy: 
\begin{equation}
\iso{}{} \left( \mathcal{E} \right) \equiv \text{EMD} \left(\mathcal{U}, \mathcal{E}\right).
\label{eq:evIsoDef}
\end{equation}
The measure of similarity is the (normalized) energy mover's distance (EMD) \cite{Komiske:2019fks,Komiske:2019jim,Komiske:2020qhg}, which is the particle physics adaptation of the earth mover's distance used in computer graphics \cite{Peleg1989AUA, Rubner:1998:MDA:938978.939133,Rubner2000,Pele2008ALT,Pele2013TheTE} and the Wasserstein metric used to compare probability distributions \cite{wasserstein1969markov}.
The EMD was introduced to quantify the distance between pairs of collider events \cite{Komiske:2019fks,Komiske:2019jim,Komiske:2020qhg}, but here we compare a single collider event $\mathcal{E}$ to an idealized isotropic (in practice, quasi-isotropic) energy distribution $\mathcal{U}$.
Since we take $\mathcal{E}$ and $\mathcal{U}$ to have the same overall energy, we can define $\iso{}{} \in [0,1]$ to be dimensionless, ranging from $0$ for a perfectly isotropic configuration to $1$ for maximally jet-like. 
We examine how the event isotropy is effective not only at identifying uniform radiation patterns at both $e^+e^-$ and $pp$ colliders, but also at separating QCD dijet samples, multi-pronged SM events, and uniform $N$-body phase space events.

There are many models that arise in the context of addressing the hierarchy problem that can produce high-multiplicity, quasi-isotropic events. 
For example, R-parity violating (RPV) supersymmetry (SUSY) models with small mass splittings yield long cascade decay chains at colliders, producing signatures with $\mathcal{O}(10)$ QCD jets distributed roughly isotropically~\cite{Barbier:2004ez, Evans:2013jna}.
Extra dimensional models \cite{Appelquist:1987nr} such as the ADD \cite{ArkaniHamed:1998rs} and Randall-Sundrum \cite{Randall:1999ee} scenarios allow for the potential formation of microscopic black holes at colliders~\cite{Giddings:2001bu,Dimopoulos:2001hw}. 
These black holes evaporate via Hawking radiation, emitting particles with a thermal distribution in isotropic directions \cite{Harris:2003db,Dai:2007ki,Cavagli2007CatfishAM} (see, however, \Ref{Meade:2007sz}).
Another general class of models arising from solutions to the hierarchy problem and string theory constructions is hidden valley scenarios~\cite{Strassler:2006im}.  
Hidden valleys can involve non-Abelian gauge theories with large 't Hooft couplings, which yield spherical radiation patterns due to rapid fragmentation of sufficiently high-multiplicity showers~\cite{Polchinski:2002jw,Hofman:2008ar,Hatta:2008tx}. 
Thus, hidden valley scenarios that are strongly coupled over a large window in the hidden sector, weakly coupled to the SM, and have a mass gap much smaller than the mass of the mediator will produce isotropic signatures~\cite{Strassler:2008bv, Strassler:2008fv}.
Assuming a substantial fraction of the hidden sector particles decay to visible final states in the detector, the signature of such models at colliders are soft, unclustered energy patterns (SUEPs)~\cite{Knapen:2016hky}. 
These are just a few examples of possible new physics scenarios that produce event shapes distinct from those arising from the SM.

Historically, several observables were developed to quantify the degree to which a collider event is isotropic versus jet-like, including thrust \cite{Brandt:1964sa,Farhi:1977sg,DeRujula:1978vmq}, sphericity \cite{Bjorken:1969wi,Ellis:1976uc}, spherocity \cite{Georgi:1977sf},%
\footnote{We briefly considered carrying on the vowel shift tradition and naming our observable ``spheracity''.}
and the $C$- and $D$-parameters \cite{Parisi:1978eg,Donoghue:1979vi,Ellis:1980wv}.
While all of these observables have provided insight into the substructure of QCD, they are not robust probes of isotropy. 
Sphericity and the $C$- and $D$-parameters can obtain their extremal values for non-isotropic events, specifically for configurations with as few as six particles as long as they are symmetric under the exchange of any axes.
Sphericity has the additional pathology of not being infrared and collinear (IRC) safe.
Spherocity was developed as an IRC-safe alternative, but it is not widely used, in part because the spherocity axis is not as well behaved as the thrust axis \cite{Larkoski:2014uqa}. 
Thrust is a well studied observable, and it is a true measure of isotropy in the sense that $T = \frac{1}{2}$ if and only if an event is perfectly spherical.
Thrust has a small dynamical range as one approaches the isotropic limit, though, and it is unable to distinguish distinct high-multiplicity, quasi-uniform samples. 
We propose event isotropy as an observable that can distinguish SM events with many hard prongs of radiation from events with genuinely isotropic signatures.

Event shape observables have been important ingredients in constructing new physics searches at the LHC.
Several recent proposals consider these observables in their high-level triggers, including dedicated searches for displaced vertices from long-lived particle decay in both the leptonic and hadronic channels \cite{Aad:2013txa,Perrotta:2015jyu,Aaij:2017mic,Aaij:2016isa}, magnetic monopoles \cite{Aad:2015kta}, black holes \cite{Sirunyan:2018njd,Sirunyan:2017anm}, and emergent jets \cite{Sirunyan:2018njd}. 
Additional search strategies have been proposed, such as dedicated triggers for emerging jets \cite{Schwaller:2015gea}, semi-visible jets \cite{Cohen:2015toa}, and SUEPs \cite{Knapen:2016hky}. 
A more complete summary of the status of partially online searches for new physics with nonstandard geometry can be found in \Ref{Alimena:2019zri}. 
The goal of this work is to develop a new, model-independent tool to identify and characterize anomalously isotropic events, to be used in conjunction with existing model-dependent search strategies for new physics.

The outline of this paper is as follows. 
In \Sec{sec:emdSM}, we review the definition of the EMD and introduce event isotropy, contrasting its behavior to traditional event shape observables.
In \Sec{sec:emdBE}, we compare the performance of event isotropy to thrust and sphericity for discriminating dijets from top-quark pairs at electron-positron colliders, as well as for characterizing uniform $N$-body phase space and quasi-isotropic samples. 
We present similar studies for proton-proton colliders in \Sec{sec:ppBE}. 
In \Sec{sec:SUEP}, we demonstrate the potential of event isotropy to characterize SUEP scenarios at the LHC. 
We conclude in \Sec{sec:conclusions} and remark on possible future applications.
%


\section{A Robust Measure of Event Isotropy}
\label{sec:emdSM}

\begin{table*}[t]
\centering
\begin{tabular}{c @ {\quad} c @ {\quad} c @ {\quad} c@ {\quad} c} 
 \hline \hline
Collider & Geometry  & Energy Weight & Ground Measure & Default $\mathcal{U}$\\
\hline
 $e^+e^-$ & Sphere & $w^\text{sph}_{i} = E_{i}/E_\text{tot}$ &$d^\text{sph}_{ij} = 2 \left( 1 - \cos \theta_{ij} \right) $ & $\mathcal{U}_{192}^\text{sph}$ \\
 $pp$ & Cylinder & $w^\text{cyl}_{i} = p_{Ti} / p_{T\text{tot}}$ & $d^\text{cyl}_{ij} = \frac{12}{\pi^2 +16 y_\text{max}^2} \left( y_{ij}^2 + \phi_{ij}^2 \right) $ & $\mathcal{U}_{160}^\text{cyl}(|y|< 2)$ \\
 $pp$ & Ring & $w^\text{ring}_{i} = p_{Ti} / p_{T\text{tot}}$ &$d^\text{ring}_{ij} = \frac{\pi}{\pi -2} \left( 1 - \cos \phi_{ij}\right)$ & $\mathcal{U}_{32}^\text{ring}$\\ 
 \hline
 \hline
 \end{tabular}
\caption{The three different event geometries used to define event isotropy in this paper, with their corresponding energy weights, ground measures, and default quasi-uniform configurations.
For the cylinder geometry, we must specify the rapidity range $|y_i| < y_\text{max}$.
Note that $p_{T\text{tot}}$ is the scalar sum of the transverse momenta.
These ground measures satisfy \Eq{eq:pwasserstein} with $\beta = 2$.
}
\label{tab:emdSpec}
\end{table*}

In this section, we define event isotropy and discuss its properties and limiting behaviors. 
We first review the mathematical definition of the earth mover's distance and its application to particle physics via the energy mover's distance.
(We use the acronym ``EMD'' in both cases, except where confusions might arise.)
We then introduce our event isotropy observable for three different choices of geometry, shown in \Tab{tab:emdSpec}, and compare its behavior with previous event shape observables.

\subsection{Review of Earth Mover's Distance}

The earth mover's distance \cite{Peleg1989AUA, Rubner:1998:MDA:938978.939133,Rubner2000,Pele2008ALT,Pele2013TheTE}, or Wasserstein metric \cite{wasserstein1969markov}, defines the distance between normalized distributions as the minimum ``work'' necessary to rearrange one distribution into another (e.g.\ moving dirt or ``earth'' from one pile to another). 
Consider two distributions $P$ and $Q$ with elements $p_i \in P$ and $q_j \in Q$ with weight $w_{p,i}$ and position $x_{p,i}$ for element $p_i$, and similarly for $q_j$.
The weights of the elements are normalized such that $\sum_{i} w_{p,i}= \sum_{j} w_{q,j} = 1$.
Given a ground measure $d_{ij}$ between the positions $x_{p,i}$ and $x_{q,j}$, the EMD to move $P$ to $Q$ is
\begin{equation}
\label{eq:EMDgeneral}
\text{EMD}(P, Q) = \min \limits_{\left\{ f_{ij} \right\}} \sum_{i j} f_{ij} \, d_{ij},
\end{equation}
where the transportation plan $\{f_{ij}\}$ is constrained by the requirements that
\begin{equation}
f_{ij} \geq 0, \qquad \sum_{j} f_{ij} = w_{p,i}, \qquad \sum_{i} f_{ij} = w_{q,j}, \qquad \sum_{i j} f_{ij} = 1.
\end{equation}
These constraints enforce that positive ``earth'' is transported and that all of the earth from $p_i \in P$ is moved to $q_j \in Q$.
For normalized distributions, the EMD is dimensionless. 
It is clear from \Eq{eq:EMDgeneral} that the EMD between distributions increases if more earth ($f_{ij}$) must be moved or if the ground metric between locations ($d_{ij}$) is large. 
Similarly, the EMD is unchanged as the distance between two points in the same distribution approaches zero.

Typically, $d_{ij}$ is a proper metric, meaning that it non-negative, symmetric, and satisfies the triangle inequality:
\begin{equation}
0 \le d_{ij} \le d_{ik} + d_{kj}.
\label{eq:beta1met}
\end{equation}
In such cases, the EMD is also a proper metric, satisfying a triangle inequality on distributions:
\begin{equation}
0 \le \text{EMD}(P,Q) \le \text{EMD}(P, R) + \text{EMD}(R,Q).
\label{eq:propmet}
\end{equation}
One can also generalize to a larger class of ground measures, characterized by a parameter $\beta > 1$, that satisfy:
\begin{equation}
\label{eq:pwasserstein}
0 \le d_{ij}^{1/\beta} \le d_{ik}^{1/\beta} + d_{kj}^{1/\beta} \quad \Rightarrow \quad \text{EMD}(P, Q)^{1/\beta} \le \text{EMD}(P,R)^{1/\beta} + \text{EMD}(R,Q)^{1/\beta}.
\end{equation}
Here, $\text{EMD}(P, Q)^{1/\beta}$ is known as a $p$-Wasserstein metric with $p=\beta$, and the usual case in \Eqs{eq:beta1met}{eq:propmet} corresponds to $\beta = 1$.
In this paper, we will find it convenient to focus on ground measures that satisfy \Eq{eq:pwasserstein} with $\beta = 2$.
A case study comparing the $\beta=1$ and $\beta=2$ ground measures is presented in \App{app:dijetSphere}.

In the context of particle physics, one can consider the distributions $P$ and $Q$ to be event radiation patterns, or more specifically, energy densities measured by an idealized calorimeter. 
As first proposed in \Ref{Komiske:2019fks}, one can then use the ``energy mover's distance'' to quantify the similarity of two collider events. 
The weights $w_{p,i}$ and $w_{q,j}$ are now the energies (or momenta) of particles, and the ground measures $d_{ij}$ are now an angular distance between particle trajectories.
In \Ref{Komiske:2019fks}, a penalty term was added to \Eq{eq:EMDgeneral} to account for the net energy difference between events, and the EMD had units of energy.
Here, we normalize all of our energy distributions to have net weight of unity, such that the EMD is dimensionless.
The EMD is IRC safe by construction, since $P$ and $Q$ are energy densities and therefore unchanged by collinear or soft emissions.

We perform all EMD calculations in this paper using the Python Optimal Transport (POT) library \cite{flamary2017pot}.
When comparing two $N$ particle configurations, the computation time scales roughly like $\mathcal{O}(N^3 \log^2 N)$, though it can be faster in practice.
As a benchmark, the event isotropy for a 50-particle event with a 192-particle reference sphere takes approximately 50 ms to compute.
The code to perform the studies below will be made available through the \textsc{Event Isotropy} repository \cite{cesarotti:2020ei}.

\subsection{Energy Mover's Distance as a Measure of Event Isotropy}

To characterize the degree of isotropy of an event $\mathcal{E}$, we use the EMD to measure its distance to a quasi-uniform radiation pattern $\mathcal{U}$ with $n$ particles.
More specifically, we define event isotropy as:
\begin{equation}
\iso{geo}{n}( \mathcal{E}) \equiv \text{EMD}_\text{geo}( \mathcal{U}^\text{geo}_n,\mathcal{E}),
\label{eq:eventIsoDef}
\end{equation}
with a notation to be explained in detail below.
We always work with normalized events and choose our distance measures such that $\iso{geo}{n} \in [0,1]$ is dimensionless.
While the EMD is well defined in the $n\rightarrow\infty$ limit, our current method of numerically computing event isotropy using the POT library requires finite $n$.

\begin{figure}[p]
\begin{centering}
\subfloat[]{
       \includegraphics[width=0.3\textwidth]{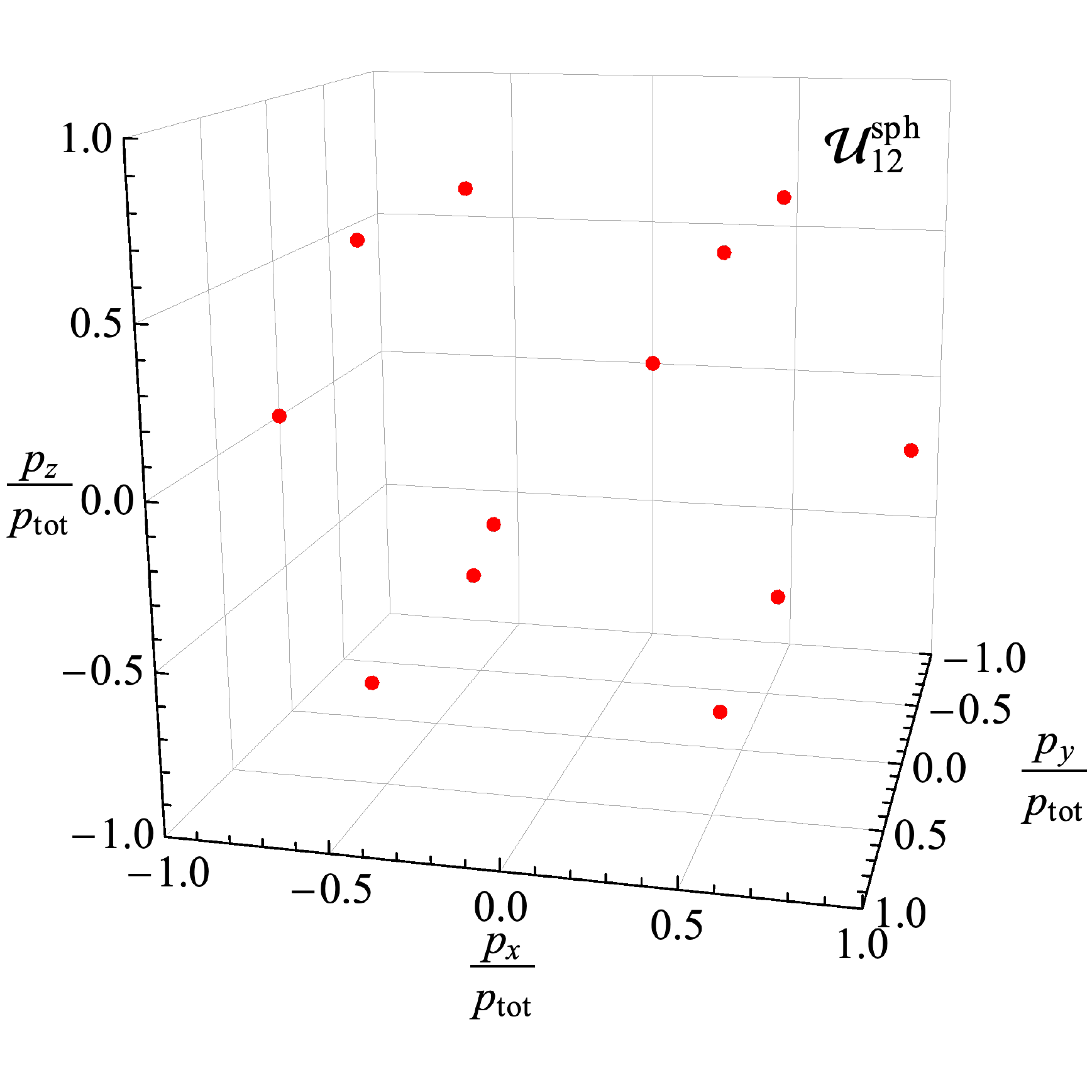}
     }
     \hfill
\subfloat[]{
       \includegraphics[width=0.3\textwidth]{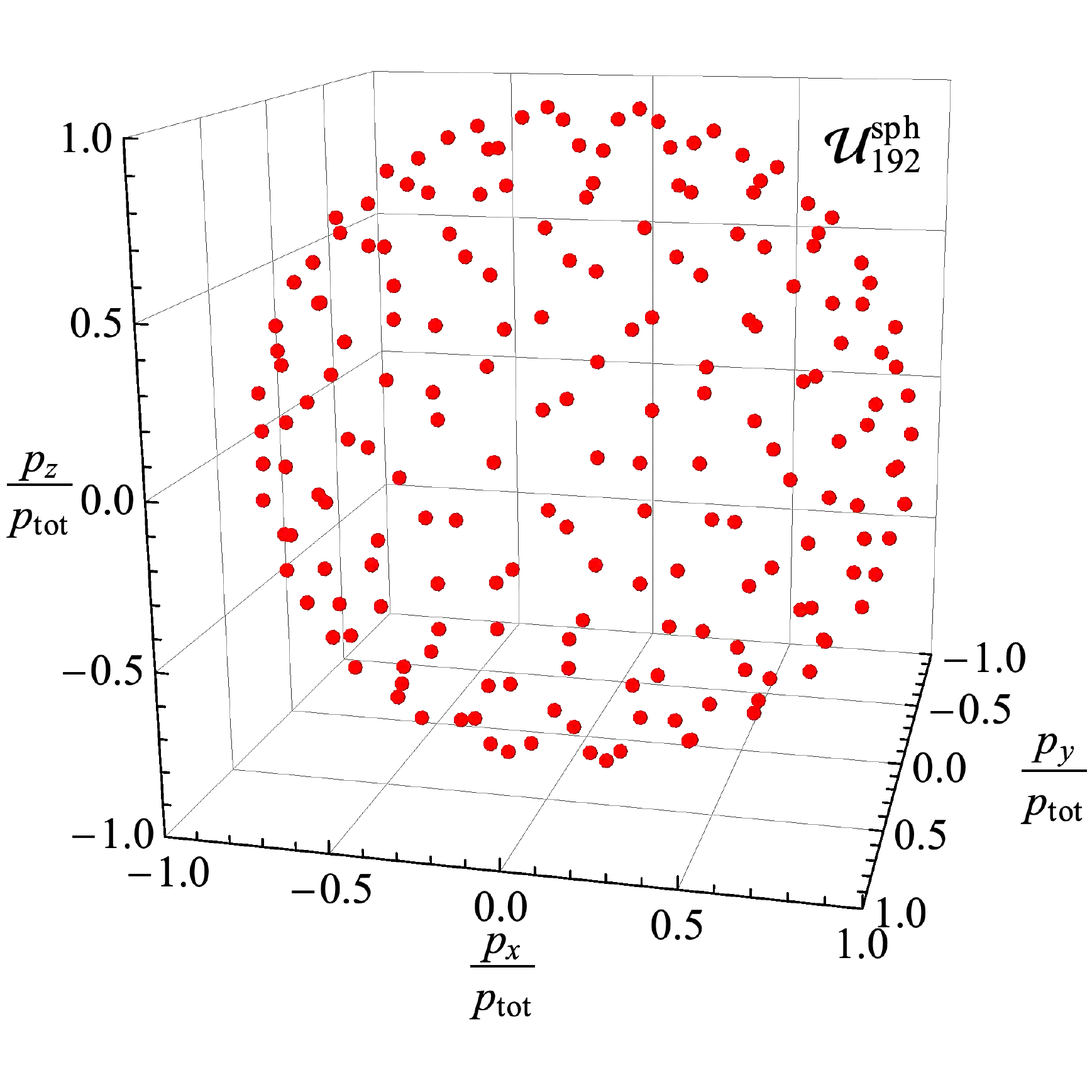}
     }
    \hfill
\subfloat[]{
	\includegraphics[width=0.3\textwidth]{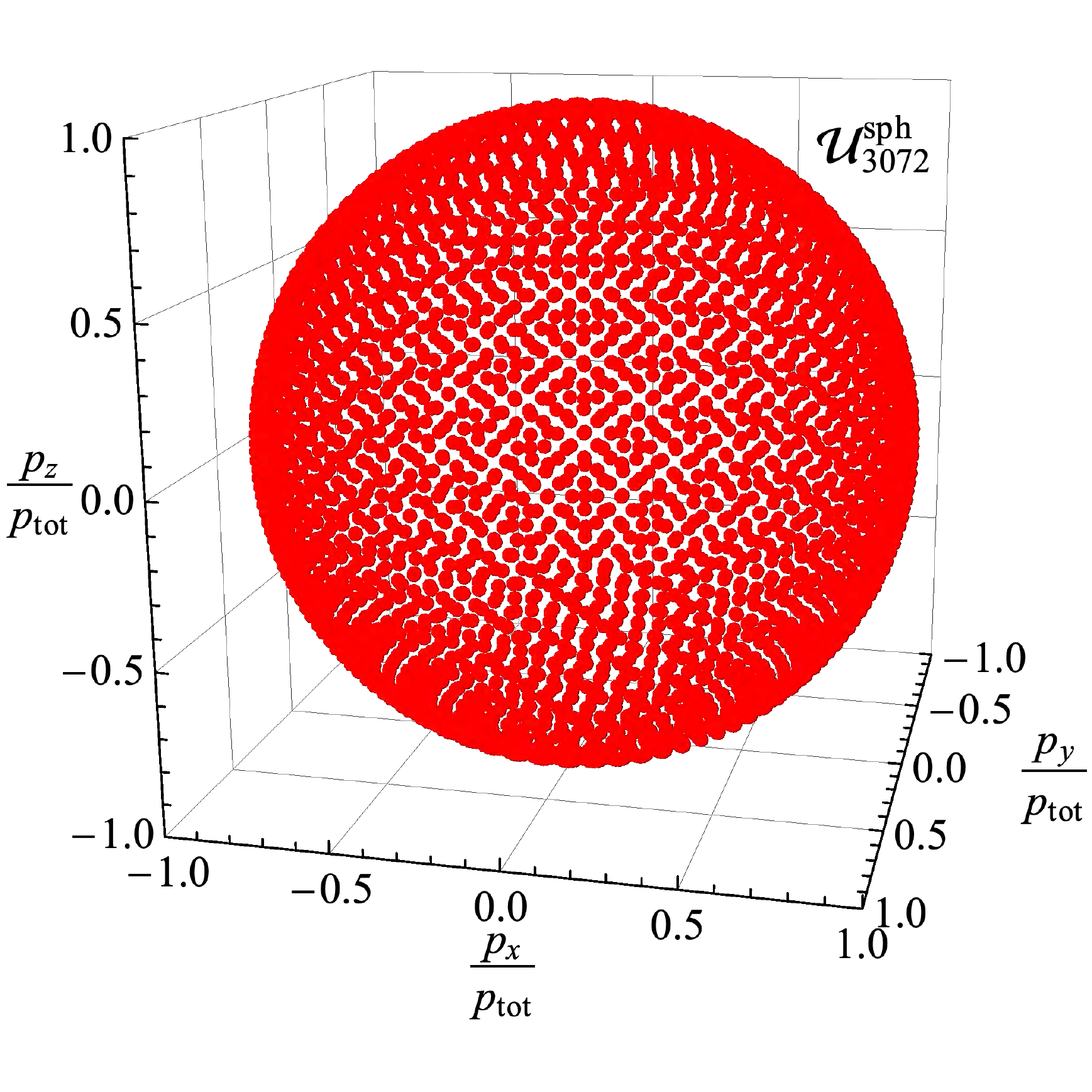}
	}
\hfill
\subfloat[]{
       \includegraphics[width=0.3\textwidth]{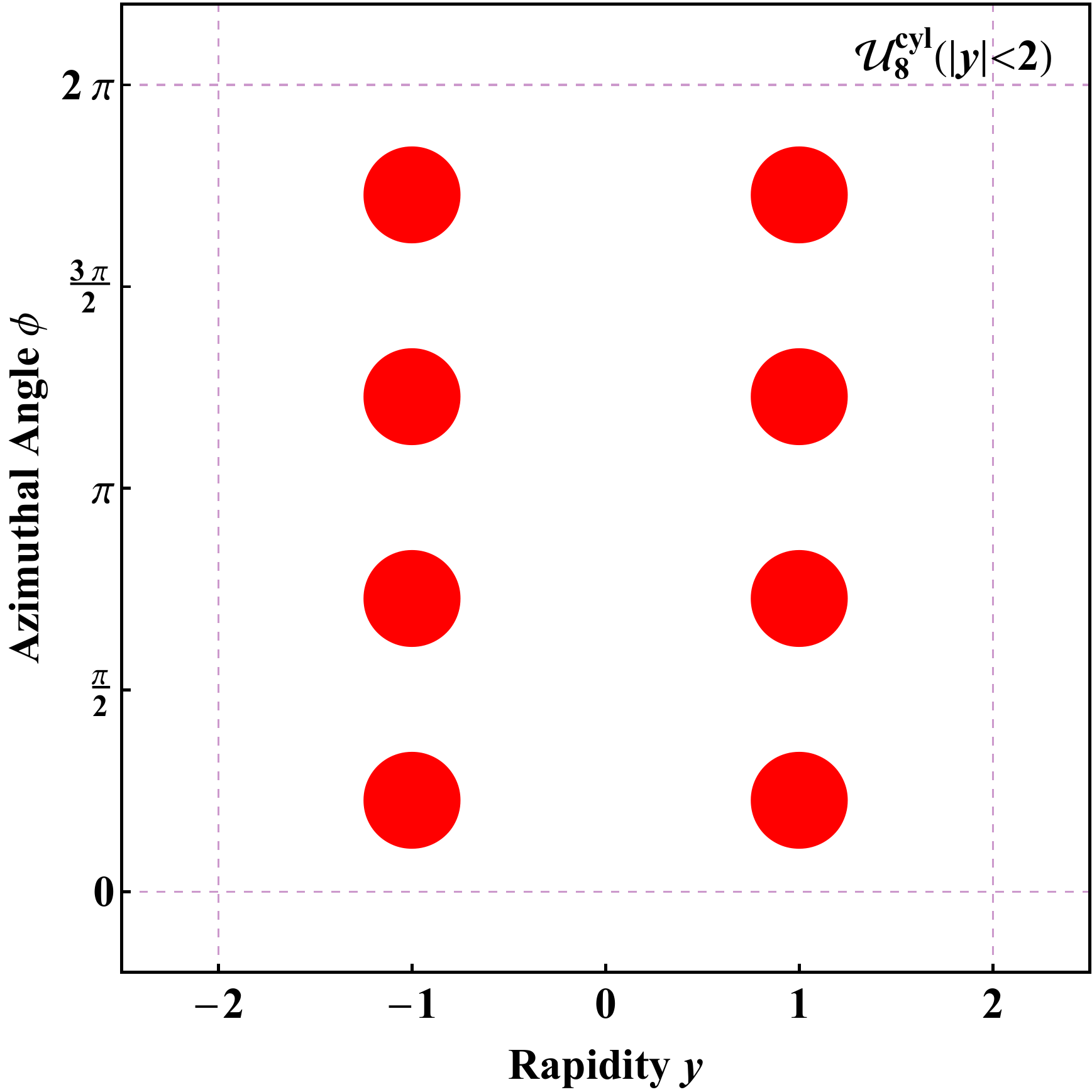}
     }
     \hfill
\subfloat[]{
       \includegraphics[width=0.3\textwidth]{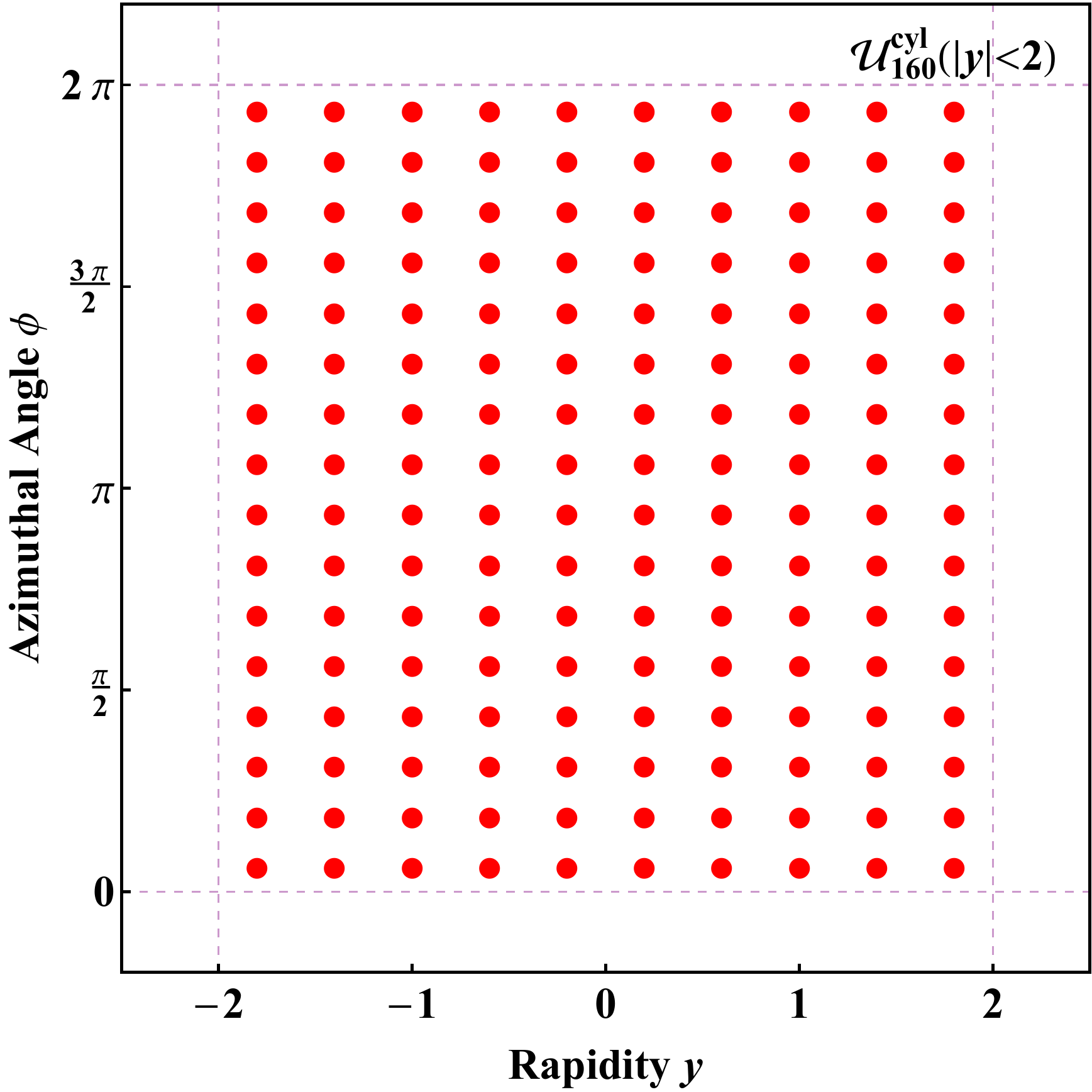}
     }
    \hfill
\subfloat[]{
       \includegraphics[width=0.3\textwidth]{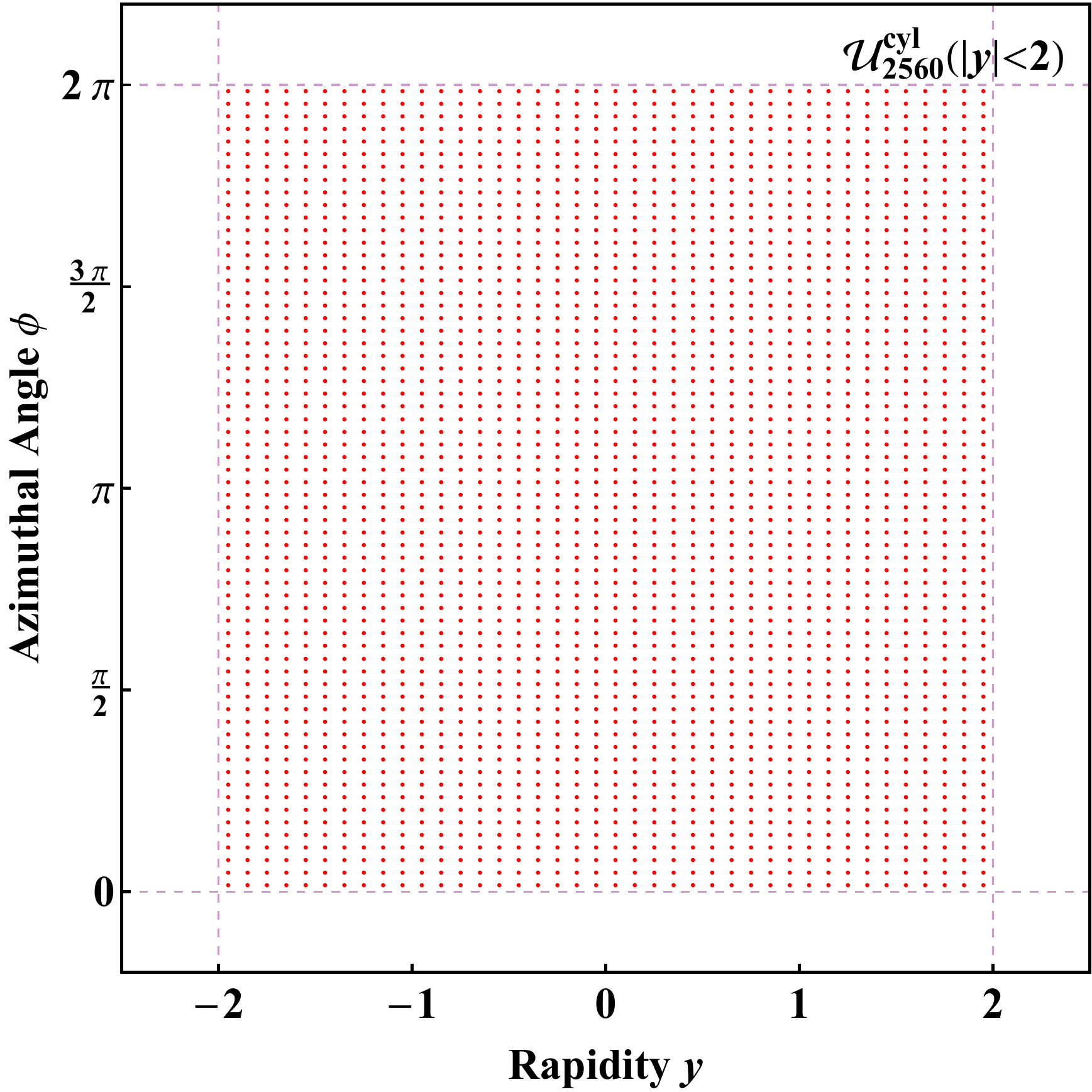}
     }
\hfill     
\subfloat[]{
       \includegraphics[width=0.3\textwidth]{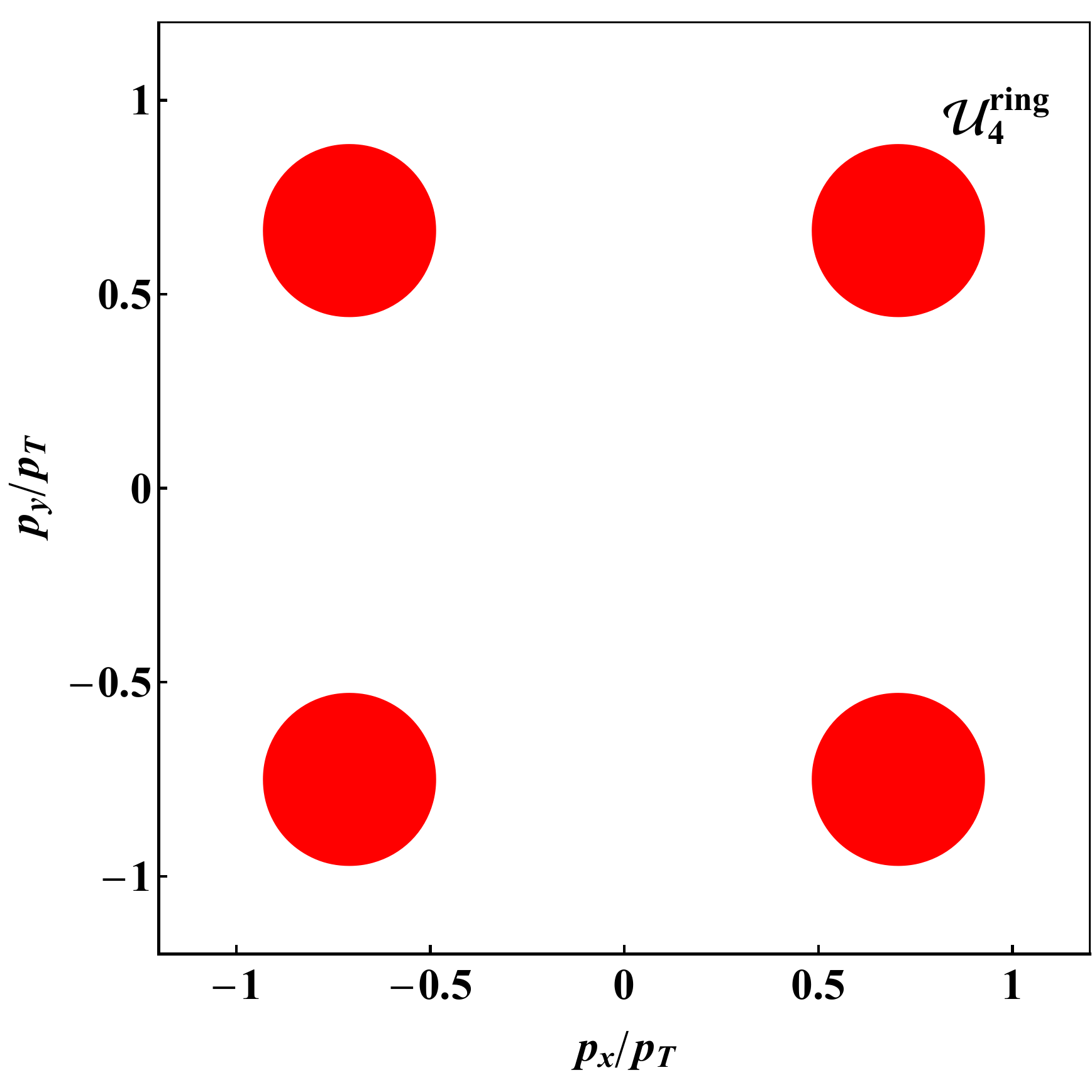}
     }
     \hfill
\subfloat[]{
       \includegraphics[width=0.3\textwidth]{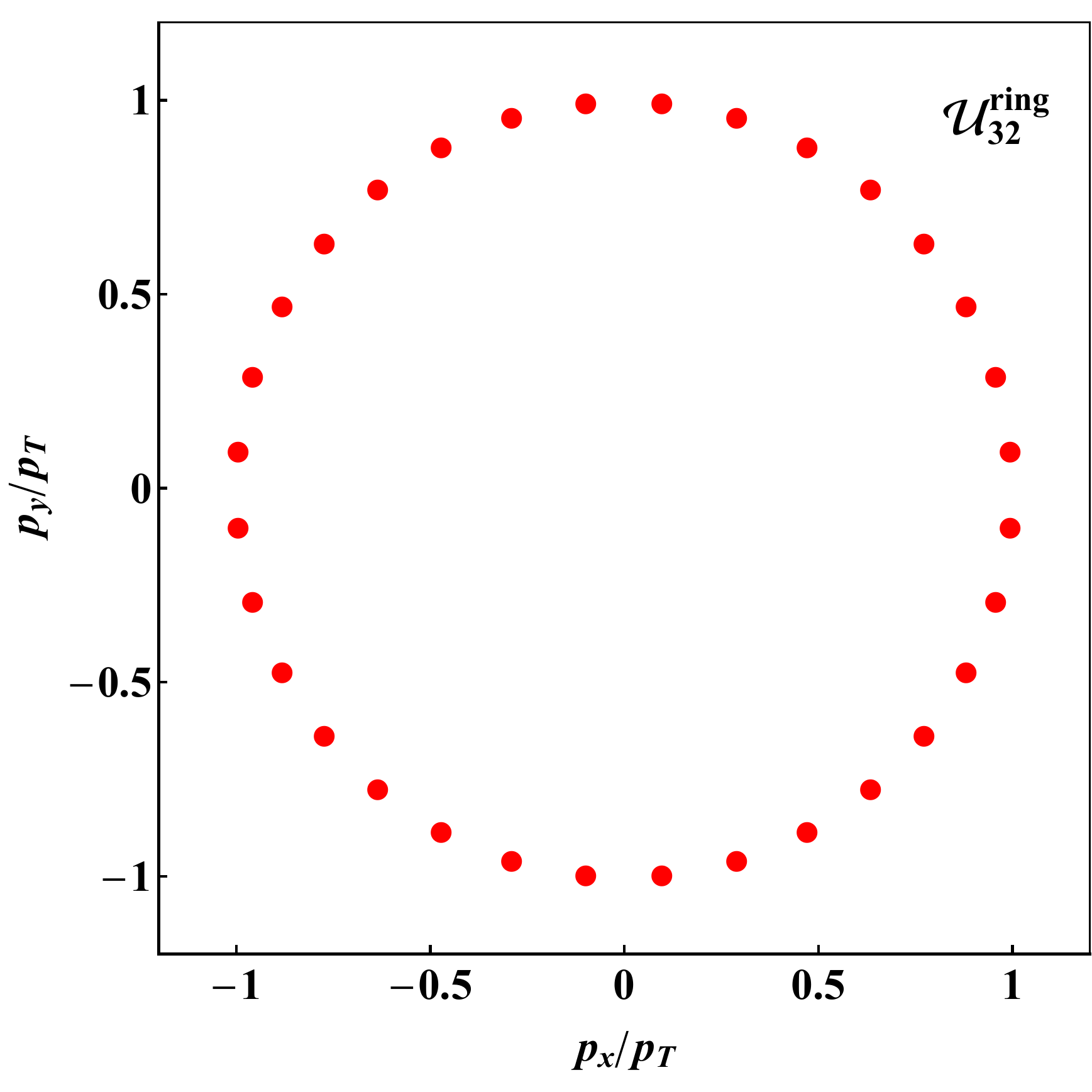}
     }
    \hfill
\subfloat[]{
	\includegraphics[width=0.3\textwidth]{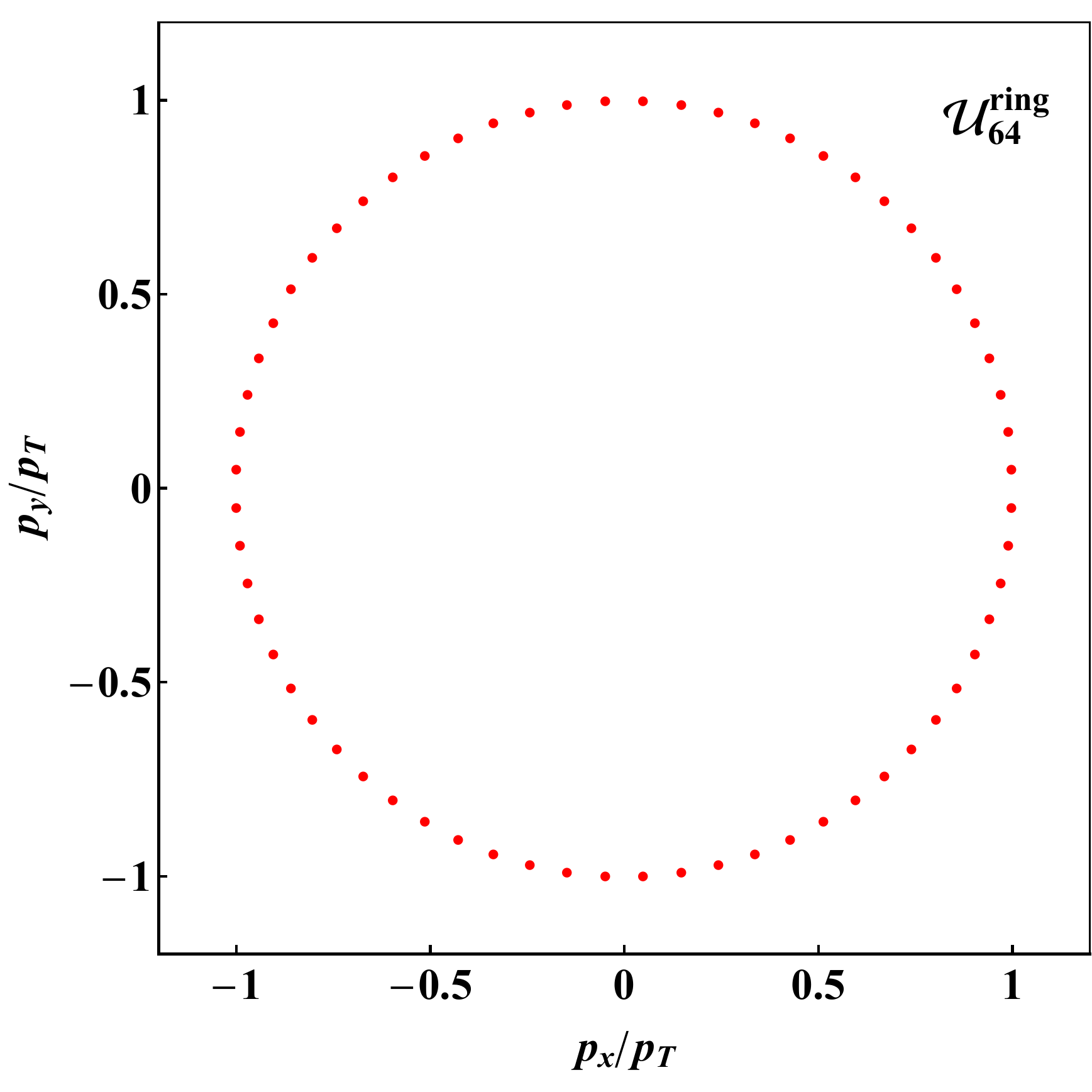}
	}
     \end{centering}
\caption{
Quasi-uniform event configurations for the three geometries considered in this paper.
{Top row} (a,b,c):  spherical configurations $\mathcal{U}^\text{sph}_n$ for $n=\{12,192,3072\}$, generated with \texttt{HEALPix}.
{Middle row} (d,e,f): cylindrical configurations $\mathcal{U}^\text{cyl}_n$ for $n=\{8,160,2560\}$.
{Bottom row} (g,h,i): ring configurations $\mathcal{U}^\text{ring}_n$ for $n=\{4,32,64\}$.
All particles have equal weights.
The middle column corresponds to our default uniform configurations for each geometry.
}
\label{fig:geoVis}
\end{figure}

\begin{figure}[t!]
\subfloat[]{
\label{fig:dijets_sph}
\includegraphics[width=0.25\textwidth]{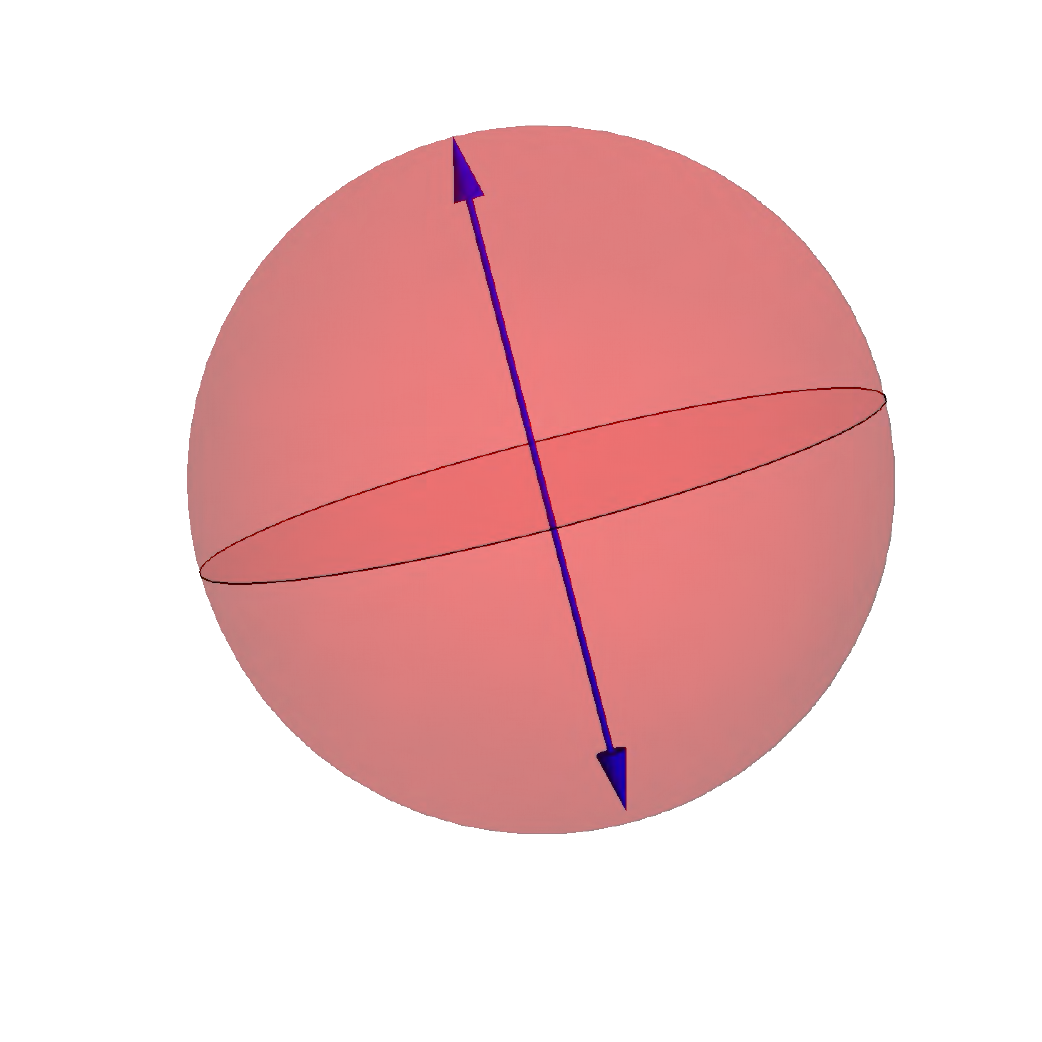}
     }
    \hfill
\subfloat[]{
\label{fig:dijets_cyl}
\includegraphics[width=0.25\textwidth]{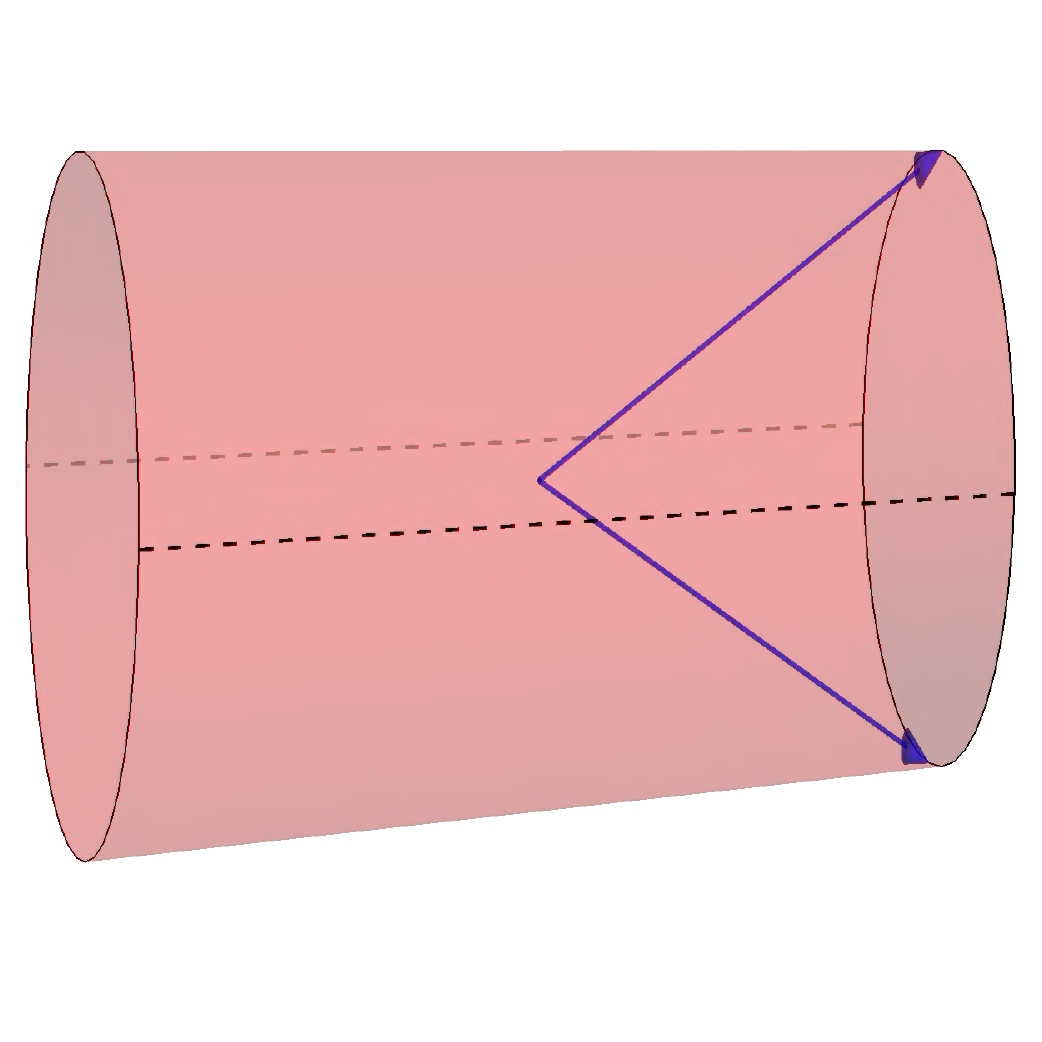} }
    \hfill
 \subfloat[]{
\label{fig:dijets_ring}
\includegraphics[width=0.25\textwidth]{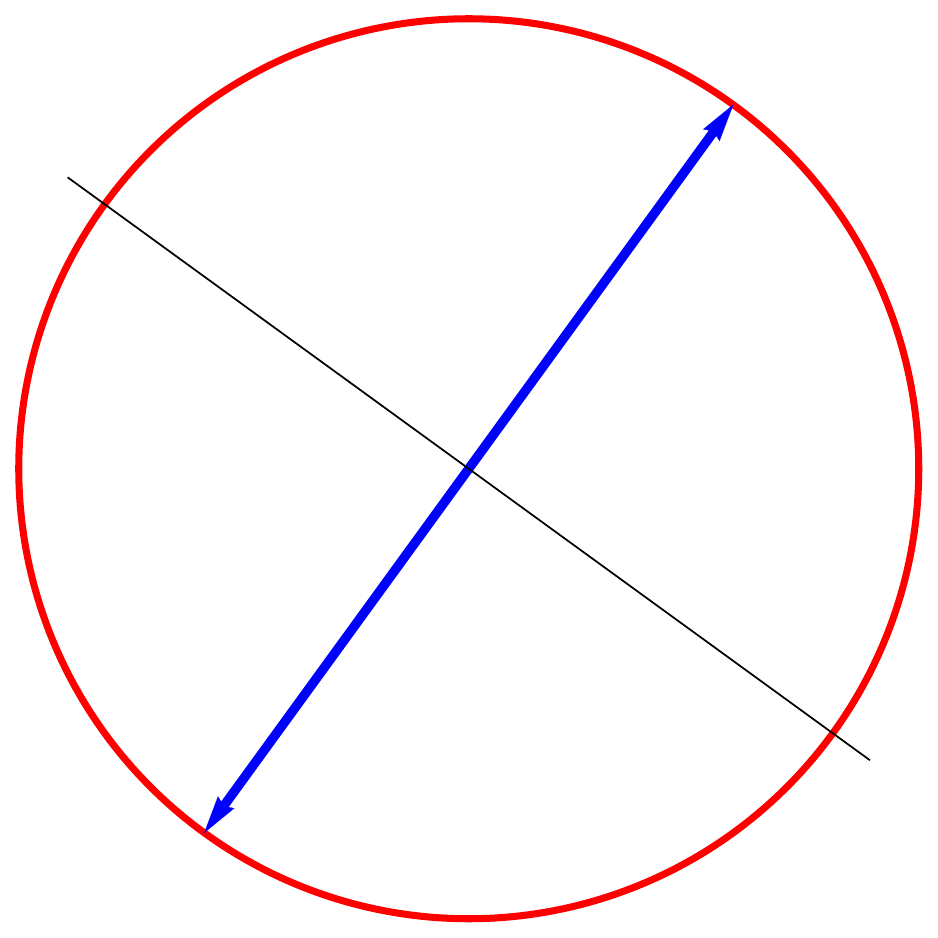}}
    \hfill
    \qquad
\caption{Event configurations that maximize event isotropy (i.e.\ least isotropic), assuming balanced (transverse) momentum.
The optimal EMD transportation plan is determined by partitioning the uniform event, where the sphere (a) is partitioned into two hemispheres, the cylinder (b) is separated along the dashed lines at fixed $\phi$ and $\phi+\pi$, and the ring (c) is divided in half.}
\label{fig:dijets}
\end{figure}

There are three components to the event isotropy observable:  the energy weight $w_i$, the ground measure $d_{ij}$, and the quasi-uniform event $ \mathcal{U}^\text{geo}_n$.
For a given geometry, $\text{EMD}_\text{geo}$ is computed via \Eq{eq:EMDgeneral}, where the energy weight determines the transportation plan $f_{ij}$ and the ground measure fixes $d_{ij}$.
The quasi-uniform event $\mathcal{U}_n^\text{geo}$ consists of $n$ massless particles of equal energies arranged as uniformly as possible according to the appropriate geometry, as shown in \Fig{fig:geoVis}.
By construction, event isotropy is 0 for a perfectly uniform event (up to finite $n$ effects), and goes to 1 for the least isotropic events shown in \Fig{fig:dijets}, where we have limited our attention to event configurations with net zero (transverse) momentum.
When making geometry- or $n$-independent statements, we drop the labels for ease of notation.

We consider three geometries in this paper---spherical, cylindrical, and ring-like---as summarized in \Tab{tab:emdSpec}. 
Although we only present results for one distance measure $d_{ij}$ per geometry, we are free to consider other measures. 
For example, an alternative distance measure for the spherical case is presented in \App{app:dijetSphere} and \Ref{cesarotti:2020mm}.
\begin{itemize}
\item \textbf{Spherical}.  The spherically symmetric geometry is intended for use at $e^+e^-$ colliders.
Detectors at $e^+e^-$ colliders are approximately hermetic in solid angle, thus the total event energy can be accurately reconstructed.
We take the weight of each particle to be its fraction of the total energy, and the ground measure to be the squared chord length on the sphere:
\begin{equation}
\label{eq:spherical_metric}
\boxed{\text{Spherical:}} \qquad  w_i = \frac{E_i}{E_\text{tot}}, \qquad d_{ij} = 2 \left( 1 - \cos \theta_{ij} \right),
\end{equation}
where $\theta_{ij}$ is the opening angle between the particles $i$ and $j$.
This distance measure was chosen to match the behavior of thrust (see \Eq{eq:thrust_def} below).
Though $d_{ij}$ is not a proper metric, it satisfies the modified triangle inequality in \Eq{eq:pwasserstein} with $\beta = 2$, and the fact that it has a quadratic (rather than linear) penalty for small $\theta_{ij}$ makes it less sensitive to finite $n$ effects.
To construct a uniform spherical event $\mathcal{U}_n^\text{sph}$, we use \texttt{HEALPix} \cite{Gorski:2004by}, which generates equal-area tilings of the sphere for specific values of $n$:
\begin{equation}
\label{eq:HEALPixN}
n = 12\times{2}^{2i}, \qquad i \in \mathbb{Z}_+.
\end{equation} 
A few examples of these tilings are shown in the top row of \Fig{fig:geoVis}, where each particle carries weight $\frac{1}{n}$.

The normalization of $d_{ij}$ has been chosen such that the maximum value of $\iso{sph}{}$ is 1.
Limiting our attention to event configuration with net zero momentum, the furthest momentum-conserving configuration from isotropic is two particles with equal momentum in opposite directions.
This idealized two-particle configuration is denoted $\mathcal{E}^\text{sph}_2$ and shown in \Fig{fig:dijets_sph}.
Using standard spherical coordinates, the event isotropy of $\mathcal{E}^\text{sph}_2$ is 1 as desired:
\begin{equation}
\label{eq:computeMaxIsoSpherical}
\iso{sph}{\infty}(\mathcal{E}^\text{sph}_2)= 2 \int_0^{\pi/2} \int_0^{2\pi} \frac{\sin \theta \, \text{d} \theta \, \text{d} \phi}{4\pi} \ 2 \left( 1 - \cos \theta \right) = 1,
\end{equation}
where we have chosen a coordinate system such that the two particles are oriented along the $z$-axis.
The overall factor of 2 is because the transportation plan maps the two particles to their corresponding hemispheres.

\item \textbf{Cylindrical}.   Detectors at $pp$ colliders typically have a cylindrical geometry such that inelastic scattering, with large particle yields in the forward direction, do not overwhelm the detector. 
This motivates axially symmetric geometries to model uniform radiation patterns.
The first $pp$ geometry we consider is a cylinder centered on the central region of a detector.
Since total longitudinal momentum cannot be accurately reconstructed, we take the weight of each particle to be its transverse momentum fraction, and we use the squared rapidity-azimuth distance as the ground measure:
\begin{equation}
\label{eq:cylindrical_metric}
\boxed{\text{Cylindrical:}} \qquad  w_i = \frac{p_{T_i}}{p_{T, \text{tot}}}, \qquad d_{ij} = \frac{12}{\pi^2 +16 \, y_\text{max}^2} \left( y_{ij}^2 + \phi_{ij}^2 \right),
\end{equation}
where $p_{T, \text{tot}} = \sum_j p_{T_j}$ includes all reconstructed particles with $|y_j| < y_\text{max}$, and $y_{ij} = |y_i - y_j|$ and $\phi_{ij} = |\phi_i -\phi_j|$ are the pairwise rapidity and azimuthal distances.  
In analogy to the spherical case, we use the squared (as opposed to absolute) $y$--$\phi$ distance in order to penalize large distances more than small ones in the EMD computation. 
To construct a uniform cylindrical event $\mathcal{U}_n^\text{cyl}(y_{\rm max})$ with equal tiling of $y$--$\phi$ space up to the maximum absolute rapidity $y_\text{max}$, we first divide the azimuthal direction of the cylinder into $2^i$ equal segments, for $i \in \mathbb{Z}_+$.
This leads to a total of $n \approx 2^{2i} y_\text{max}/\pi$ equal-weight particles, as shown in the middle row of \Fig{fig:geoVis}.

Restricting our attention to configurations that conserve transverse momentum, the radiation pattern furthest from the uniform cylindrical event is the two-particle configuration $\mathcal{E}^\text{cyl}_2$ with equal and opposite transverse momenta at the same extremal rapidity value, as shown in \Fig{fig:dijets_cyl}. 
For two particles located at $\left(y_\text{max},\pm \frac{\pi}{2}\right)$, we can compute the event isotropy by splitting the cylinder in half along the $\phi$ direction:
\begin{equation}
\iso{cyl}{\infty}( \mathcal{E}^\text{cyl}_2)= 2 \int_{0}^\pi \int_{- y_\text{max}}^{y_\text{max}} \frac{\text{d} \phi \,  \text{d} y}{4\pi y_\text{max}} \frac{12}{\pi^2 + 16 \, y_\text{max}^2}  \left( \left(\frac{\pi}{2}-\phi\right)^2 + \left(y_\text{max}-y\right)^2 \right) = 1,
\end{equation}
such that cylindrical isotropy is bounded by 1 as desired.

\item \textbf{Ring-like}.   Hard scattering processes at $pp$ colliders are typically longitudinally boosted, explicitly breaking the translational symmetry along the beamline.
We therefore consider a ring-like geometry that marginalizes over the rapidity direction: 
\begin{equation}
\label{eq:ring_metric}
\boxed{\text{Ring-like:}} \qquad  w_i = \frac{p_{T_i}}{p_{T, \text{tot}}}, \qquad d_{ij} = \frac{\pi}{\pi -2} \left( 1 - \cos \phi_{ij}\right),
\end{equation}
where $\phi_{ij}$ is the transversely projected opening angle.
We have chosen to use the squared chord length on the ring (instead of just $\phi_{ij}^2$) in order to match the behavior of transverse thrust \cite{Bertram:2002sv,Nagy:2003tz,Banfi:2004nk,Banfi:2010xy} (see \Eq{eq:transthrust_def} below).
As shown in the bottom row of \Fig{fig:geoVis}, a uniform ring-like event $\mathcal{U}_n^\text{ring}$ is tiled in the same way as the cylinder: $2^i$ equally spaced particles around the $\phi$ direction. 
Any two particle event $\mathcal{E}^\text{ring}_2$ that conserves transverse momentum will maximize the event isotropy, as illustrated in \Fig{fig:dijets_ring}. 
For two particles oriented along $\phi = 0$, the event isotropy is saturated by 1 as desired:
\begin{equation}
\iso{ring}{\infty}( \mathcal{E}^\text{ring}_2)= 2 \times \int_{-\frac{\pi}{2}}^{\frac{\pi}{2}} \frac{\text{d} \phi}{2\pi} \frac{\pi}{\pi-2} \left( 1 - \cos \phi \right) = 1.
\end{equation}

\end{itemize}

For practical event isotropy calculations with $n$-particle uniform configurations, we need to choose $n$ large enough to mitigate the effect of finite tiling.
Unless otherwise specified, we take the default of $n=192$ for the spherical geometry, to balance computation time against performance. 
For the default cylindrical case, we set $y_\text{max} = 2$ and slice the $\phi$ direction into 16 segments, which translates to $n=160$.
The default ring-like case has $n=32$.%
\footnote{As noted in \Ref{Komiske:2020qhg}, there are efficient algorithms to solve one-dimensional optimal transport problems~\cite{Rabin:2011jd}.
Though we use the POT library for this paper, faster algorithms could enable the implementation of ring-like event isotropy at the trigger level.}
Because our $d_{ij}$ are based on squared distances, the results we present are largely insensitive to the precise choice of $n$.
For events that do not conserve transverse momentum due to the finite rapidity range of the selection, we add in an extra ``particle'' corresponding to the missing $p_T$ vector before computing cylindrical or ring-like event isotropy.


\subsection{Comparison to Existing Event Shape Observables}
\label{subsec:compToEventShape}

\begin{table*}[!t]
\centering
\begin{tabular}{ c @{\quad} c @{\quad} c} 
\hline \hline
Collider & Observable & Rescaled Definition \\
\hline 
$e^+e^-$ & Thrust & $\widetilde{T} \equiv 2\,T-1$ \\
$e^+e^-$ &Sphericity & $\widetilde{S} \equiv 1 - S$ \\
$e^+e^-$ &$C$-parameter & $\widetilde{C} \equiv 1 - C$ \\
\hline
$pp$ & Transverse Thrust &  $\widetilde{T}_\perp \equiv \frac{\pi}{\pi-2}\left(T_\perp-\frac{\pi}{2}\right)$ \\
 \hline
 \hline
 \end{tabular}
\caption{
Previous event shape observables and their rescaled variants.
Like event isotropy $\mathcal{I}$, the rescaled event shapes range from [0,1], where $0$ is perfectly isotropic.
}
\label{tab:obsSum}
\end{table*}

To quantify the performance of event isotropy to identify quasi-uniform particle configurations, it is instructive to compare its behavior to existing event shape observables.
We focus primarily on the spherical geometry, where we compare to thrust, sphericity, and the $C$-parameter.
We also consider transverse thrust for the ring-like geometry.
To compare more directly to event isotropy, it is convenient to define rescaled event shapes that range from [0,1], with 0 corresponding to most isotropic.
These rescaling relations are summarized in \Tab{tab:obsSum}.

We begin by considering thrust $T$, which is defined as~\cite{Brandt:1964sa,Farhi:1977sg,DeRujula:1978vmq}: 
\begin{equation}
\label{eq:thrust_def}
T(\mathcal{E}) = \max_{\hat{n}} \frac{\sum_i | \vec{p}_i \cdot \hat{n}| }{\sum_j |\vec{p}_j|},
\end{equation} 
where $\vec{p}_i$ is the three-momentum of each particle in the event, and $\hat{n}$ is the unit vector that maximizes the value of $T$, called the thrust axis.
The values of $T$ fall in the range $[\frac{1}{2},1]$, where a perfectly spherical event ($\mathcal{U}^\text{sph}_{\infty}$) has thrust $T=\frac{1}{2}$ and two back-to-back particles ($\mathcal{E}^\text{sph}_2$) has thrust $T=1$. 
The rescaled thrust $\widetilde{T}$ in \Tab{tab:obsSum} is:
\begin{equation}
\widetilde{T} \equiv 2\, T-1,
\end{equation}
such that its range is [0,1].

As recently shown in \Ref{Komiske:2020qhg}, thrust can be written in the language of the EMD.\footnote{We thank Patrick Komiske and Eric Metodiev for useful discussions on this point.}
Using the spherical geometry from \Eq{eq:spherical_metric}:
\begin{equation}
\label{eq:thrust_as_EMD}
\tilde{T}(\mathcal{E}) = 1- \min_{\mathcal{E}' \in \mathcal{P}_2^{\rm BB}} \Big[\text{EMD}_\text{sph} \left(\mathcal{E}, \mathcal{E}' \right) \Big],
\end{equation}
where $\mathcal{P}_2^{\rm BB}$ is the set of all back-to-back massless two-particle events.%
\footnote{One potential confusion is that $\mathcal{P}_2^{\rm BB}$ includes unphysical configurations in which the two massless particles have unequal energies.  As discussed in \Ref{Komiske:2020qhg}, there is an alternative formulation based on time-like (instead of light-like) directions that avoids this confusion, though it does not exactly mimic the behavior of thrust.}
This remarkable definition of thrust in terms of the EMD motivates our choice of spherical ground measure for this paper.

For a perfectly isotropic event, $T(\mathcal{U}^\text{sph}_\infty) = \frac{1}{2}$, so thrust offers a useful test for isotropic configurations.
In fact, thrust is a true measure of isotropy, since $T$ only equals $\frac{1}{2}$ (i.e.\ $\widetilde{T} = 0$) for isotropic configurations.
To prove this, consider an observable $t$ depending on an arbitrary unit vector $\hat{n}$: 
\begin{equation}
t\left(\hat{n}; \mathcal{E}\right) = \frac{\sum_i | \vec{p}_i \cdot \hat{n}| }{\sum_j |\vec{p}_j|}.
\end{equation}
The thrust of an event $T$ is equal to $t$ maximized over all $\hat{n}$.
If we average over $\hat{n}$ instead, we find
\begin{equation}
\langle t(\hat{n}; \mathcal{E}) \rangle_{\hat{n}} = \frac{\sum_i  \langle | \vec{p}_i \cdot \hat{n}| \rangle_{\hat{n}}}{\sum_j |\vec{p}_j|} = \frac{1}{2},
\end{equation}
where we have used the fact that the average projected length of a vector $\vec{p}_i$ in three dimensions is $|\vec{p}_i|/2$.
Since the average value $\langle t \rangle =  \frac{1}{2}$, then if $t(\hat{n}; \mathcal{E}) <  \frac{1}{2}$ for some $\hat{n}$, there exists an $\hat{n}'$ for which $t(\hat{n}';  \mathcal{E}) >  \frac{1}{2}$, and therefore $\hat{n}$ is not the thrust axis. 
The only way $T$ can equal $\frac{1}{2}$ is if $t =  \frac{1}{2}$ for all $\hat{n}$, which implies that the event configuration is perfectly isotropic.

\begin{figure}[t!]
\centering
 \includegraphics[width=0.55\textwidth]{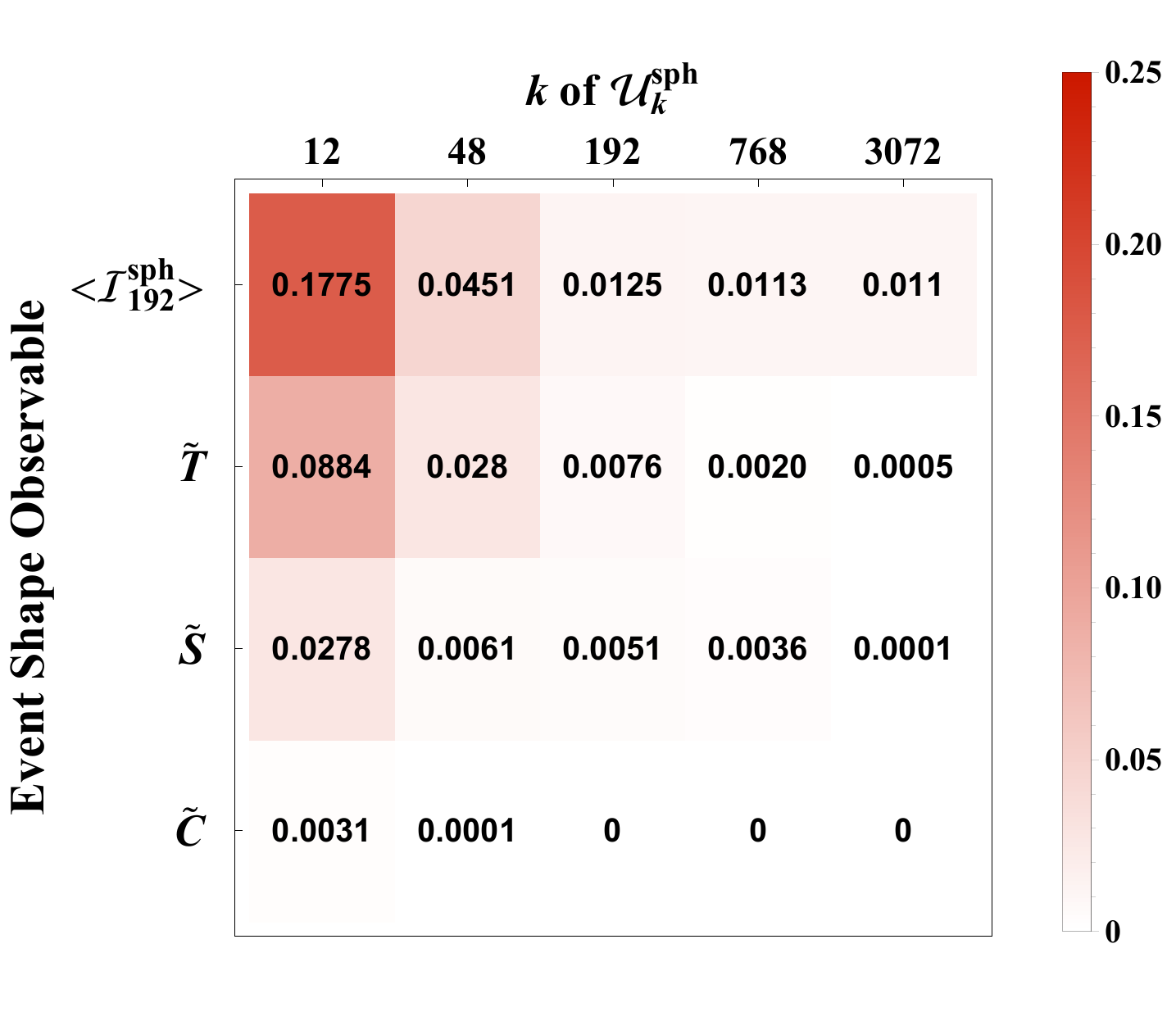}
\caption{Event shapes evaluated on the \texttt{HEALPix} quasi-isotropic configurations with $k=\{12,48,192,768,3072\}$.
We show the rescaled observables from \Tab{tab:obsSum} such that a value near zero indicates a quasi-isotropic event.
Observables with darker shading indicate more dynamic range.
We average the value of event isotropy over $10^3$ random orientations.
While thrust has a larger dynamic range than sphericity, note that it counterintuitively ranks a ring plus back-to-back particle configuration (\Fig{subfig:ringDijet}, $\mathcal{T} = 0.074$) as slightly more isotropic than an $k = 12$ sphere tiling ($\mathcal{T} = 0.088$).
For additional values $\iso{sph}{n}$ beyond $n = 192$, see \Fig{fig:eeSphereToSphere}.}
\label{fig:obsOnSpher}
\end{figure}

\begin{figure}[t!]
\centering
 \subfloat[]{
 \label{subfig:ringDijet}
\includegraphics[width=0.35\textwidth]{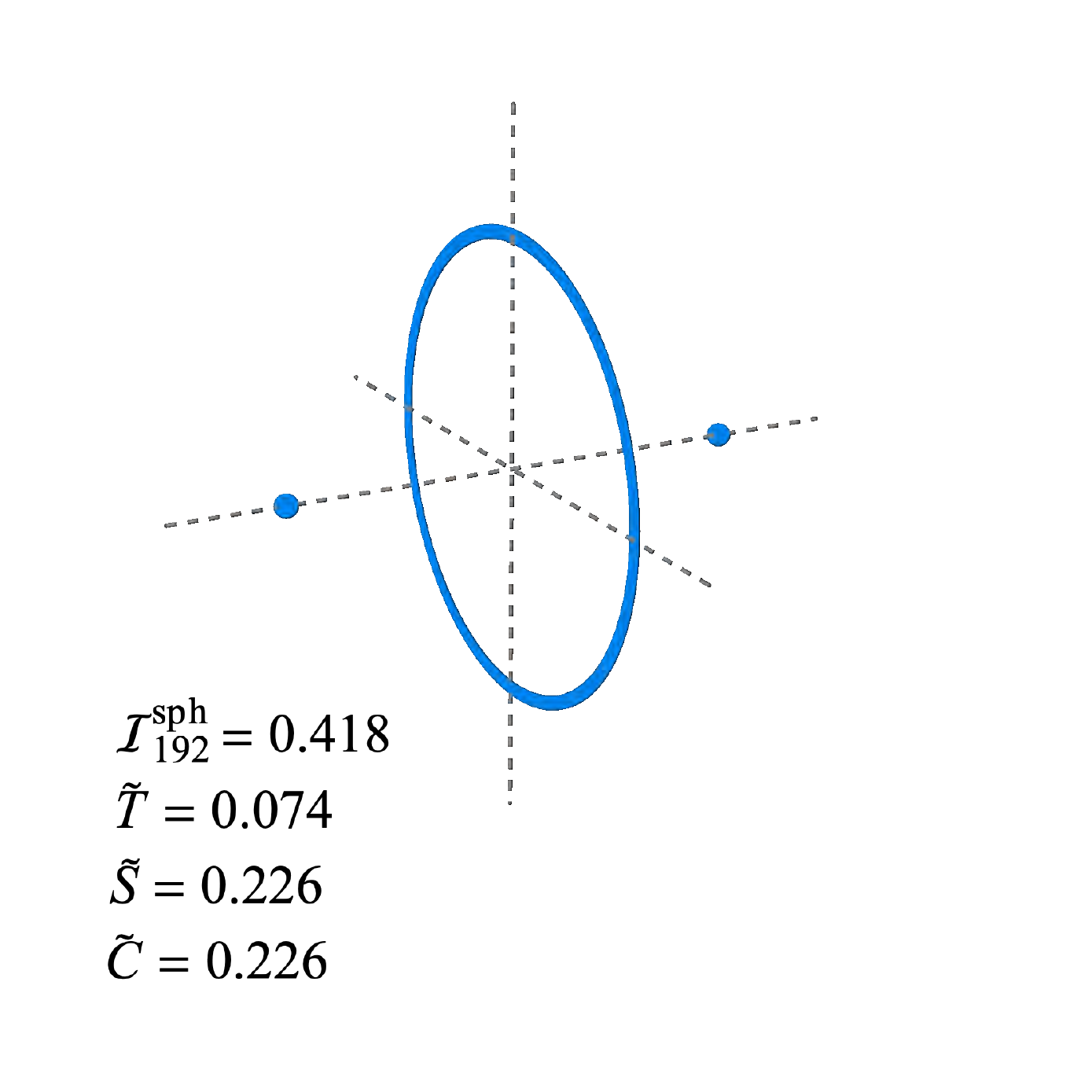}
 }
$\qquad$
\subfloat[]{
\label{fig:dijetThrust_6particle}
       \includegraphics[width=0.35\textwidth]{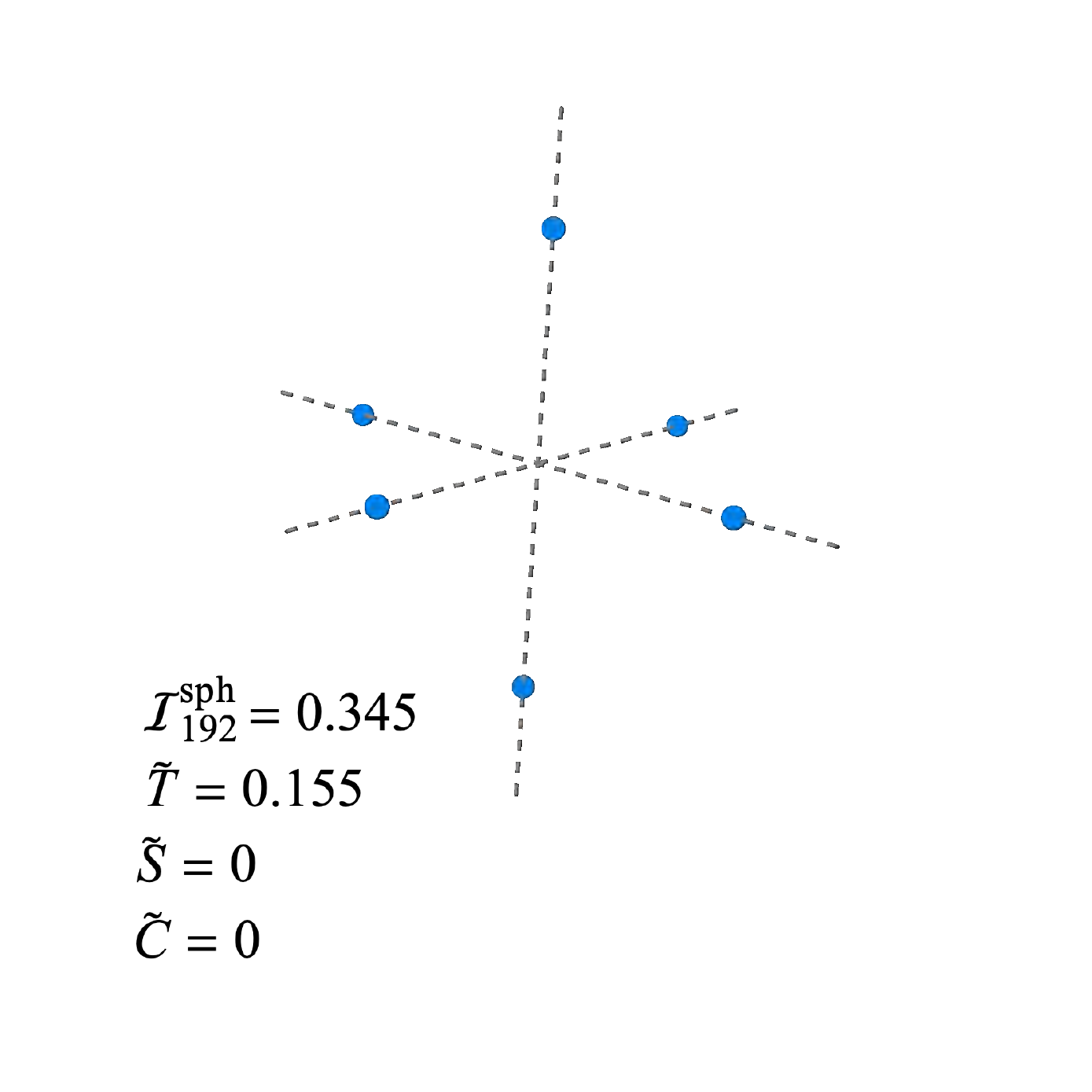}
     }
\caption{Event configurations that (nearly) saturate traditional event shape observables.
The blue points represent the particles in momentum space.
(a) An anisotropic event with small thrust.  The thrust is minimized when the summed momentum of the ring ($p_{\rm ring}$) is related to the momenta of each of the individual particles $p_{\rm part}$ by $p_{\rm part} = \frac{4}{\pi} p_{\rm ring}$.
(b) A six particle configuration that saturates the spherical bound for both sphericity and the $C$-parameter.}
\label{fig:dijetThrust}
\end{figure}
Although $\widetilde{T} = 0$ only for exactly isotropic events, thrust has a relatively small dynamic range for quasi-isotropic configurations, meaning that the same value of thrust can arise from many different event topologies in this regime.
More specifically, distributions in scaled thrust $\widetilde{T}$ for different quasi-isotropic samples generally overlap, which results in a loss of discrimination power compared to using event isotropy $\mathcal{I}^\text{sph}$.
As an example, the values of rescaled thrust for the first five tilings of a sphere from \texttt{HEALPix} are shown in \Fig{fig:obsOnSpher}, where a 12-particle tiling has $\widetilde{T}=0.0884$. 
Another configuration that yields a comparable value of rescaled thrust is shown in \Fig{subfig:ringDijet}, where two back-to-back particles plus a uniform ring of radiation can get as small as $\widetilde{T}=0.0740$.
However, the event isotropy of these configurations are $\mathcal{I}^\text{sph}_{192} = 0.178$ and $\mathcal{I}^\text{sph}_{192} = 0.418$, respectively, which are numerically quite distinct.
As we explore further in \Secs{sec:emdBE}{sec:ppBE}, these raw numerical differences between individual events translate into improved discrimination power between ensembles of events.

Two other well-known $e^+e^-$ event shape observables are sphericity ($S$)~\cite{Bjorken:1969wi,Ellis:1976uc} and the $C$-parameter~\cite{Parisi:1978eg,Donoghue:1979vi,Ellis:1980wv}.
They are defined such that an idealized dijet event has $S, C = 0$ and a perfectly spherical event has $S, C = 1$.  This is the opposite behavior to event isotropy, so we introduce flipped versions in \Tab{tab:obsSum}:
\begin{equation}
\widetilde{S} \equiv 1 - S, \qquad \widetilde{C} \equiv 1 - C.
\end{equation}
Both of these observables are computed via the generalized sphericity tensor:
\begin{equation}
S^{(r) \alpha \beta} = \frac{\sum_i |p_i|^{r-2}p_i^\alpha p_i^\beta }{\sum_i |p_i|^r}, \qquad \alpha, \beta \in \{1,2,3\},
\label{eq:spherMat}
\end{equation}
where sphericity is traditionally computed with $r=2$, and the $C$-parameter with $r=1$. 
Let $\lambda_1 \geq \lambda_2 \geq \lambda_3$ be the eigenvalues of the tensor in \Eq{eq:spherMat}. 
The sphericity and $C$-parameter of an event are defined as:
\begin{align}
r=2: \qquad S &= \frac{3}{2}\left( \lambda_2 + \lambda_3 \right), \nonumber \\
r=1: \qquad C &= 3 \left(\lambda_1 \lambda_2 + \lambda_1 \lambda_3 + \lambda_2 \lambda_3 \right).
\label{eq:sandc}
\end{align}
We see that $S$ and $C$ are maximized whenever there is symmetry between all orthogonal directions, i.e.\ when the eigenvalues are equal. 
This means that an event configuration as simple as \Fig{fig:dijetThrust_6particle}, with six particles aligned along the coordinate axes, can have $\widetilde{S}, \widetilde{C} = 0$. 
In this sense, sphericity and $C$-parameter are not true measures of isotropy, but rather measures of symmetry along the axes. 
The dynamic range of $\widetilde{C}$ is very poor, as shown in \Fig{fig:obsOnSpher}, so we only include $\widetilde{S}$ in our comparisons in \Sec{sec:emdBE}.

Finally, for $pp$ colliders, a well-known event shape is transverse thrust $T_\perp$~\cite{Bertram:2002sv,Nagy:2003tz,Banfi:2004nk,Banfi:2010xy}, which is a probe of ring-like configurations.
The transverse thrust of an event is:
\begin{equation}
\label{eq:transthrust_def}
T_\perp(\mathcal{E}) = \max_{\hat{n}_\perp} \frac{\sum_i | \vec{p}_{T,i} \cdot \hat{n}_\perp | }{\sum |\vec{p}_{T,i}|},
\end{equation}
where $\hat{n}_\perp$ is a unit vector constrained to the transverse plane. 
The value of transverse thrust ranges from $[2/\pi, 1]$, so the rescaled version in \Tab{tab:obsSum} is
\begin{equation}
\widetilde{T}_\perp \equiv \frac{\pi}{\pi-2}\left( T_\perp - \frac{2}{\pi}\right). 
\end{equation}

\section{Benchmark Scenarios: Electron-Positron Colliders}
\label{sec:emdBE}

In this section, we study the performance of event isotropy at separating various signal and background processes, benchmarking it against traditional event shape observables.
We consider electron-positron ($e^+e^-$) collisions, such that the dominant QCD background process consists mostly of two or three pronged events from light-quark pair production ($e^+ e^- \to q\bar{q}$ plus gluon radiation), including possible photon initial state radiation (ISR) prior to the hard scattering process. 
These QCD events are collimated, so we expect event isotropy to be effective at rejecting these events in favor of more uniform radiation patterns.

We study three benchmark scenarios that generate signals with quasi-uniform energy distributions.
First, we consider top-quark pair production ($e^+ e^- \to t\bar{t}$) as a SM process that yields 6-prong event configurations.
Second, we consider events distributed according to uniform $N$-body phase space via the RAMBO algorithm \cite{Kleiss:1985gy} as a model of new physics with relatively high multiplicity events. 
Finally, we consider the \texttt{HEALPix} equal area configurations shown already in \Fig{fig:geoVis} as an idealized model for isotropic new physics scenarios.

\subsection{Top Pair Production vs.\ QCD Dijet}
\label{subsec:top_ee}

\begin{figure}[t]
\centering
\subfloat[\label{subfig:qq}]{
       \includegraphics[width=0.3\textwidth]{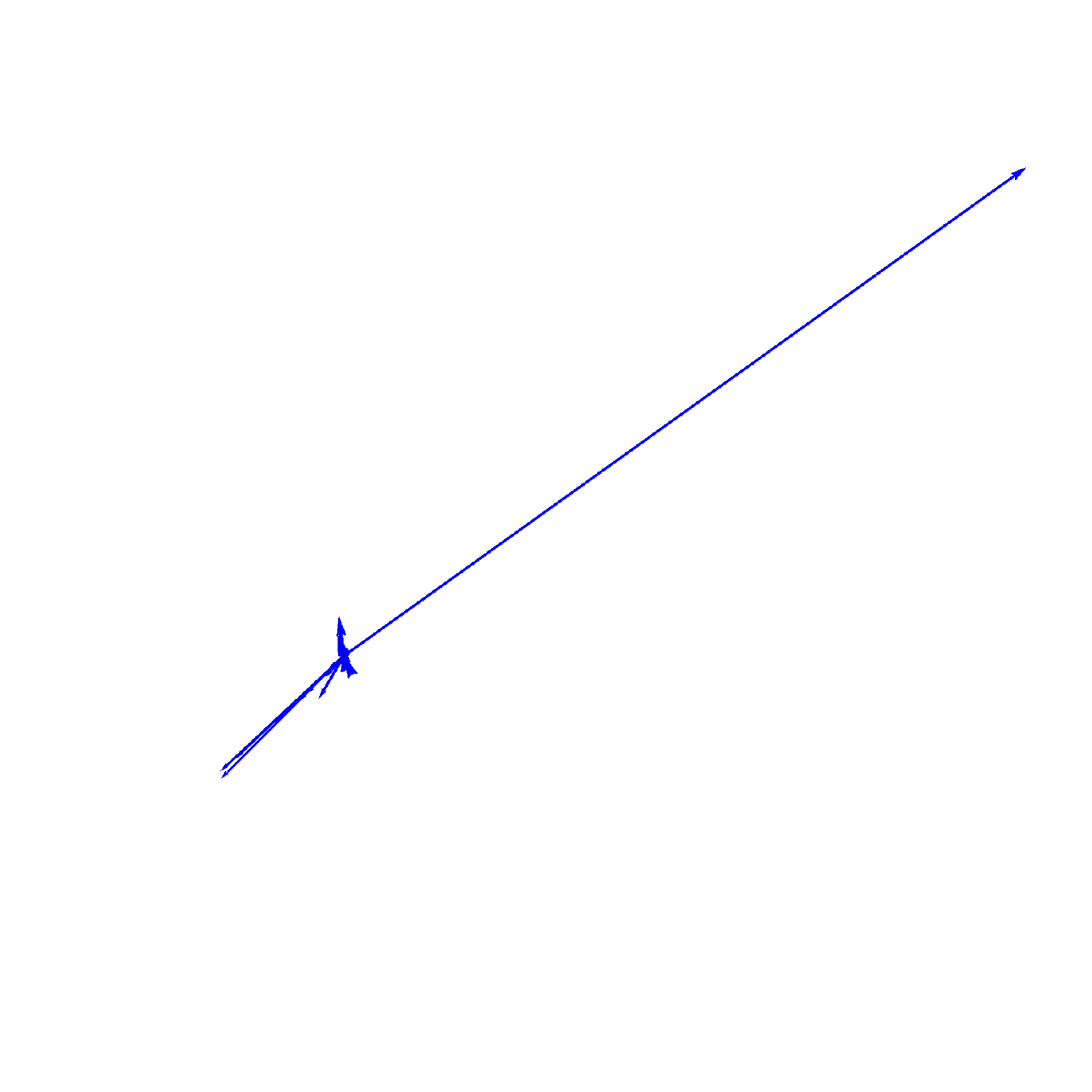}
     }  
$\qquad$
\subfloat[\label{subfig:tt}]{
       \includegraphics[width=0.3\textwidth]{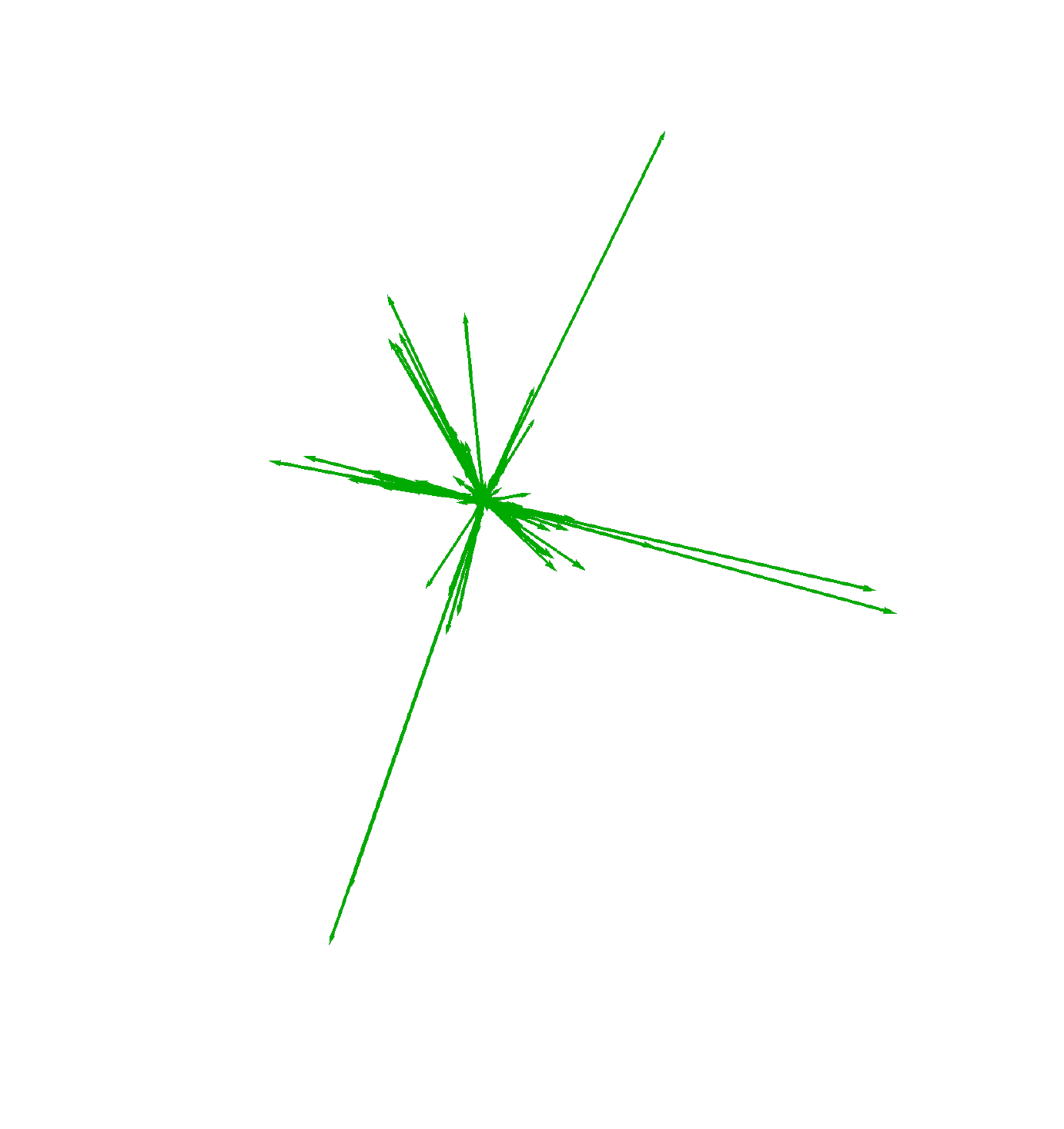}
     }
          \hfill
\caption{Examples of $e^+e^-$ collisions at $\sqrt{s} = 350$ GeV yielding (a) a $q\bar{q}$ event plus a hard ISR photon and (b) a $t\bar{t}$ event.
The events are shown in momentum space, and the magnitude of each vector is proportional to the energy of the corresponding particle.}
\label{fig:eeVis}
\end{figure}

Let us begin with top-quark pair production via the hard process $e^+e^- \rightarrow t \bar{t}$.
This is a SM process that often produces six jets, since in the all-hadronic channel, each top quark decays to three partons via $t\rightarrow b W^+ \rightarrow b q \bar{q}'$. 
Both the $t\bar{t}$ and $q\bar{q}$ samples are generated with \texttt{Pythia 8.243} \cite{Sjostrand:2014zea}, including ISR, FSR, and hadronization. 
We turn off leptonic decays of the $W$ boson to remove the subset of $t\bar{t}$ events easily tagged with missing energy or electron/muon signatures. 
Events are generated at a center-of-mass collision energy near the $t\bar{t}$ threshold of $\sqrt{s} = 350$ GeV, a proposed run configuration of a future circular collider \cite{Abada:2019zxq}. 
At this collision energy, the tops are unboosted, thus the direction of their decay products are quasi-isotropic. 
This process is therefore a good benchmark for moderate multiplicity and relatively isotropic signal processes.
Visualizations of typical $q\bar{q}$ and $t\bar{t}$ events are shown in \Fig{fig:eeVis}.

\begin{figure}[t!]
\subfloat[]{
       \includegraphics[width=0.45\textwidth]{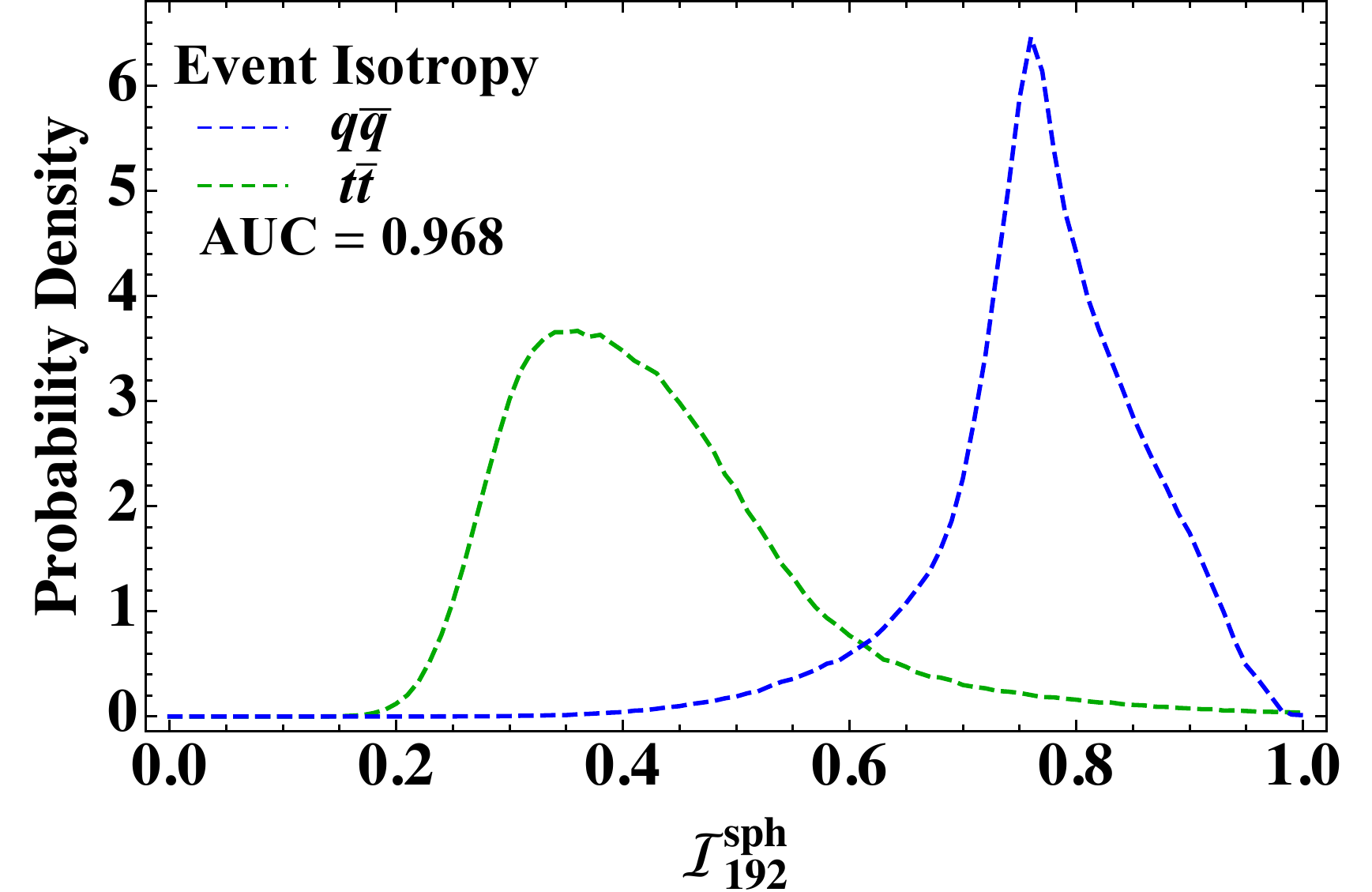}
       	\label{fig:eeSpec_iso}
     }
     \hfill
     \subfloat[]{
       \includegraphics[width=0.45\textwidth]{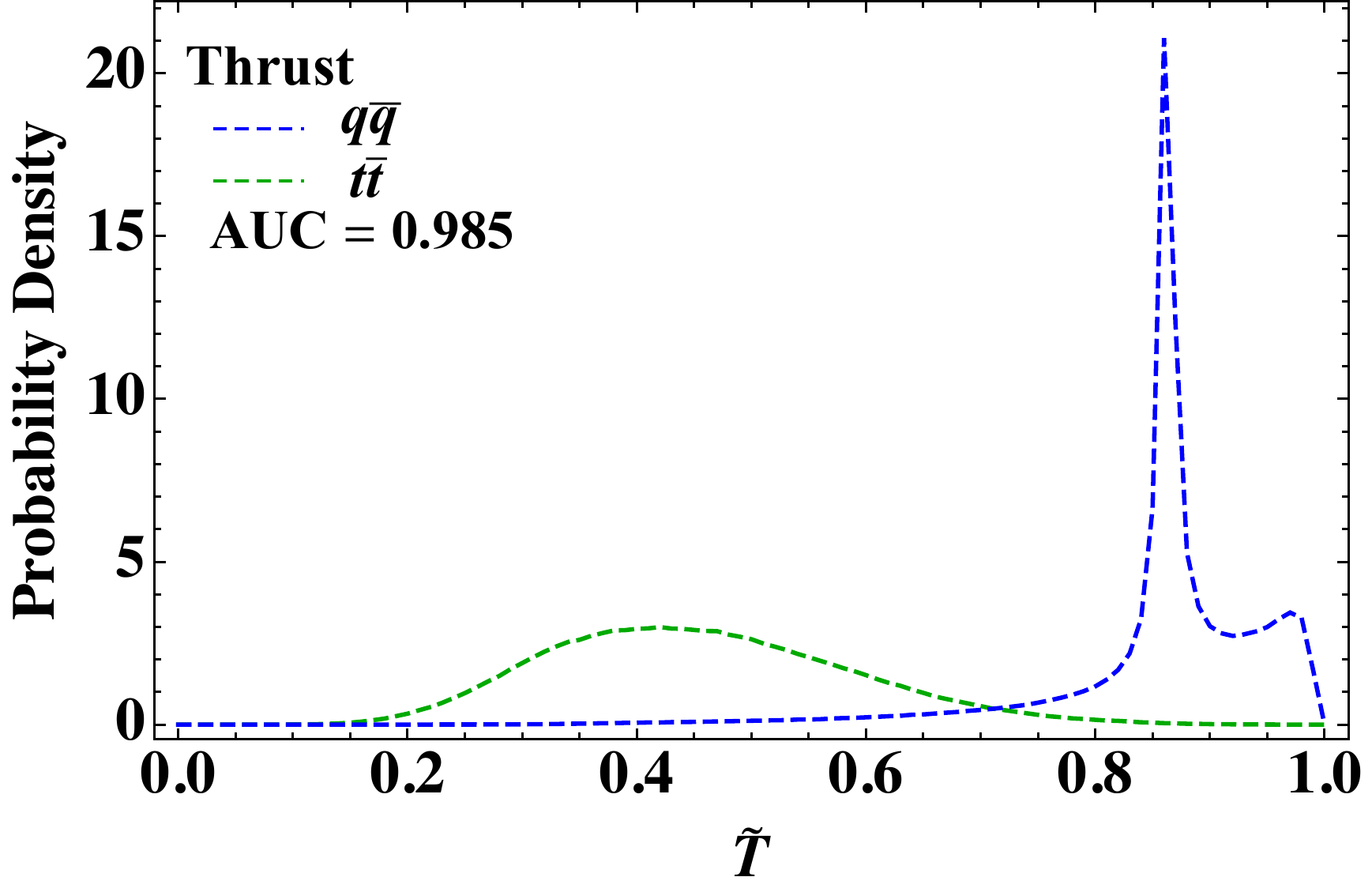}
     }
     \hfill
     \subfloat[]{
       \includegraphics[width=0.45\textwidth]{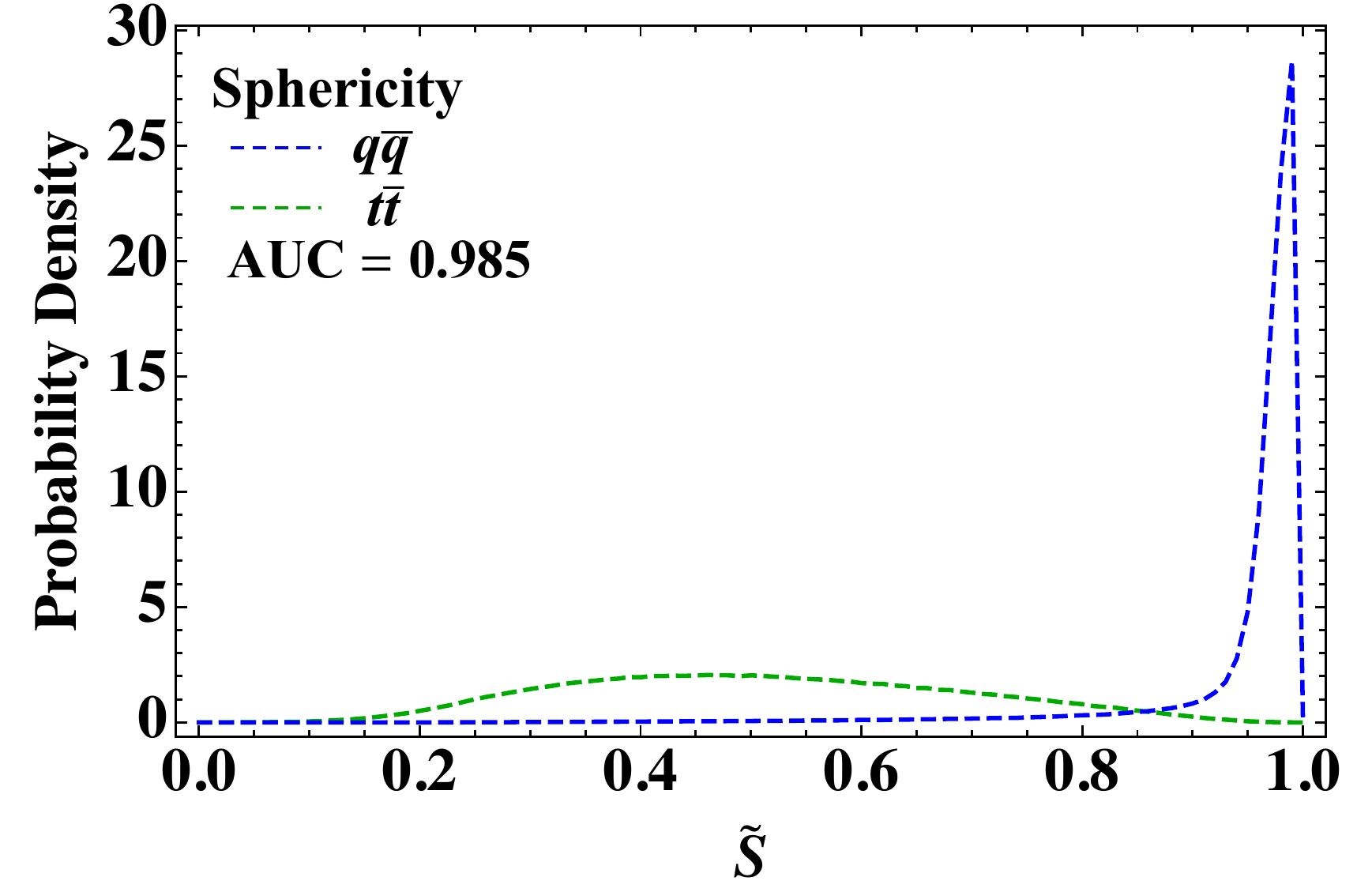}
     }
     \hfill
     \subfloat[]{
       \includegraphics[width=0.45\textwidth]{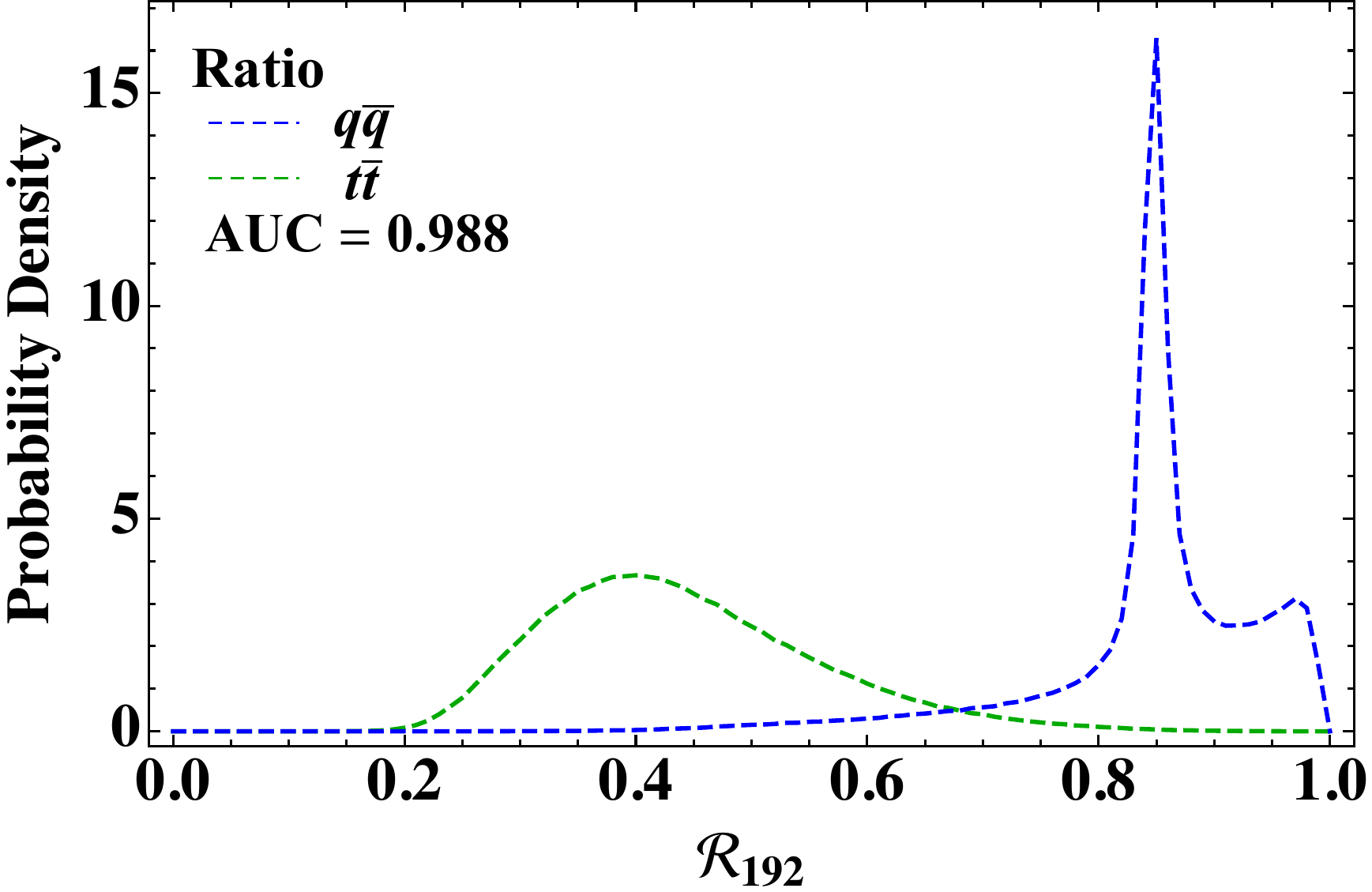}
     }
\caption{The distributions of four event shape observables in $e^+e^-$ collisions at $\sqrt{s} = 350$ GeV:  (a) event isotropy with $n = 192$, (b) thrust, (c) sphericity, and (d) the isotropy/thrust ratio. 
For all observables, the $q\bar{q}$ (blue) and $t\bar{t}$ (green) distributions are well separated, with the AUC summarized in the plot legend.}
\label{fig:eeSpec}
\end{figure}

In the following study, we compare the discrimination power of several event shape observables:  event isotropy $\iso{sph}{192}$, thrust $\widetilde{T}$, and sphericity $\widetilde{S}$.
We also introduce an observable $\remd$ based on a combination of thrust and event isotropy:
\be
\label{eq:remd}
\remd \equiv \frac{\iso{sph}{n}}{\iso{sph}{n} + (1-\widetilde{T})}.
\ee
The motivation for \Eq{eq:remd} is that event isotropy measures the distance to a perfectly isotropic event while thrust measures the distance to an idealized dijet configuration with two back-to-back particles, so $\remd$ quantifies the relative distance to these idealized configurations. 
All four observables we consider run from 0 (for an isotropic event) to 1 (for two back-to-back particles).

The distributions of these event shape observables are shown in \Fig{fig:eeSpec}, comparing $10^6$ events each from the $q\bar{q}$ and $t\bar{t}$ samples.
The spike near $\widetilde{T} \simeq 0.85$ in the $q \bar{q}$ sample is due to the radiative return process, since the collision energy $\sqrt{s}$ is above the $Z$ resonance.
All four observables show good qualitative separation power between the $q\bar{q}$ and $t\bar{t}$ samples.

To quantify the overall discrimination power, we calculate the area under the curve (AUC).
The AUC is the area under the Receiver Operating Characteristic (ROC) curve, but more simply, it is the probability that a binary classifier will correctly order a random signal event and a random background event:
\begin{equation}
\text{AUC} = P\left(\mathcal{O}_{\rm signal} < \mathcal{O}_{\rm background} \right),
\end{equation}
where $\mathcal{O}_{\rm signal}$ ($\mathcal{O}_{\rm background}$) is the observable value for the signal (background) event.
An AUC value of 1 indicates perfect separation, while $\text{AUC} = 0.5$ corresponds to random guessing.
Since the AUC is insensitive to monotonic rescalings of an observable, it is a reliable measure of an observable's dynamic range for distinguishing event samples.
As shown in the \Fig{fig:eeSpec} labels, all four observables exhibit $\text{AUC} > 0.96$.
While thrust and sphericity outperform event isotropy, the performance is still comparable.
The best discrimination performance comes from the $\remd$ observable that combines thrust and isotropy information with $\text{AUC} = 0.988$.

While the $q\bar{q}$ and $t\bar{t}$ spectra are well separated by event isotropy, neither sample comes close to saturating the isotropic limit of $\iso{sph}{192} = 0$. 
This therefore suggests that the event isotropy will have sufficient dynamic range to not only separate $t\bar{t}$ signals from dijet backgrounds, but also separate $t\bar{t}$-like events from quasi-isotropic or fully isotropic events.

\subsection{Toy Model: Uniform $N$-body Phase Space}
\label{subsec:uniformNbody_ee}

\begin{figure}[p]
\subfloat[]{
       \includegraphics[width=0.3\textwidth]{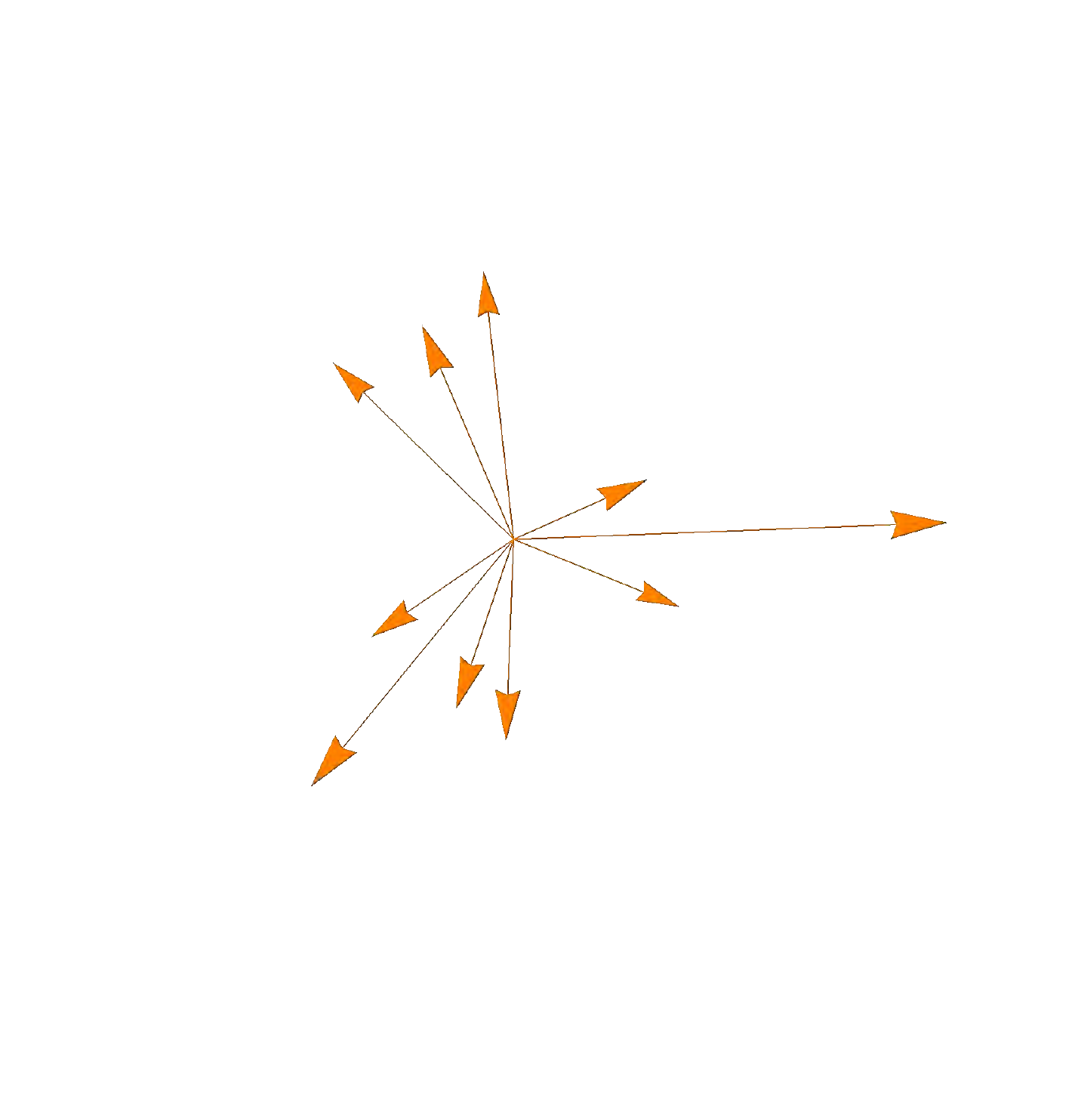}
     }
     \hfill
     \subfloat[]{
       \includegraphics[width=0.3\textwidth]{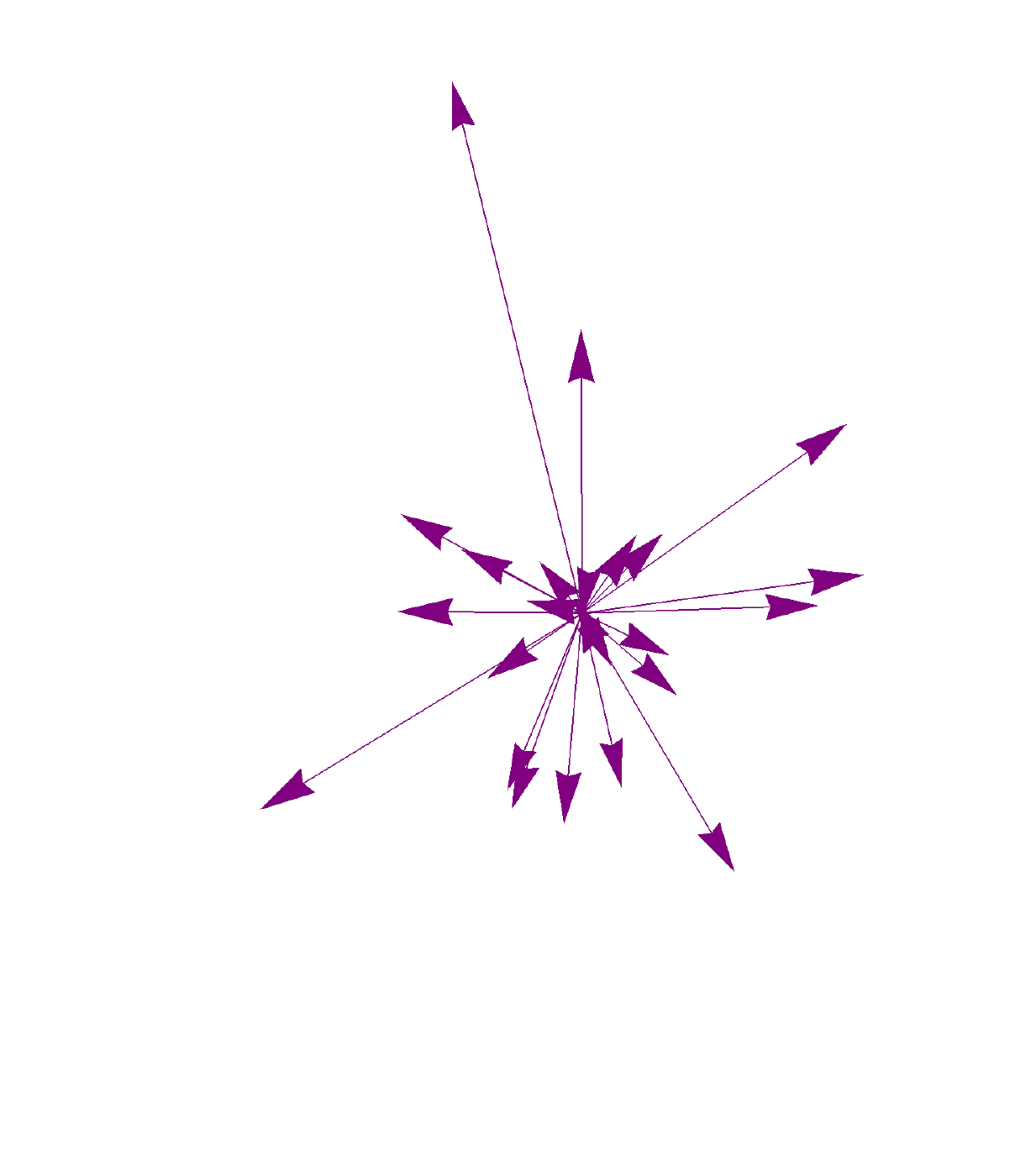}
     }
      \hfill
     \subfloat[]{
       \includegraphics[width=0.3\textwidth]{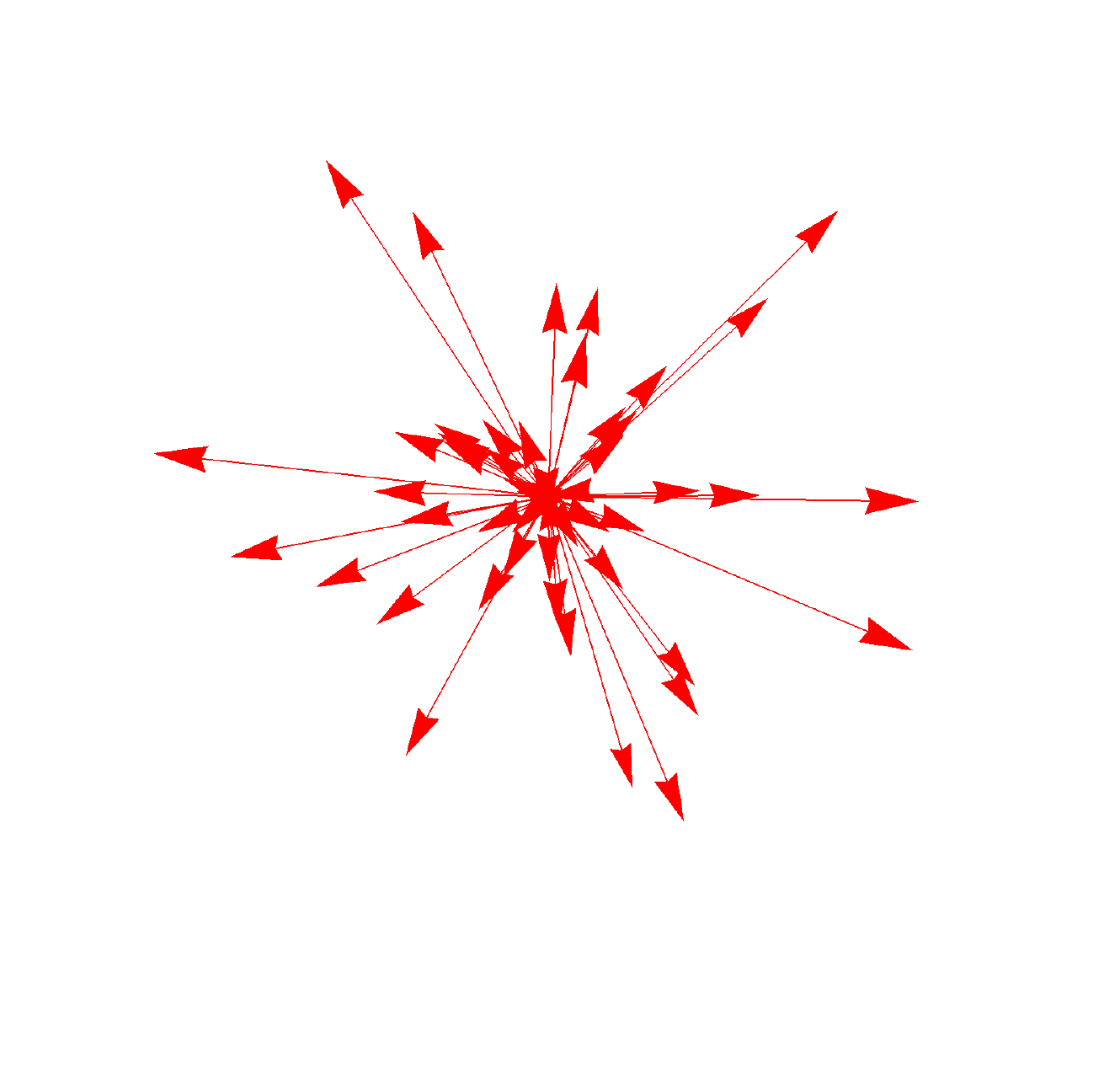}
     }
\caption{Examples of events generated according to uniform $N$-body phase space for (a) $N = 10$, (b) $N = 25$, and (c) $N = 50$.  The magnitude of each vector is proportional to the energy of the corresponding particle.}
\label{fig:nBody}
\end{figure}

Although the $t\bar{t}$ signal is a useful benchmark, it is not a particularly isotropic or high-multiplicity scattering process. 
We now consider high-multiplicity events as a toy model for new physics scenarios like RPV SUSY, where many particles with similar momenta are produced due to long cascade decays and small mass splittings. 
To generate events according to uniform $N$-body phase space, we use the RAMBO algorithm~\cite{Kleiss:1985gy} for $N=\{10, 25, 50\}$, and compare to both the $q\bar{q}$ and $t\bar{t}$ processes from \Sec{subsec:top_ee}.
The $q\bar{q}$ process is an example of a generic QCD background process.
The $t\bar{t}$ process allows us to investigate the discrimination power of event isotropy against moderate multiplicity backgrounds.
Some visualizations of typical uniform $N$-body events are shown in \Fig{fig:nBody}.

\begin{figure}[p]
\centering
\subfloat[]{
       \includegraphics[width=0.45\textwidth]{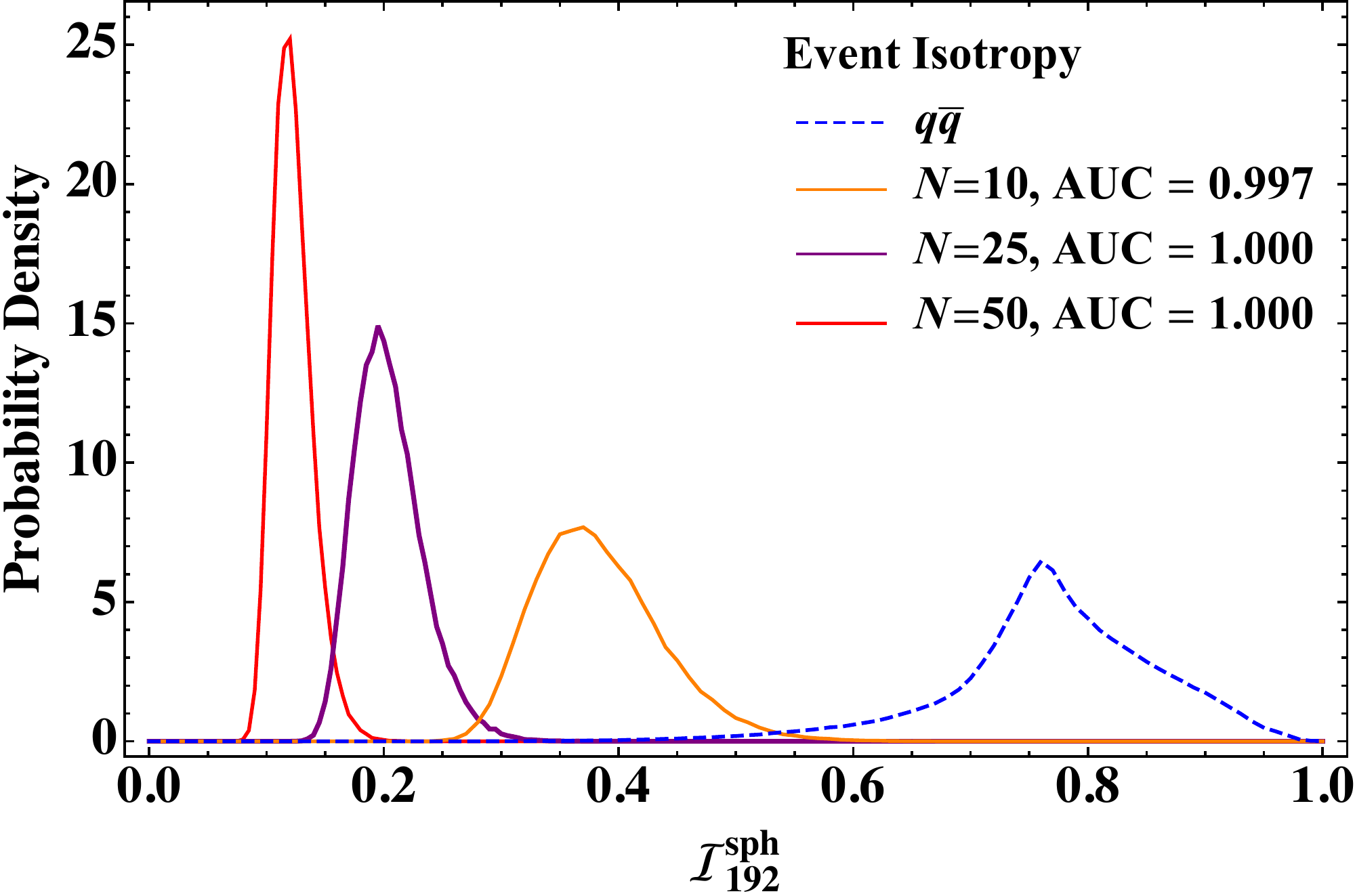}
       	\label{fig:eeRamboSpec_iso}
     }
     \hfill
     \subfloat[]{
       \includegraphics[width=0.45\textwidth]{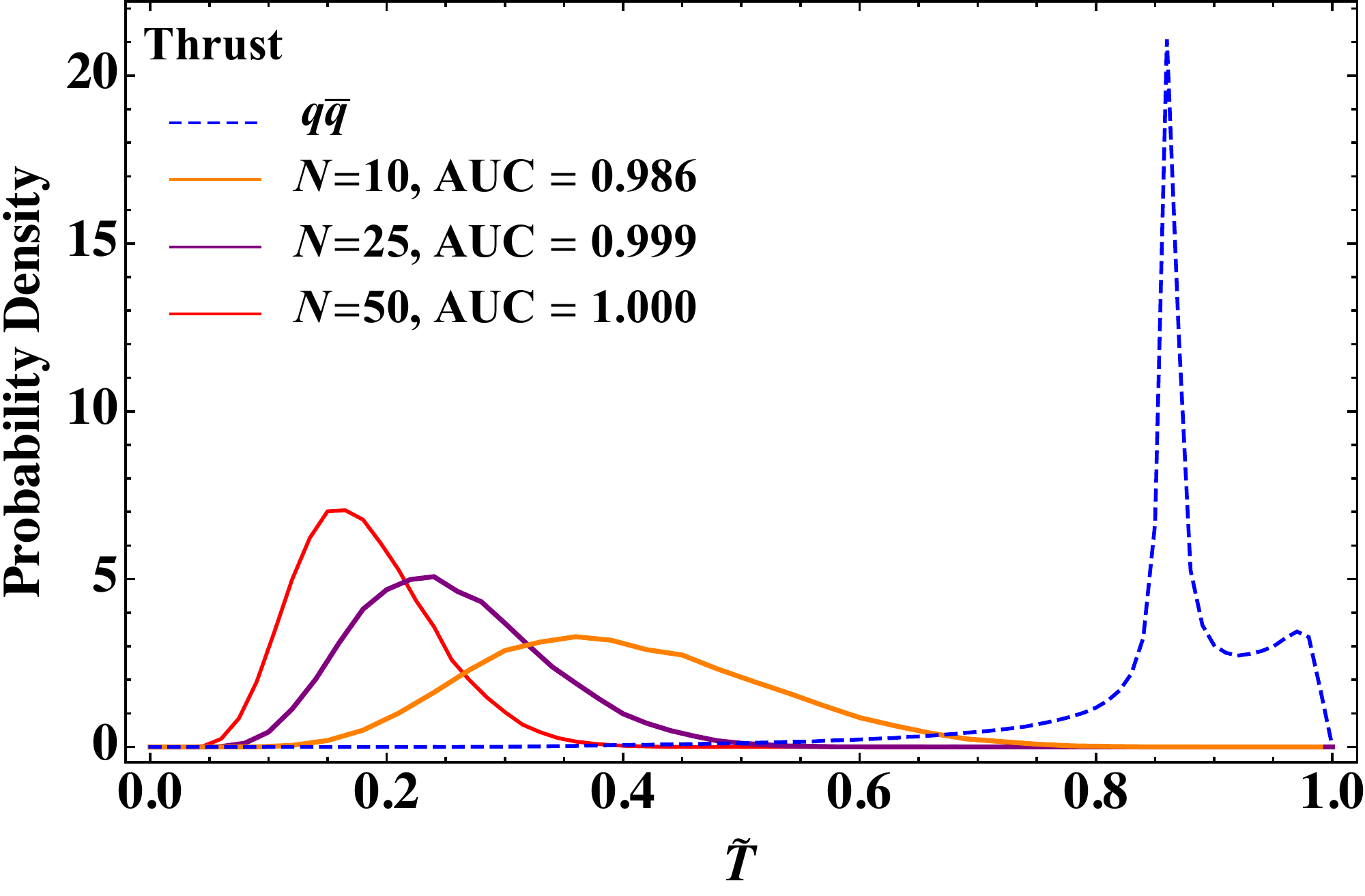}
     }
      \hfill
     \subfloat[]{
       \includegraphics[width=0.45\textwidth]{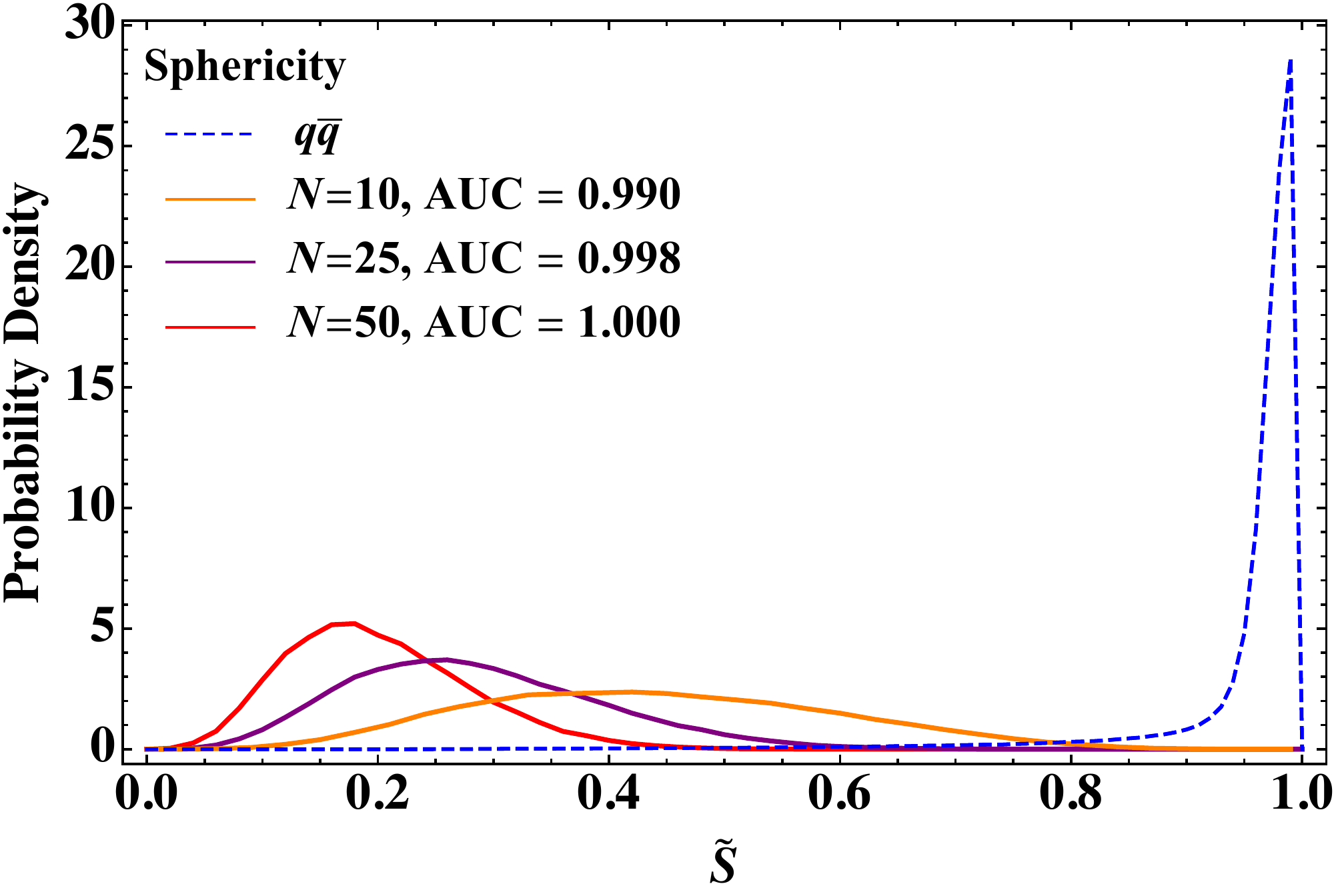}
     }
      \hfill
     \subfloat[]{
       \includegraphics[width=0.45\textwidth]{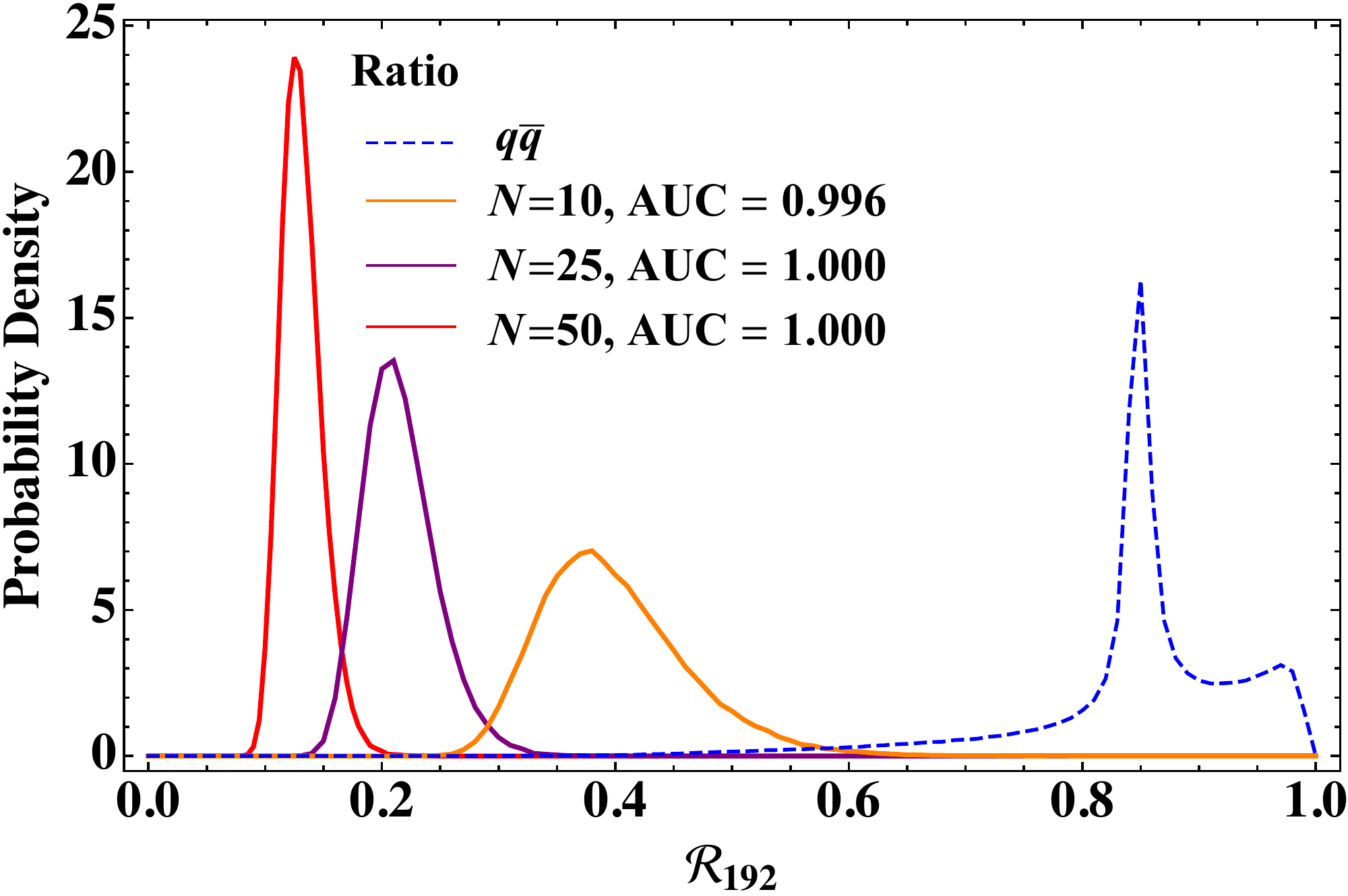}
     }
\caption{Comparing the $q\bar{q}$ background to uniform $N$-body samples with $N = \{10,25,50\}$, using the same observables as \Fig{fig:eeSpec} in $e^+e^-$ collisions at $\sqrt{s} = 350$ GeV.
The AUC values in the legend correspond to the separation power with respect to the $q\bar{q}$ sample.}
\label{fig:eeRamboSpec}
\end{figure}

\begin{figure}[t!]
\subfloat[]{
       \includegraphics[width=0.45\textwidth]{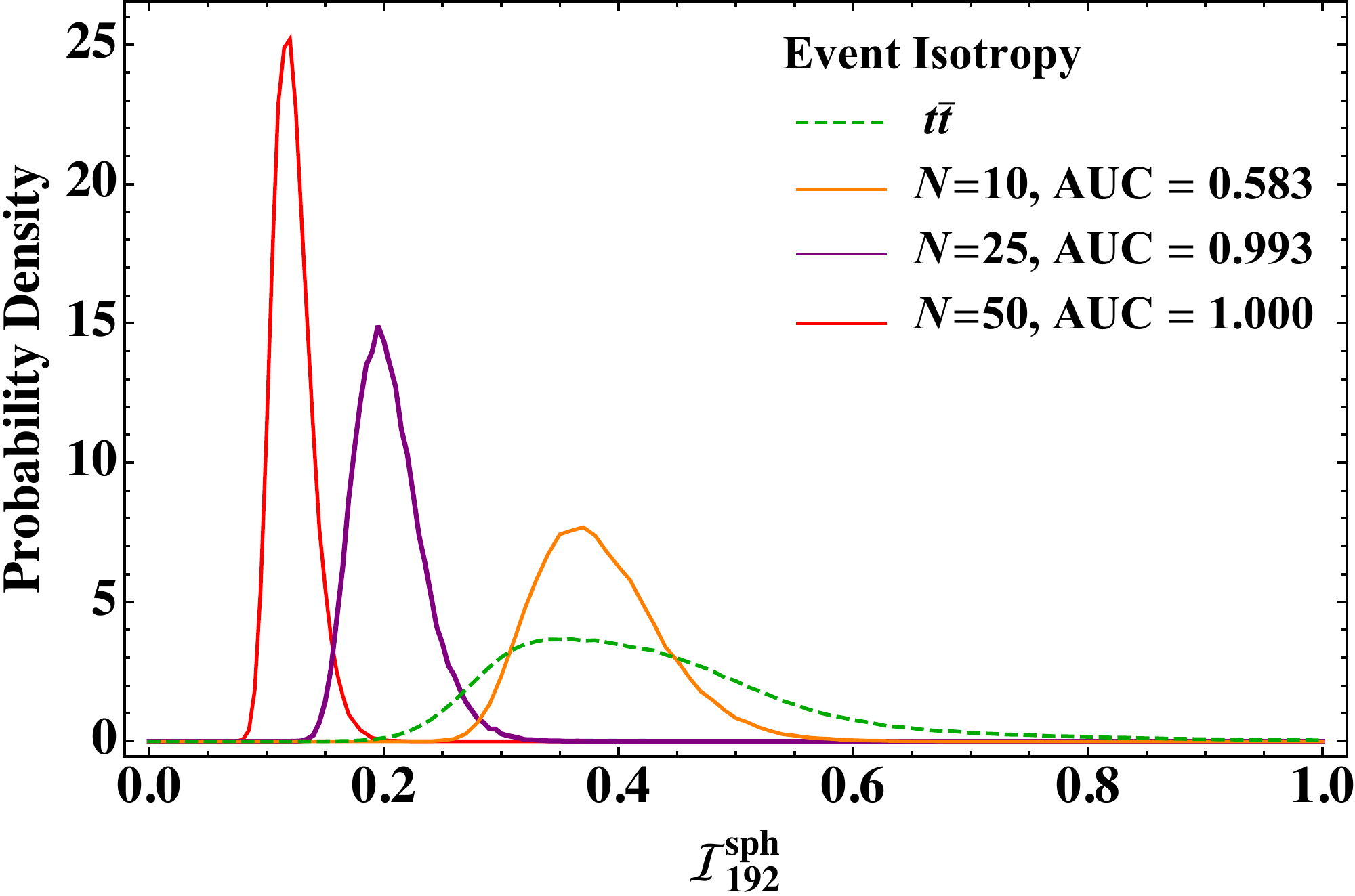}
       	\label{fig:eeRambottSpec_iso}
     }
     \hfill
     \subfloat[]{
       \includegraphics[width=0.45\textwidth]{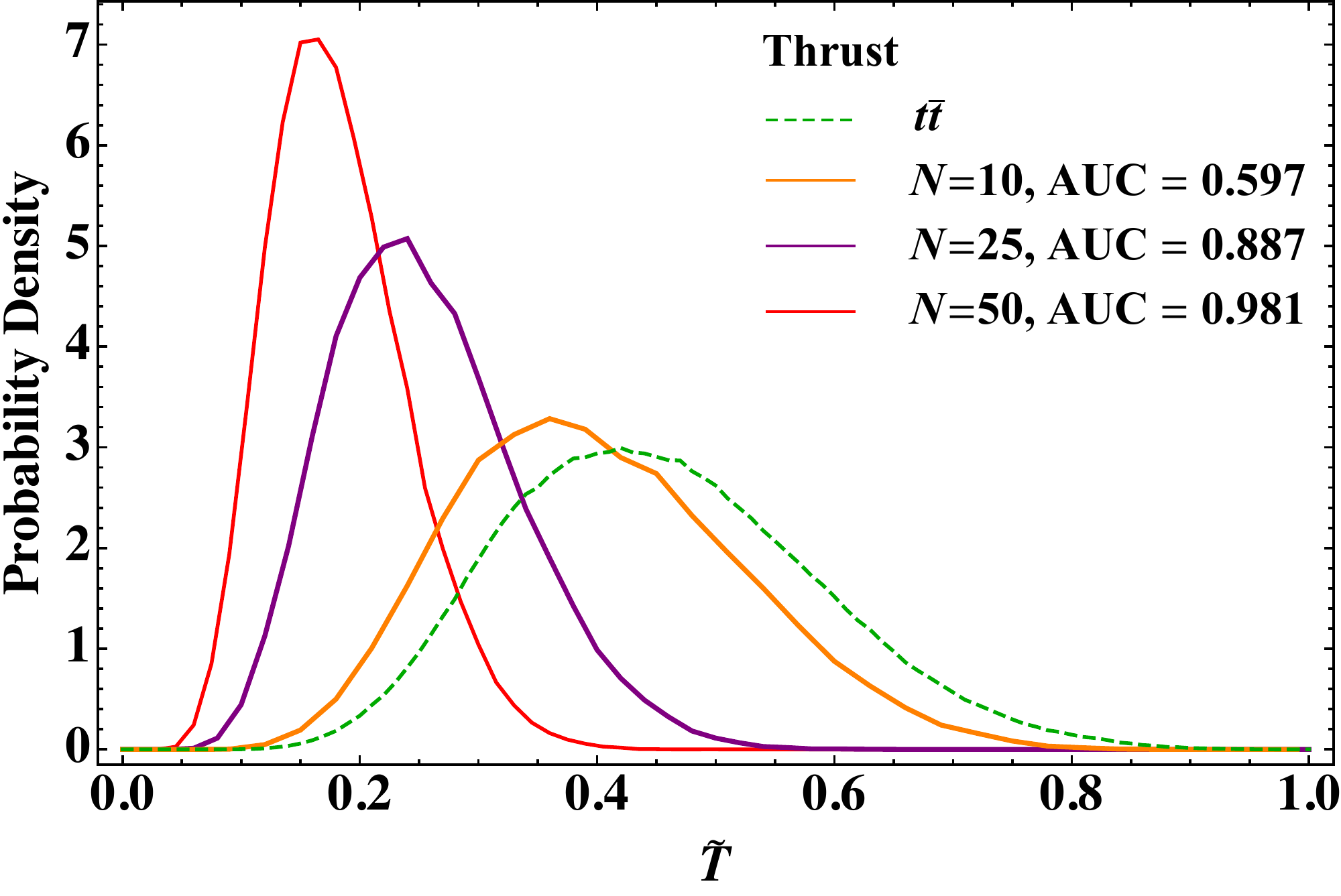}
     }
      \hfill
     \subfloat[]{
       \includegraphics[width=0.45\textwidth]{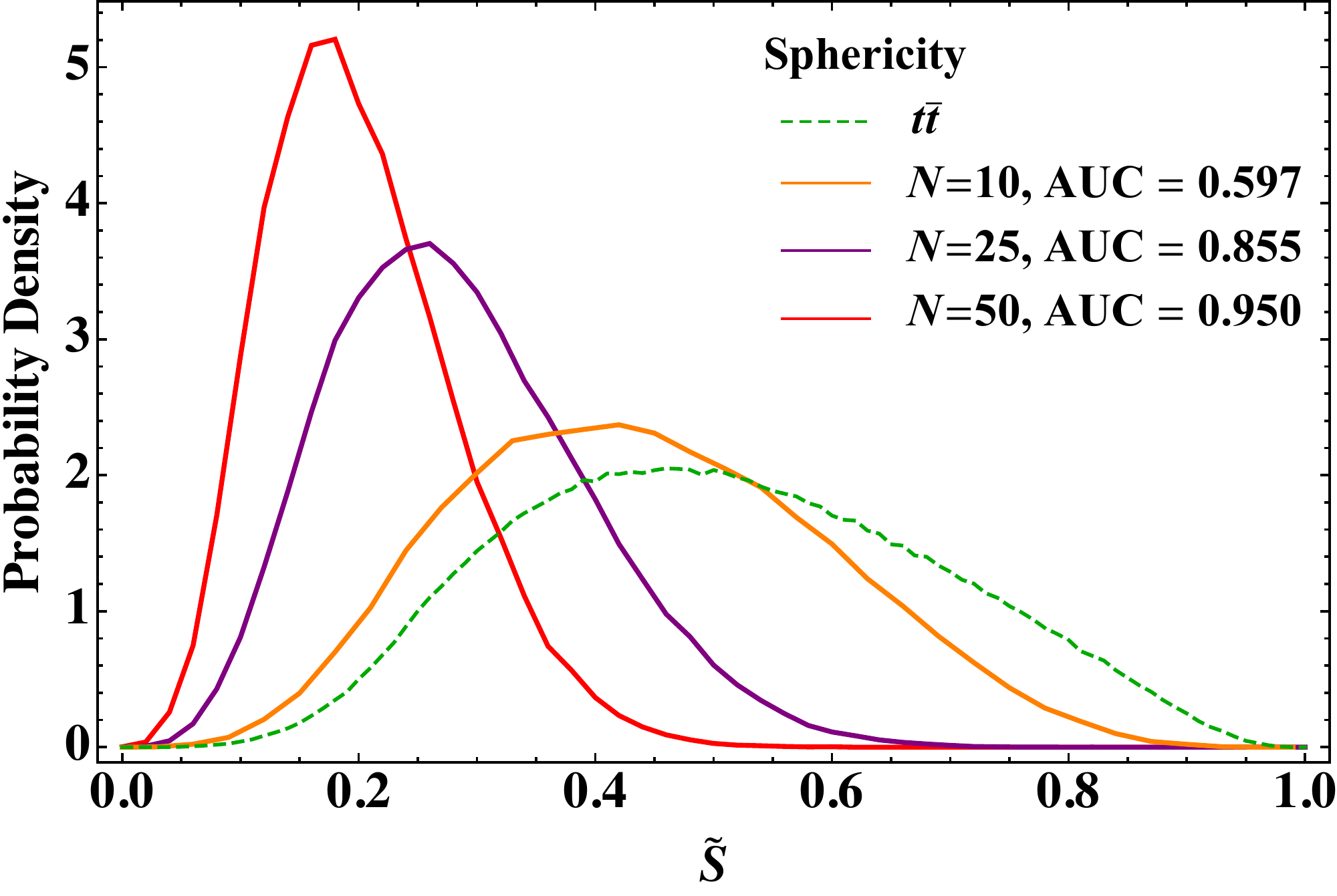}
     }
      \hfill
     \subfloat[]{
       \includegraphics[width=0.45\textwidth]{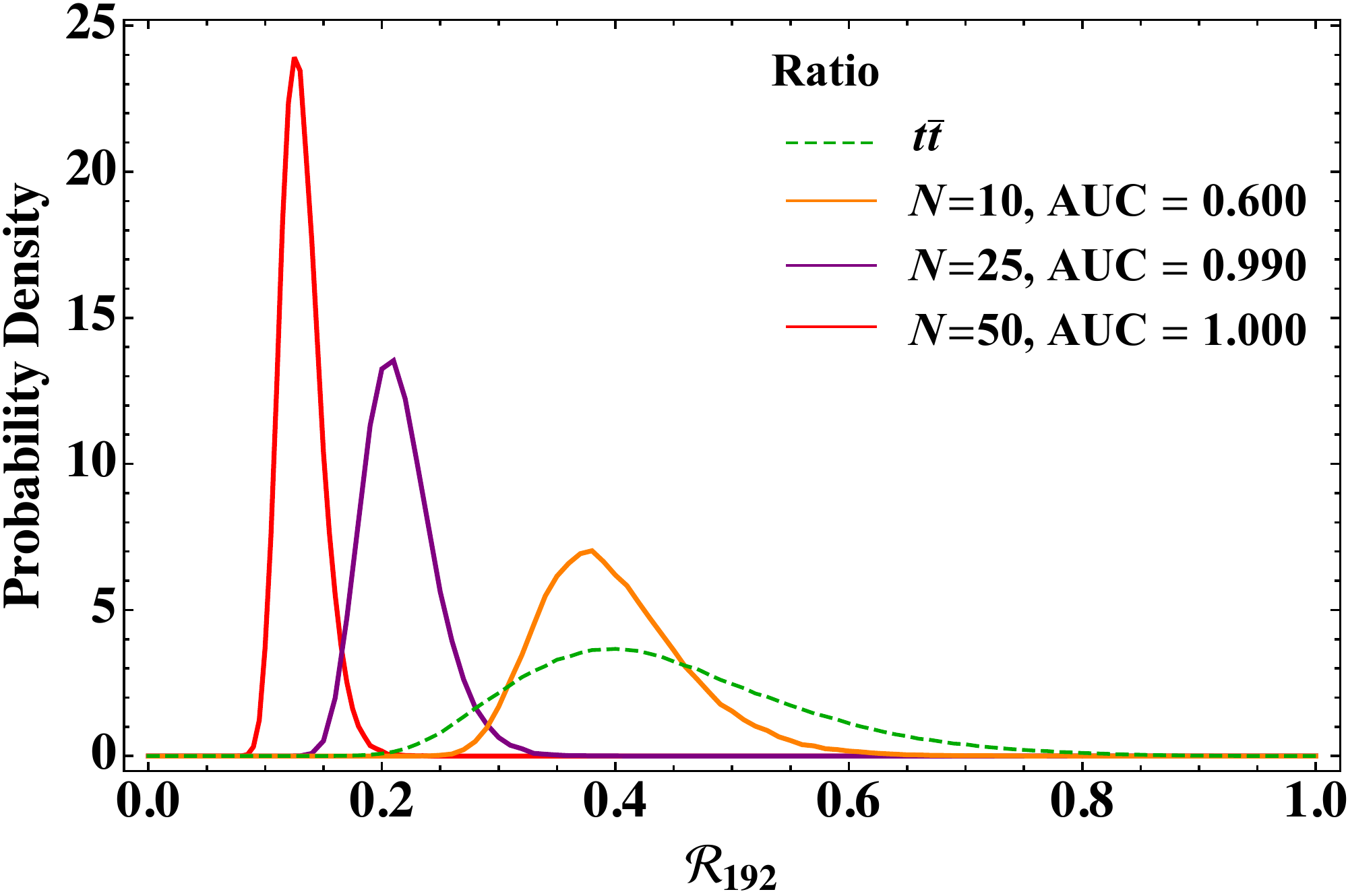}
     }
\caption{The same as \Fig{fig:eeRamboSpec}, but now compared to the $t\bar{t}$ background.
For sufficiently large $N$, event isotropy is the most effective discriminant. }
\label{fig:eeRambottSpec}
\end{figure}

\begin{figure}[t!]
\subfloat[]{
       \includegraphics[width=0.45\textwidth]{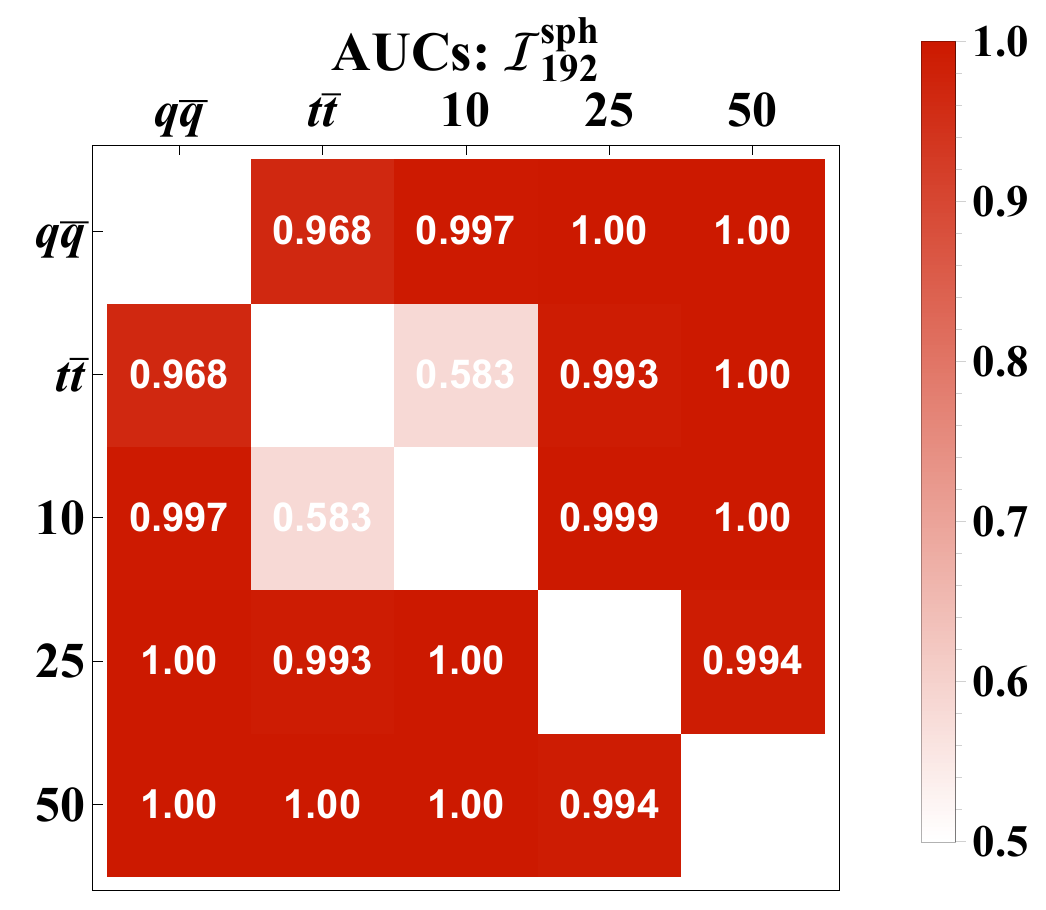}
     }
     \hfill
     \subfloat[]{
       \includegraphics[width=0.45\textwidth]{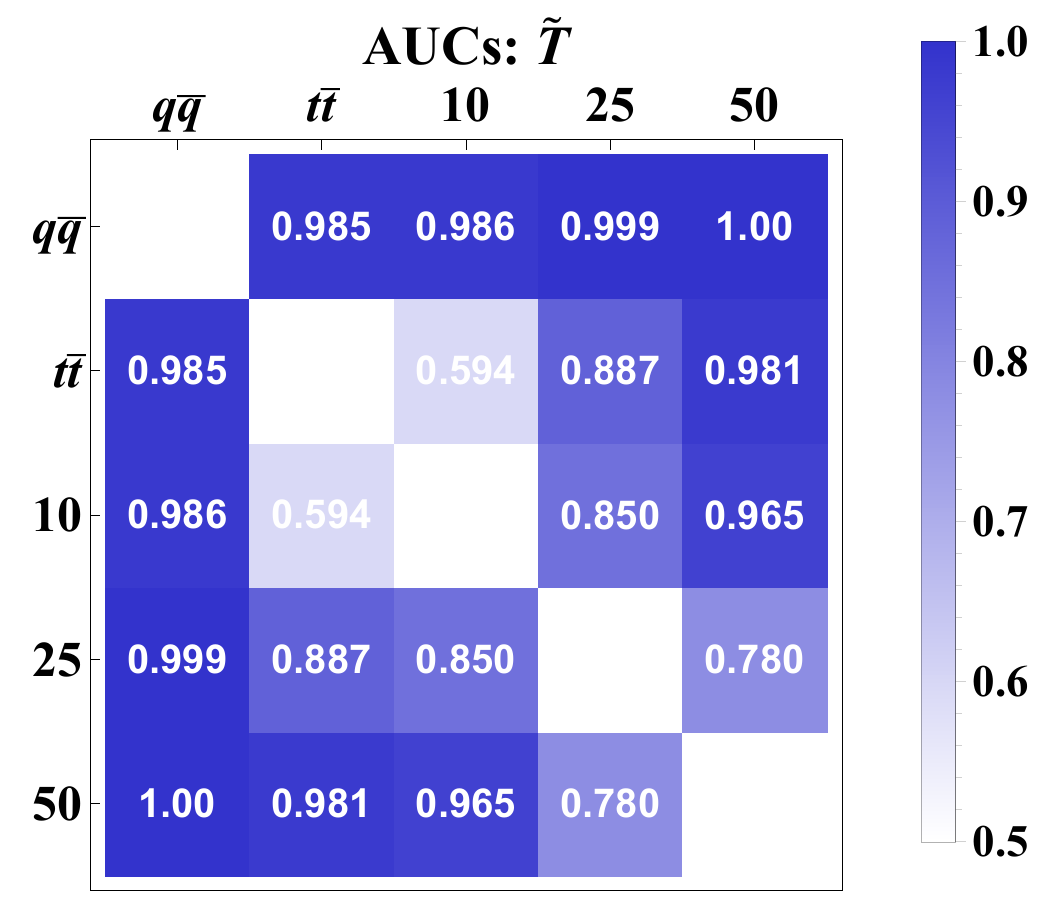}
     }
      \hfill
     \subfloat[]{
       \includegraphics[width=0.45\textwidth]{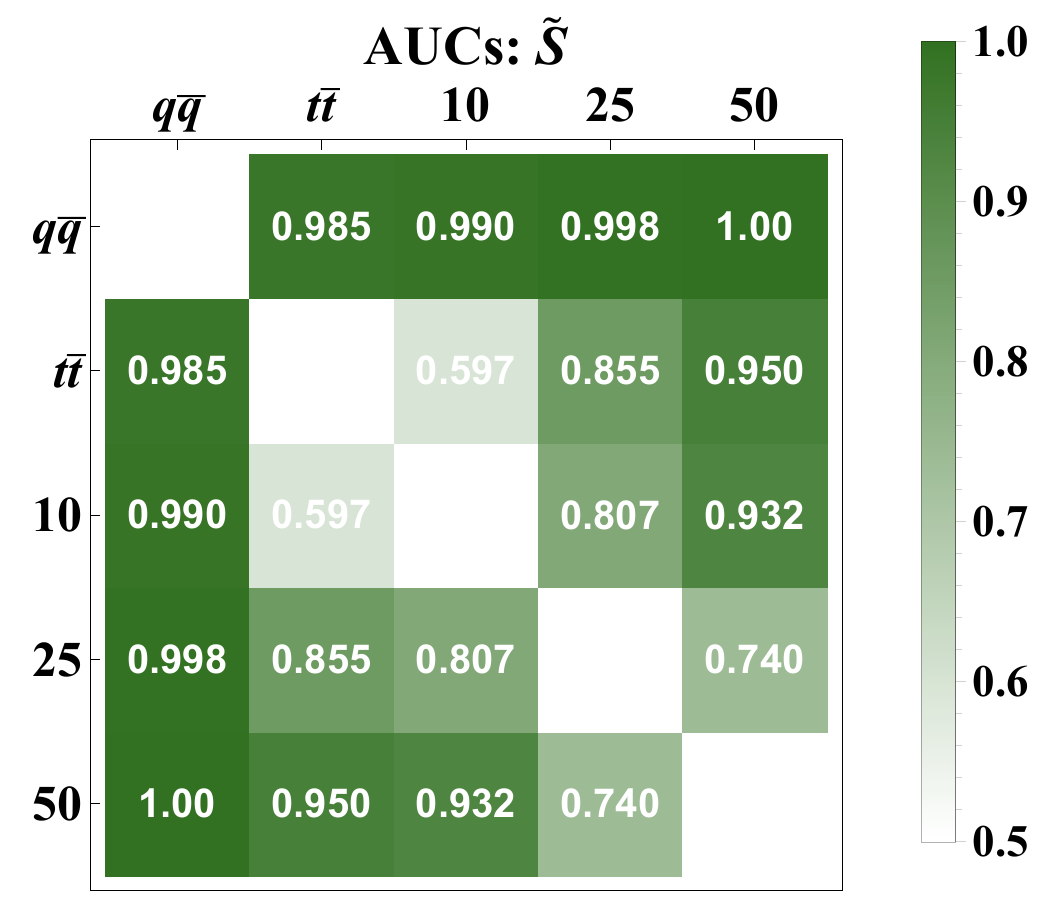}
     }
      \hfill
     \subfloat[]{
       \includegraphics[width=0.45\textwidth]{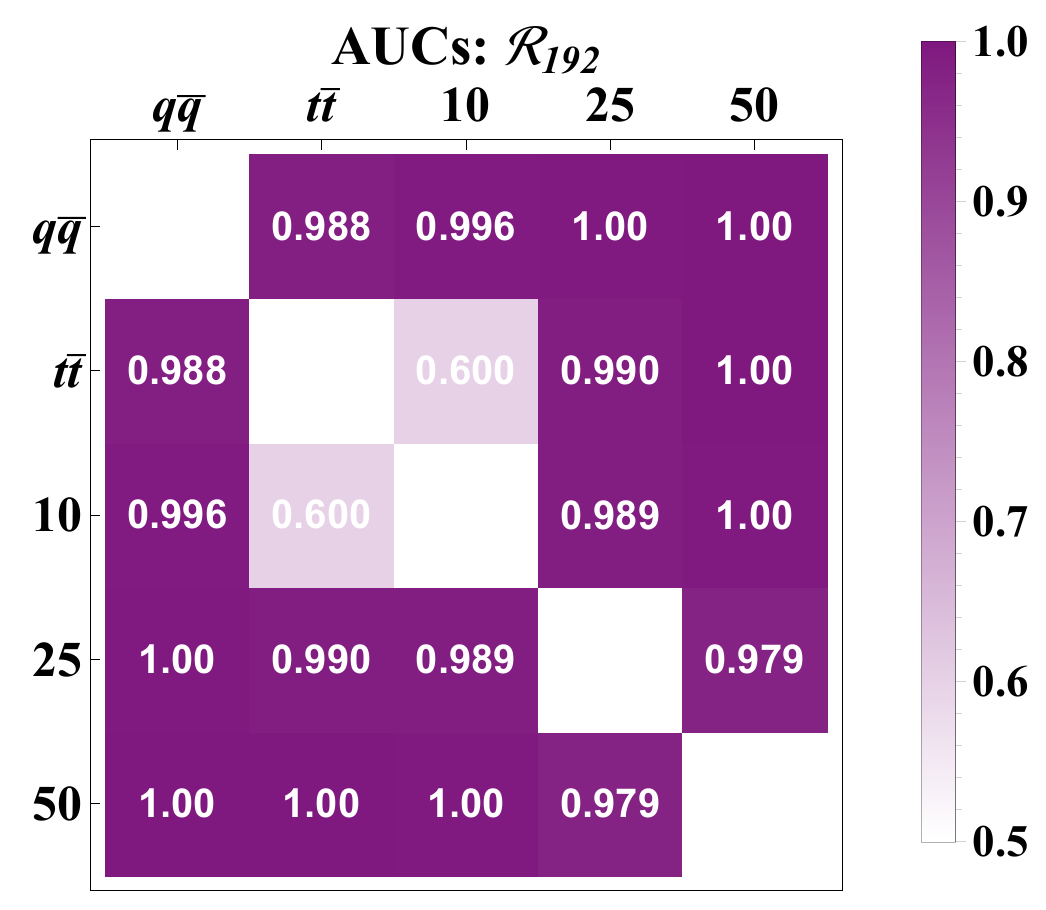}
     }
\caption{The discrimination power between each pair of $e^+ e^-$ samples from \Figs{fig:eeRamboSpec}{fig:eeRambottSpec}, as summarized by the AUC for (a) event isotropy with $n = 192$, (b) thrust, (c) sphericity, and (d) the isotropy/thrust ratio. 
Darker shades indicate more efficient discrimination, with the darkest indicating complete separation.
Apart from $t\bar{t}$ vs.\ $N = 10$, event isotropy has excellent discrimination power between any pair of samples.
Sphericity and thrust are nearly saturated for the larger $N$-body samples, and therefore have smaller AUC values.}
\label{fig:eeAUCtabs}
\end{figure}

Repeating the analysis of \Sec{subsec:top_ee}, we compare the effectiveness of our four event shape observables in separating uniform $N$-body events from SM $q\bar{q}$ events, with results shown in \Fig{fig:eeRamboSpec}. 
For all observables, the separation between the $N$-body and $q\bar{q}$ samples is very good for $N = 10$, and nearly perfect for $N = 25$ and $N = 50$. 
The same $N$-body samples are compared to the $t\bar{t}$ background in \Fig{fig:eeRambottSpec}.
For all of the event shapes, the $N=10$ and $t\bar{t}$ distributions overlap. 
This is expected since there are $\mathcal{O}(10)$ prongs in a $t\bar{t}$ event.
For $N=25$ and $N=50$, the event isotropy and the ratio observable perform better than thrust and sphericity at discriminating the $N$-body sample from the $t\bar{t}$ background.

Intriguingly, event isotropy---and by extension the isotropy/thrust ratio---is further able to separate the $N$-body samples from each other; thrust and sphericity are much less efficient at this task. 
To quantify the ability of these observables at distinguishing different values of $N$, we plot the AUC for discriminating each pair of samples in \Fig{fig:eeAUCtabs}. 
Event isotropy excels at discriminating between the various high multiplicity $N$-body samples, with noticeably better performance than traditional event shape observables.

\begin{figure}[t!]
\centering
   \subfloat[]{
   \label{fig:aucOfRambo_qq}
\includegraphics[width=0.45\textwidth]{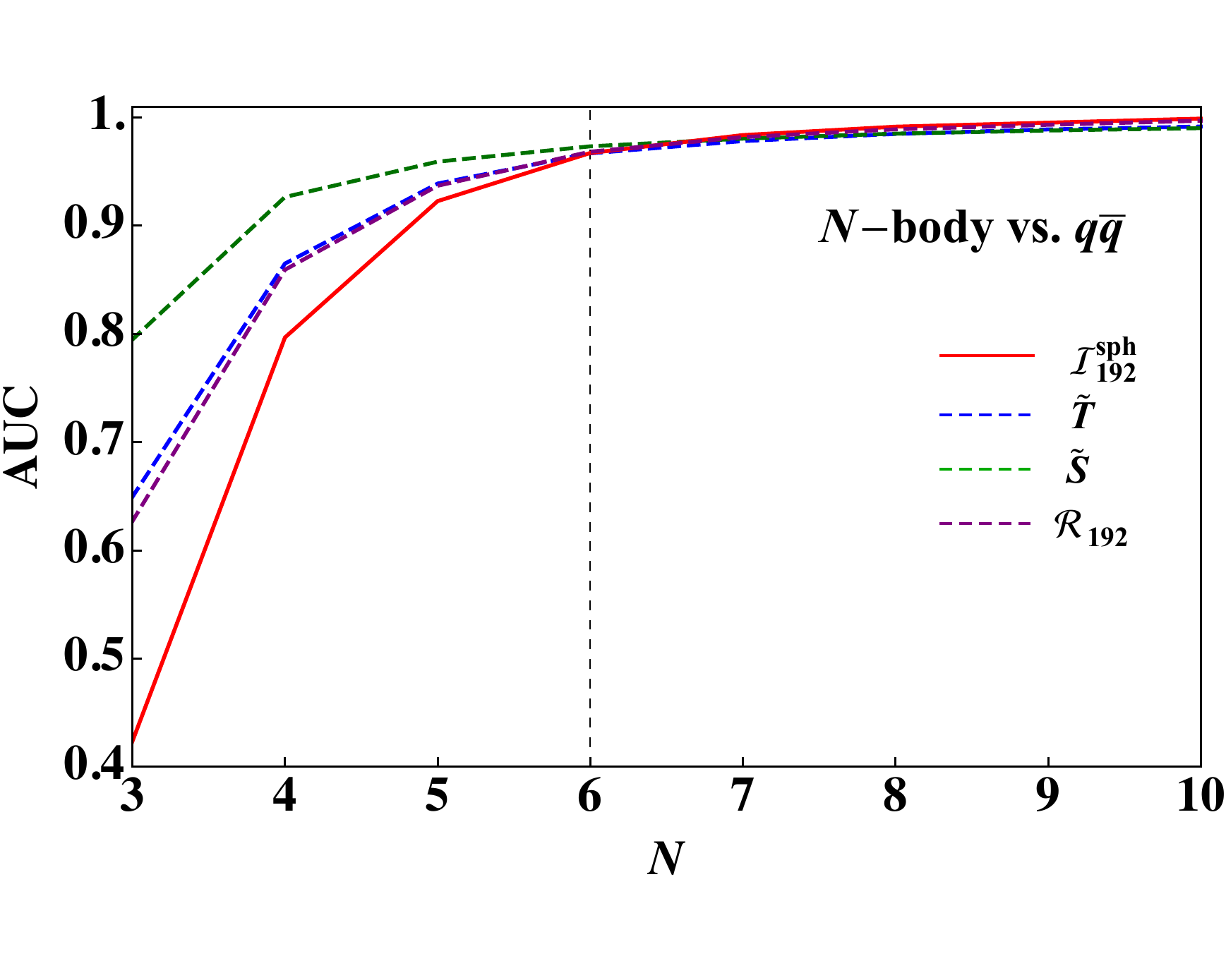}
}
\hfill
   \subfloat[]{
   \label{fig:aucOfRambo_tt}

\includegraphics[width=0.45\textwidth]{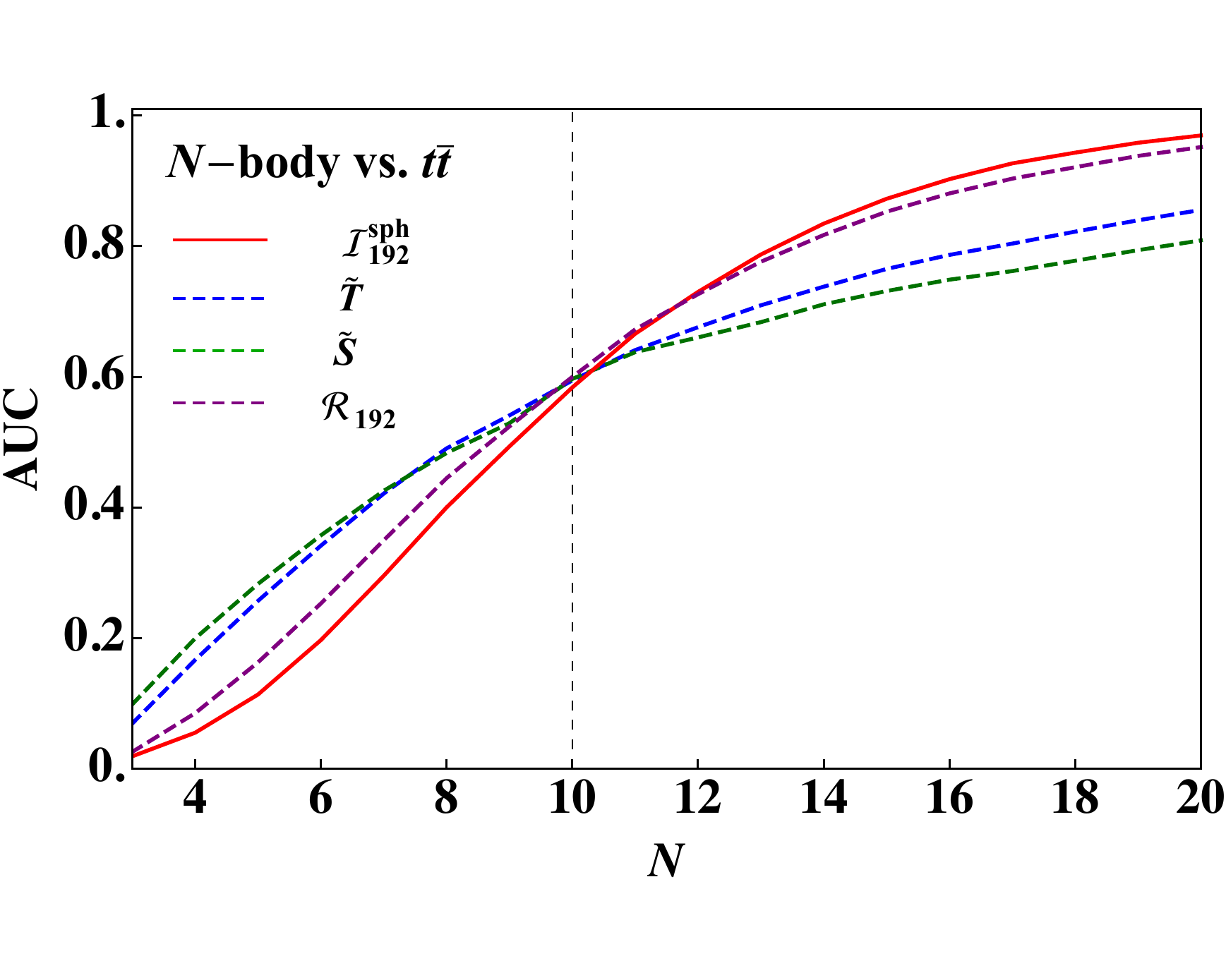}
}
\caption{The AUC values as a function of $N$ for discriminating uniform $N$-body samples from (a) $q\bar{q}$ events and (b) $t\bar{t}$ events.
Shown are event isotropy, thrust, sphericity, and the ratio observable as computed for $e^+e^-$ collisions at $\sqrt{s} = 350$ GeV.
The crossover where event isotropy outperforms the other observables occurs after $N=6$ for the $q\bar{q}$ sample and after $N=10$ for the $t\bar{t}$ sample.
AUC values less than 0.5 indicate when the $N$-body sample is less isotropic than the SM sample.
}
\label{fig:aucOfRambo}
\end{figure}

As discussed in \Sec{sec:emdSM}, thrust measures the distance of an event to a back-to-back two particle configuration (effectively $N = 2$), whereas event isotropy measures the distance to an isotropic radiation pattern (effectively $N = \infty$). 
It is natural to wonder above what value of $N$ does event isotropy outperform thrust as a discriminant.
To address this question, we compute the AUC for separating the $q\bar{q}$ and $t\bar{t}$ samples from uniform $N$-body phase space for $N \geq 3$, comparing event isotropy, thrust, sphericity, and the ratio observable.
The AUC values relative to the $q\bar{q}$ sample are plotted in \Fig{fig:aucOfRambo_qq}, where crossover in performance happens right after $N=6$.
This could have been anticipated from the study in \Sec{subsec:top_ee}, since $t \bar{t}$ events are similar to $N= 6$ events.%
\footnote{Because of the hierarchy $m_t > m_W$, the $W$ bosons are quasi-relativistic and the jets from the $W$ decay are slightly boosted, so $t \bar{t}$ events are a bit less isotropic than uniform 6-body phase space.}
Thus, for low multiplicity SM backgrounds, we expect event isotropy will be most effective in identifying new physics scenarios that typically produce seven or more final state partons of comparable energies.

The performance crossover for the $t\bar{t}$ sample compared to the $N$-body sample occurs for $N \sim 10$, as shown in \Fig{fig:aucOfRambo_tt}.
To understand this, note that samples with $N\lesssim6$ are less isotropic than the $t\bar{t}$ sample, so 2-prong discriminants like thrust are expected to be more sensitive in this regime. 
For $\mathcal{O}(15)$ or more hard prongs, the event isotropy is able to distinguish the $N$-body sample from the $t\bar{t}$ sample considerably more efficiently than thrust and sphericity. 
Thus, event isotropy is a useful probe of new physics even when considering moderately isotropic SM backgrounds.

\subsection{Toy Model: Quasi-Isotropic Distributions}
\label{subsec:ee_kbody}

As a final benchmark for the $e^+e^-$ case, we consider $k$-body discretized isotropic events. 
This is the idealized limit of a strongly-coupled quasi-conformal new physics scenario, where the lightest modes in the hidden sector have a 100\% branching ratio back to SM particles. 
A cartoon of this new physics signal would be a uniformly tiled sphere, as already shown in \Fig{fig:geoVis} for events generated by \texttt{HEALPix}. 

\begin{figure}[t!]
\centering
 \includegraphics[width=0.55\textwidth]{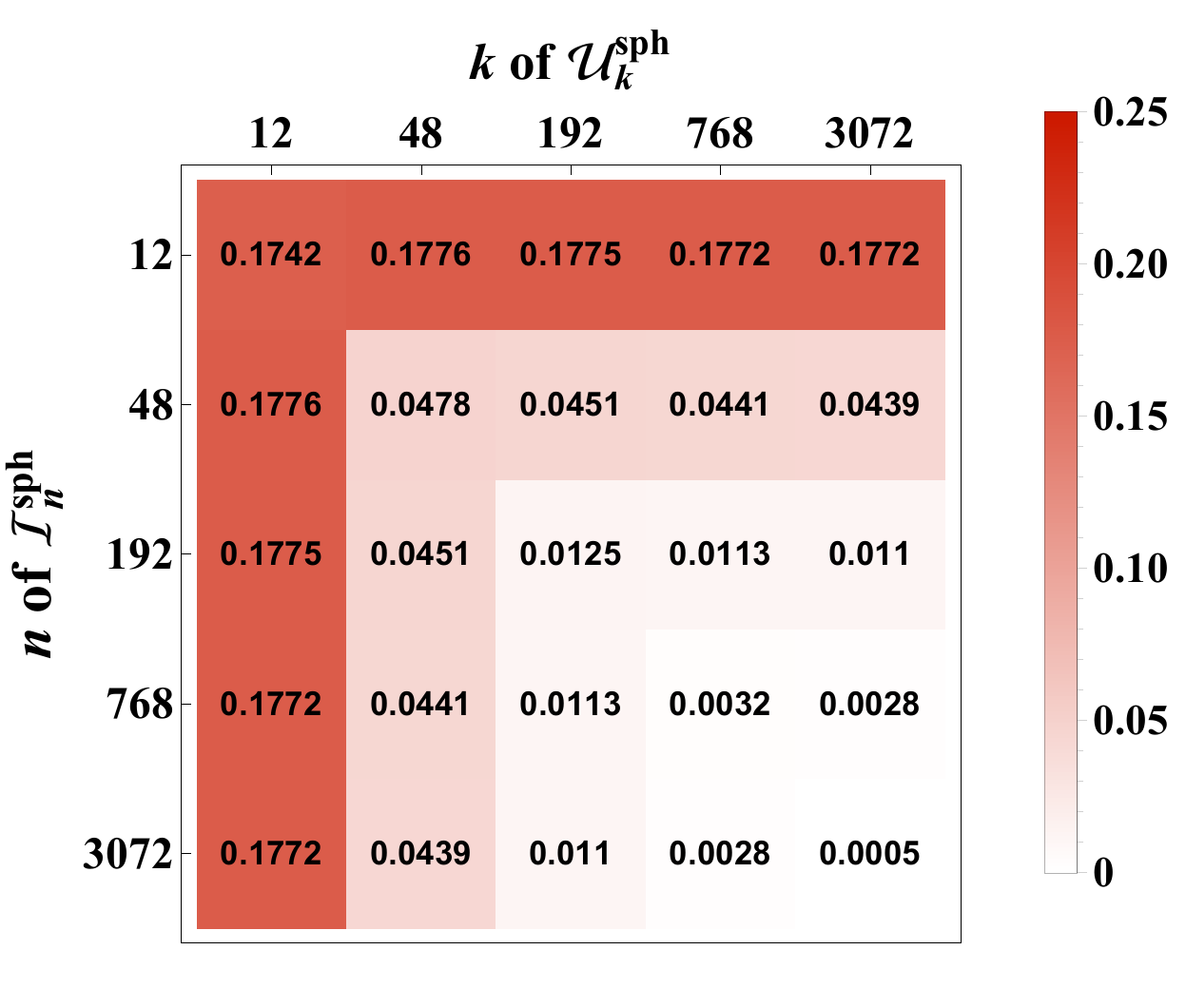}
\caption{The event isotropy $\iso{sph}{n}$ for a $k$-particle sphere, averaged over random orientations.
The table is symmetric because the EMD is a metric, and therefore symmetric between the function arguments.
The distance from the coarsest sphere to the finest sphere ($\iso{sph}{3072}(\mathcal{U}^\text{sph}_{12})$ =  $\iso{sph}{12}(\mathcal{U}^\text{sph}_{3072}) \simeq 0.18$) is useful benchmark for when an event should be considered ``isotropic.''
See \Fig{fig:obsOnSpher} to compare the behavior of other event shape observables on spherical configurations.}
\label{fig:eeSphereToSphere}
\end{figure}

In \Fig{fig:eeSphereToSphere}, we calculate $\iso{sph}{n}(\mathcal{U}^\text{sph}_{k})$ for the five smallest $n$ and $k$ values permitted by \texttt{HEALPix}.
To avoid pathological behavior when the \texttt{HEALPix} events happen to be aligned, we average over $10^4$ random orientations of the spheres when computing event isotropy.
This averaging explains why $\iso{sph}{n}(\mathcal{U}^\text{sph}_{n})$ is not exactly zero.
An analytic approximation to $\iso{sph}{\infty}(\mathcal{U}^\text{sph}_{k})$ is presented in \App{app:eiApp}.

From the definition in \Eq{eq:EMDgeneral}, it is clear that $\iso{sph}{n}(\mathcal{U}^\text{sph}_{k})$ should be small for large values of $n, k$ (i.e.\ events that better approximate isotropy). 
Smaller values of $n$ and/or $k$ provide a useful benchmark for what it means for an event to be ``isotropic.''
The largest values of event isotropy seen in \Fig{fig:eeSphereToSphere} are for the coarsest tiling, where $\iso{sph}{n}(\mathcal{U}^\text{sph}_{12}) \simeq 0.18$. 
Notably, this benchmark value is smaller than what was seen in the $t\bar{t}$ sample in \Fig{fig:eeSpec_iso}, and nearly aligned with the peak of $N=50$ distribution in \Fig{fig:eeRamboSpec_iso}. 
For $k>12$, the average event isotropy is below any of the $N$-body distributions considered. 
This demonstrates the ability of event isotropy to distinguish events composed of several quasi-uniform hard prongs from truly isotropic events.

\section{Benchmark Scenarios: Proton-Proton Colliders}
\label{sec:ppBE}

We now turn our attention to proton-proton ($pp$) collisions.
Because of the partonic substructure of protons, the products from the hard scattering process have nonzero longitudinal boosts along the beam axis. 
We therefore consider the cylindrical and ring-like geometries for event isotropy, since these are more relevant for studying axially symmetric radiation patterns.
As a baseline for comparison, we use transverse thrust from \Eq{eq:transthrust_def}, which a commonly studied $pp$ event shape~\cite{Bertram:2002sv,Nagy:2003tz,Banfi:2004nk,Banfi:2010xy}.

As in \Sec{sec:emdBE}, we consider three benchmark signals---$t\bar{t}$ production, uniform $N$-body phase space, and quasi-isotropic distributions---and quantify discrimination power against SM backgrounds.
To match the conditions of the LHC, we consider $pp$ collisions at a center-of-mass energy $\sqrt{s} = 14$ TeV. 
The events are generated in \texttt{Pythia 8.243} with multi-parton interactions turned on to model the underlying event.
Any missing transverse momentum lost to invisible final state particles is added to the event as a single vector when calculating transverse event shape observables. 
We do not include the effect of pileup, namely multiple $pp$ collisions per beam crossing, which would tend to make the events more isotropic without some form of pileup mitigation~\cite{Soyez:2018opl}.

\subsection{Top Pair Production vs.\ QCD Dijet}
\label{subsec:toppair_pp}
\begin{figure}[t!]
\centering
\subfloat[]{
       \includegraphics[width=0.42\textwidth]{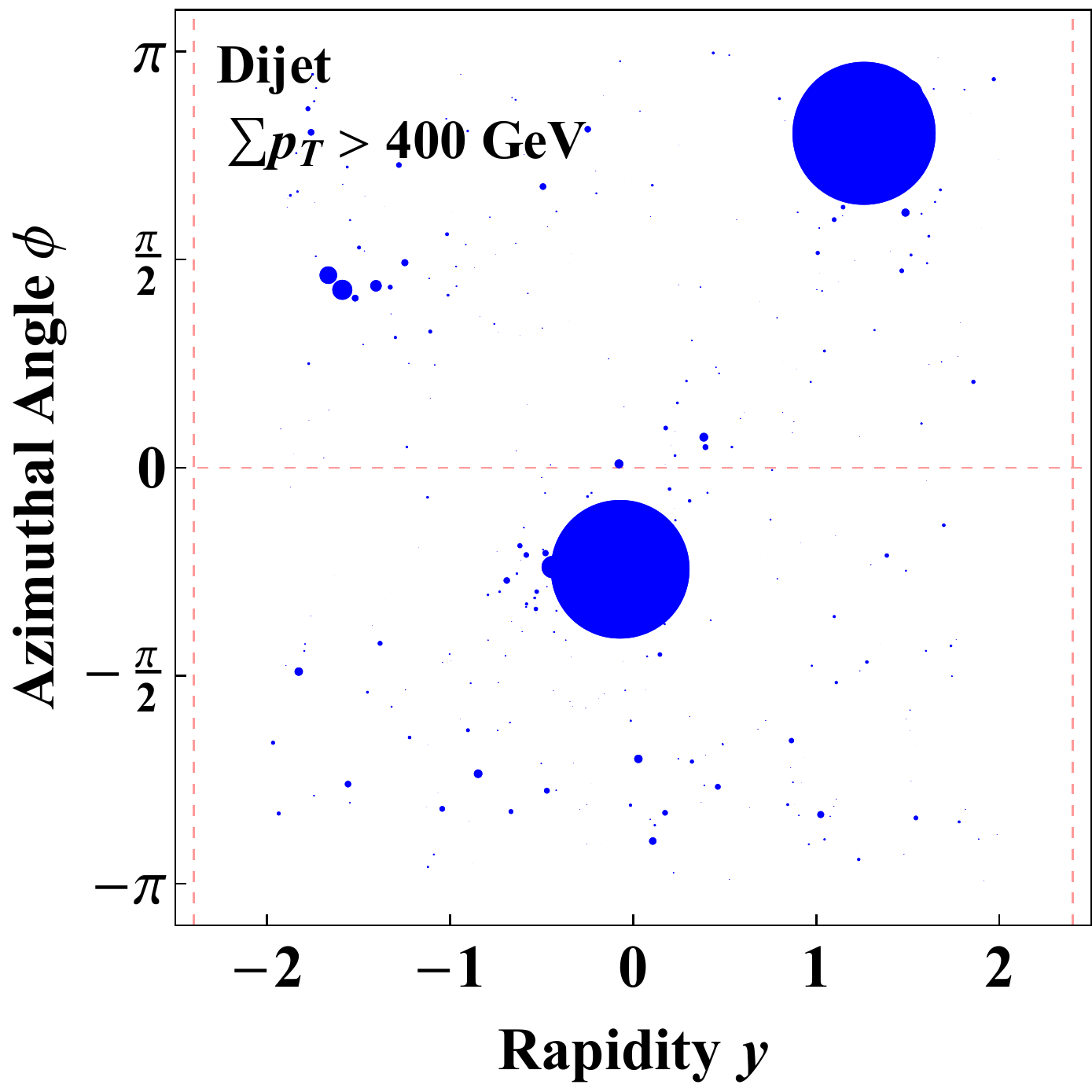}
     }
     \hfill
     \subfloat[]{
       \includegraphics[width=0.42\textwidth]{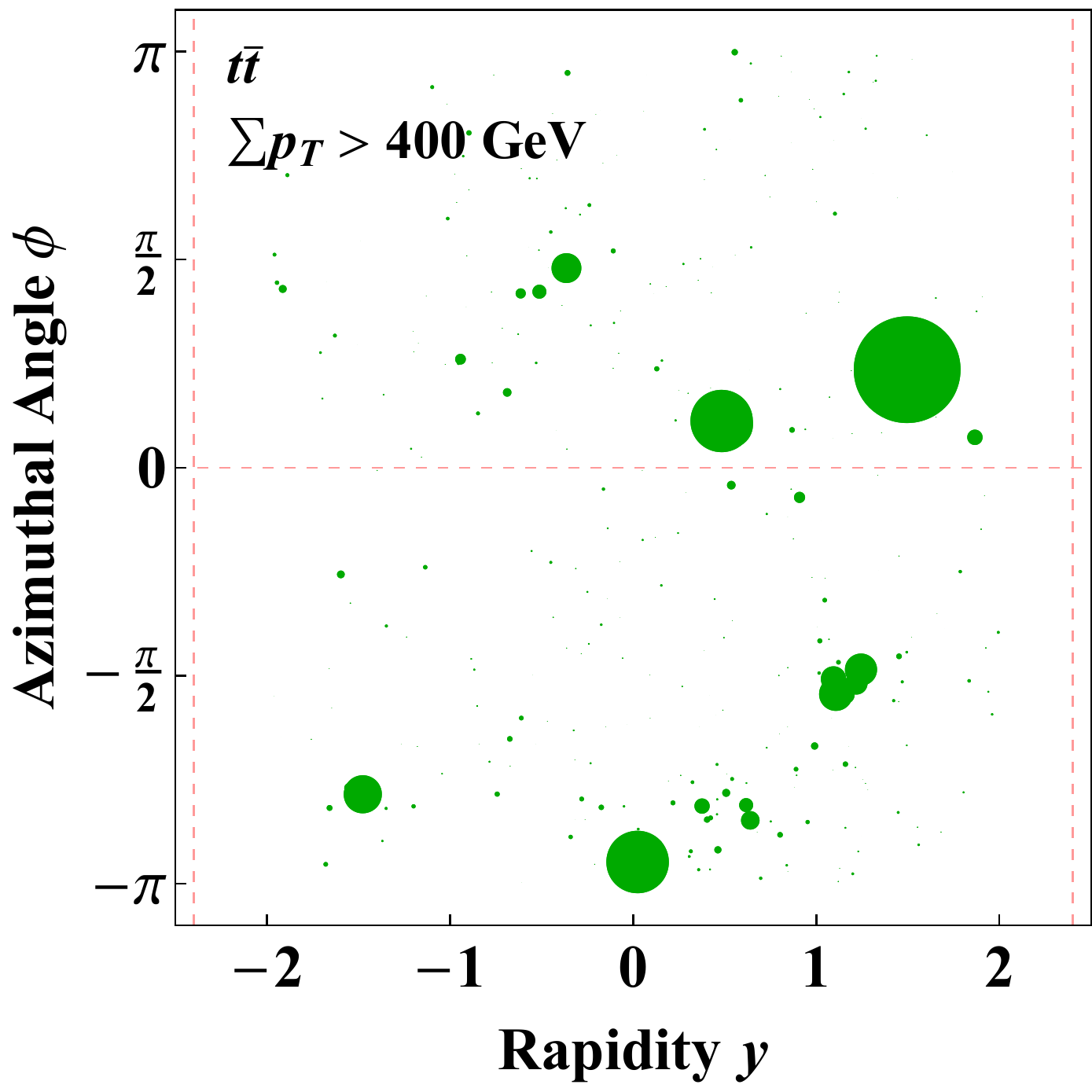}
     }
     \hfill
\caption{Examples of (a) a QCD dijet event and (b) a $t\bar{t}$ event in $pp$ collisions at $\sqrt{s} =14$ TeV. 
The events selection is $\sum p_T > 400$ GeV for particles within $|y| < 2$. 
The size of the marker is proportional to the $p_T$ of the particle.
We can clearly see the two hard prongs of momentum in the QCD dijet event compared to roughly six hard prongs in the $t\bar{t}$ event.
}
\label{fig:ppVis}
\end{figure}

Our first $pp$ benchmark signal process is SM top-quark pair production, which we want to identify against QCD dijet backgrounds.
We select events where the scalar transverse momentum sum satisfies $\sum_i p_{T,i} > 400$ GeV for all visible particles in the rapidity range $|y| < 2$.%
\footnote{There is a generator cut on the minimum transverse momentum of the hard process of $\hat{p}_T >100$ GeV for the QCD dijet process and $\hat{p}_T > 70$ GeV for the $t\bar{t}$ process, though this has a negligible impact on our results.
We also considered alternative rapidity ranges of $|y|<2.4$ and $|y| < 4$, finding that the impact of rapidity cuts on event isotropy is relatively mild, even for the cylindrical event isotropy which depends explicitly on $y_{\rm max}$.} 
Visualizations of these events are shown in \Fig{fig:ppVis}.
We compare the efficiencies of three event shape observables: cylindrical event isotropy $\iso{cyl}{160}$, ring-like event isotropy $\iso{ring}{32}$, and rescaled transverse thrust $\widetilde{T}_{\perp}$.

\begin{figure}[t!]
\centering
\subfloat[]{
       \includegraphics[width=0.45\textwidth]{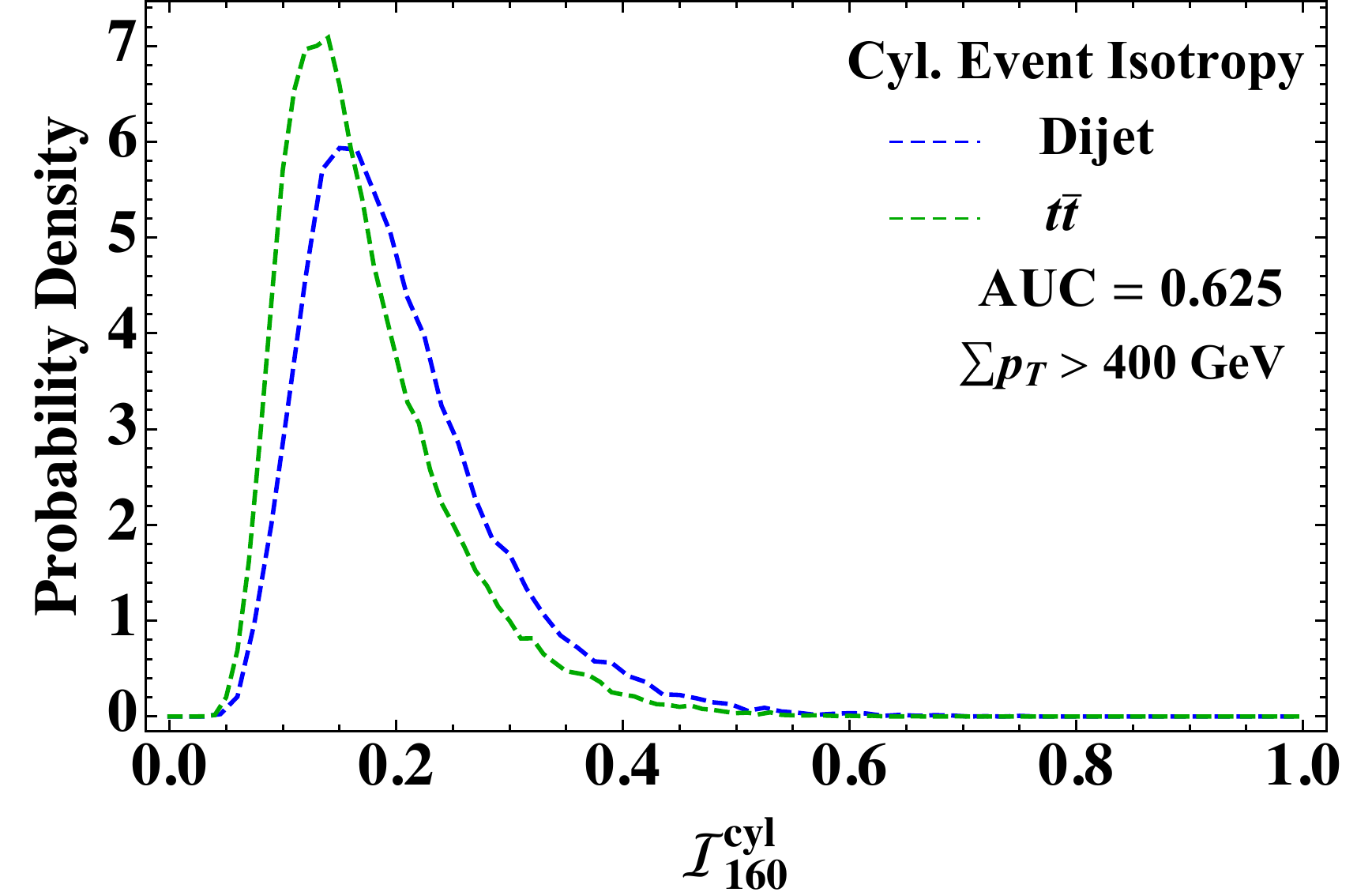}
     }
     \hfill
     \subfloat[]{
       \includegraphics[width=0.45\textwidth]{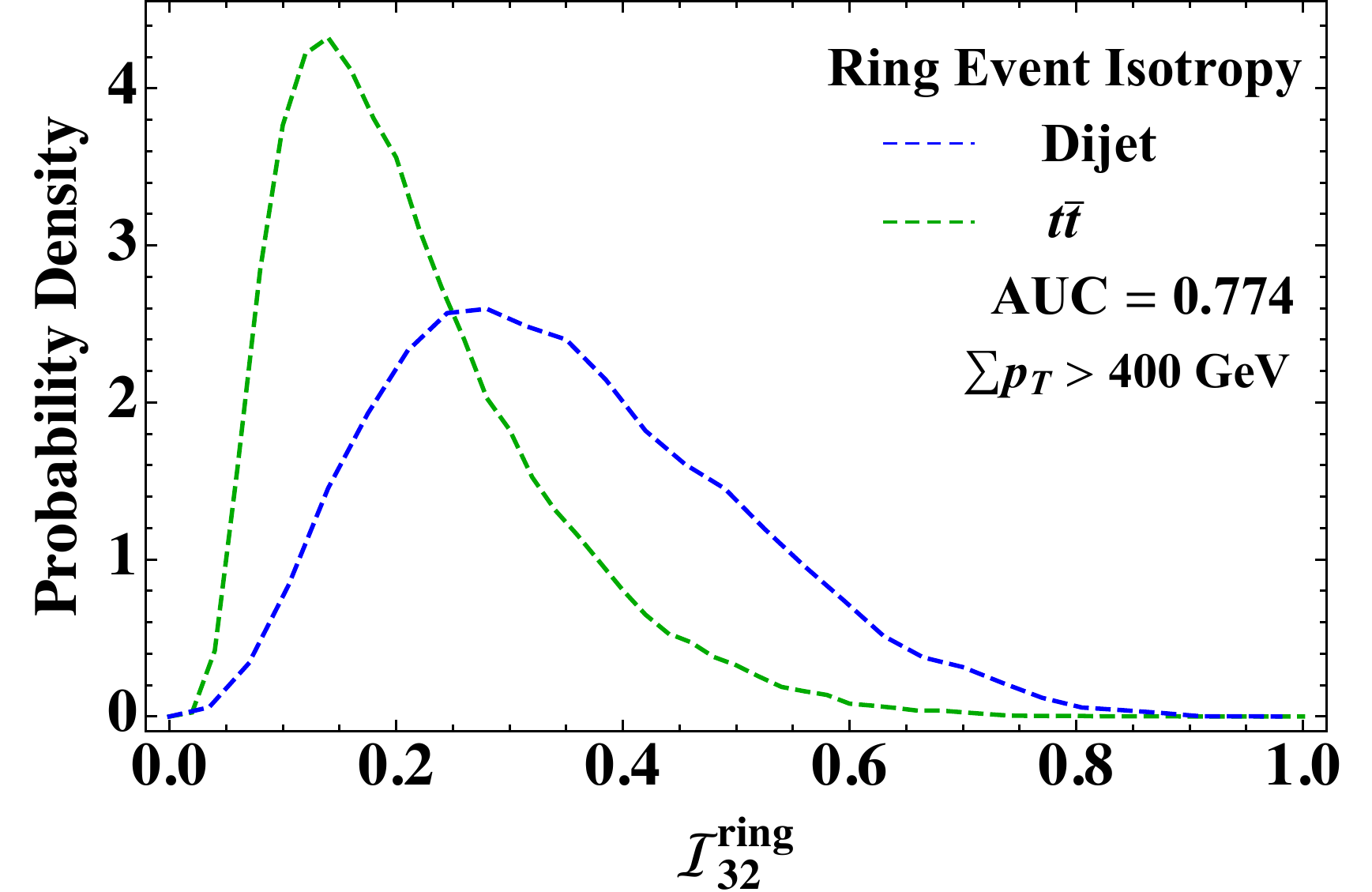}
     }
     \hfill
     \subfloat[]{
       \includegraphics[width=0.45\textwidth]{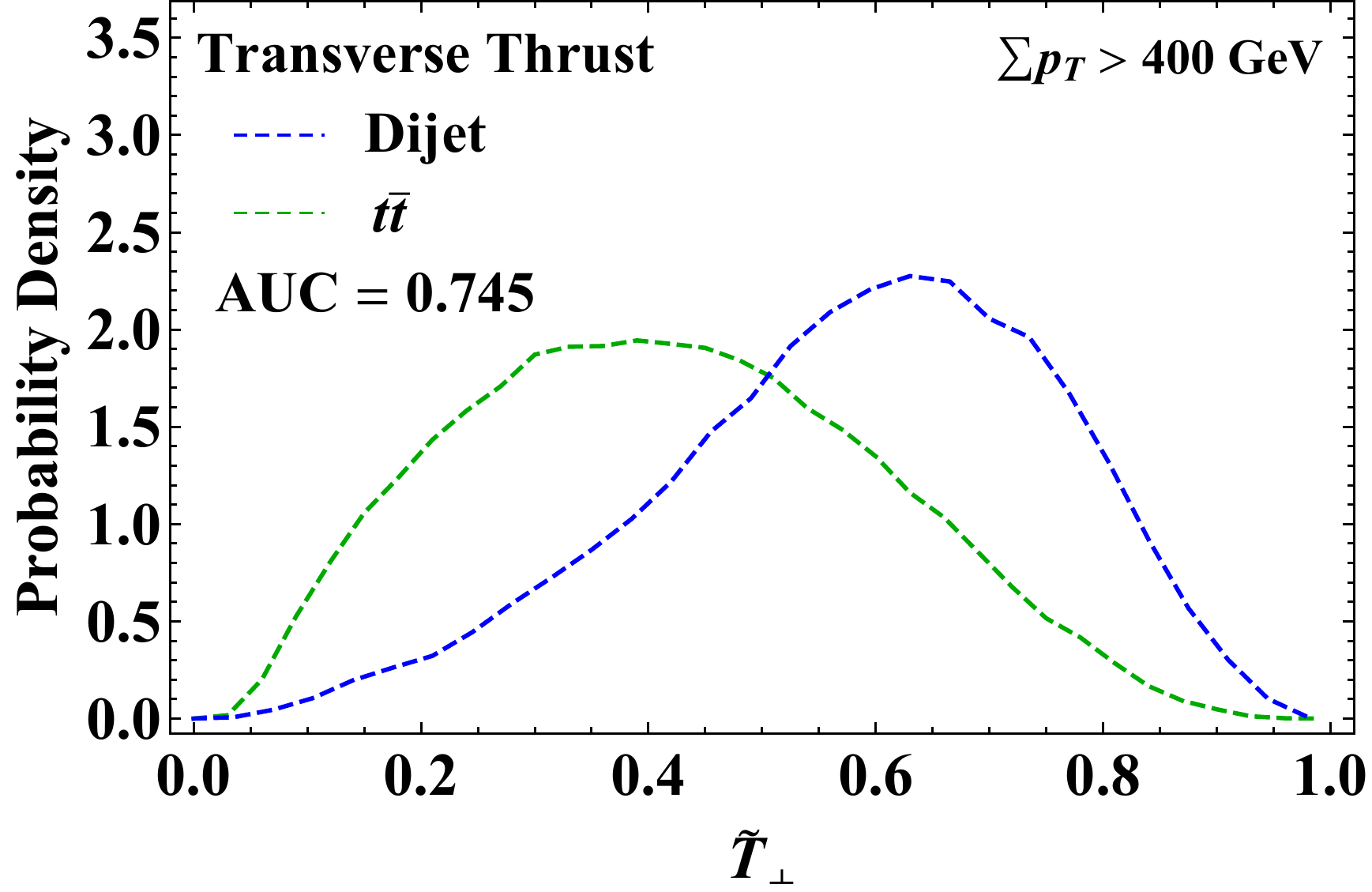}
     }
\caption{The distributions of (a) cylindrical event isotropy, (b) ring-like event isotropy, and (c) transverse thrust in the QCD dijet and $t\bar{t}$ samples produced in $pp$ collisions at $\sqrt{s} = 14$ TeV.}
\label{fig:ppSpec}
\end{figure}

The distributions of these three observables for $10^5$ events, along with their corresponding AUC values, are shown in \Fig{fig:ppSpec}.
Overall, the dijet and $t\bar{t}$ samples are less well separated compared to the $e^+e^-$ study from \Fig{fig:eeSpec}.
This is due to a variety of factors, including the fact that the dijet sample now includes gluon jets, which are more isotropic than quark jets, and the fact that $pp$ collisions have a large contribution to their radiation pattern from ISR.

The best discrimination power is achieved by ring-like event isotropy, with transverse thrust performing only marginally worse.
These two observables are the ones that project out the rapidity information from the event.
We also tested the analog of the ratio observable from \Eq{eq:remd} combining ring-like event isotropy and transverse thrust, though we did not find significant gains.

While cylindrical event isotropy captures more information about the overall uniformity of the event, $pp$ collision events are often boosted to one side of the cylinder.
Thus, unless the event is centered near $y = 0$, cylindrical event isotropy will penalize events that are uniform in the hard scattering frame but longitudinally boosted to the lab frame.
Such central events are less generic, though they could be a more prevalent in new physics models where an on-shell particle is produced close to the kinematic limit.

\subsection{Toy Model: Uniform $N$-body Phase Space}
\label{subsec:uniformNbody_pp}

We now study the event isotropy in the context of uniform $N$-body phase space events.
To make the $N$-body sample into a toy model for a generic hidden valley scenario with a heavy mediator, we first generate a $1$ TeV $Z'$ vector boson from $pp$ collisions in \texttt{Pythia}, and then use the RAMBO algorithm to decay the $Z'$ into $N = \{10, 25, 50\}$ configurations.
The full event then includes the $N$-body decay products, boosted from the $Z'$ rest frame, as well as ISR from the incoming proton beams.
For a more realistic benchmark, we could have implemented the massive version of the RAMBO algorithm, with the mass scale set by hidden sector mesons that subsequently decay into massless SM states, but we use the massless algorithm here for conceptual simplicity and generation efficiency, leaving a more realistic study to \Sec{sec:SUEP}.
As in \Sec{subsec:toppair_pp}, we set the acceptance as all visible particles with $|y|<2$,
but we impose a higher net transverse momentum cut of $\sum p_T > 650$ GeV due to the higher mass scale of the $Z'$ resonance.%
\footnote{Correspondingly, we raise the generator cut on the QCD dijet hard process to $\hat{p}_T > 200$ GeV and on the $t\bar{t}$ hard process to $\hat{p}_T>120$ GeV.}
We study the discrimination power against QCD dijet and $t\bar{t}$ backgrounds as well as between $N$-body configurations with different values of $N$.

\begin{figure}[t!]
\centering
\subfloat[]{
       \includegraphics[width=0.45\textwidth]{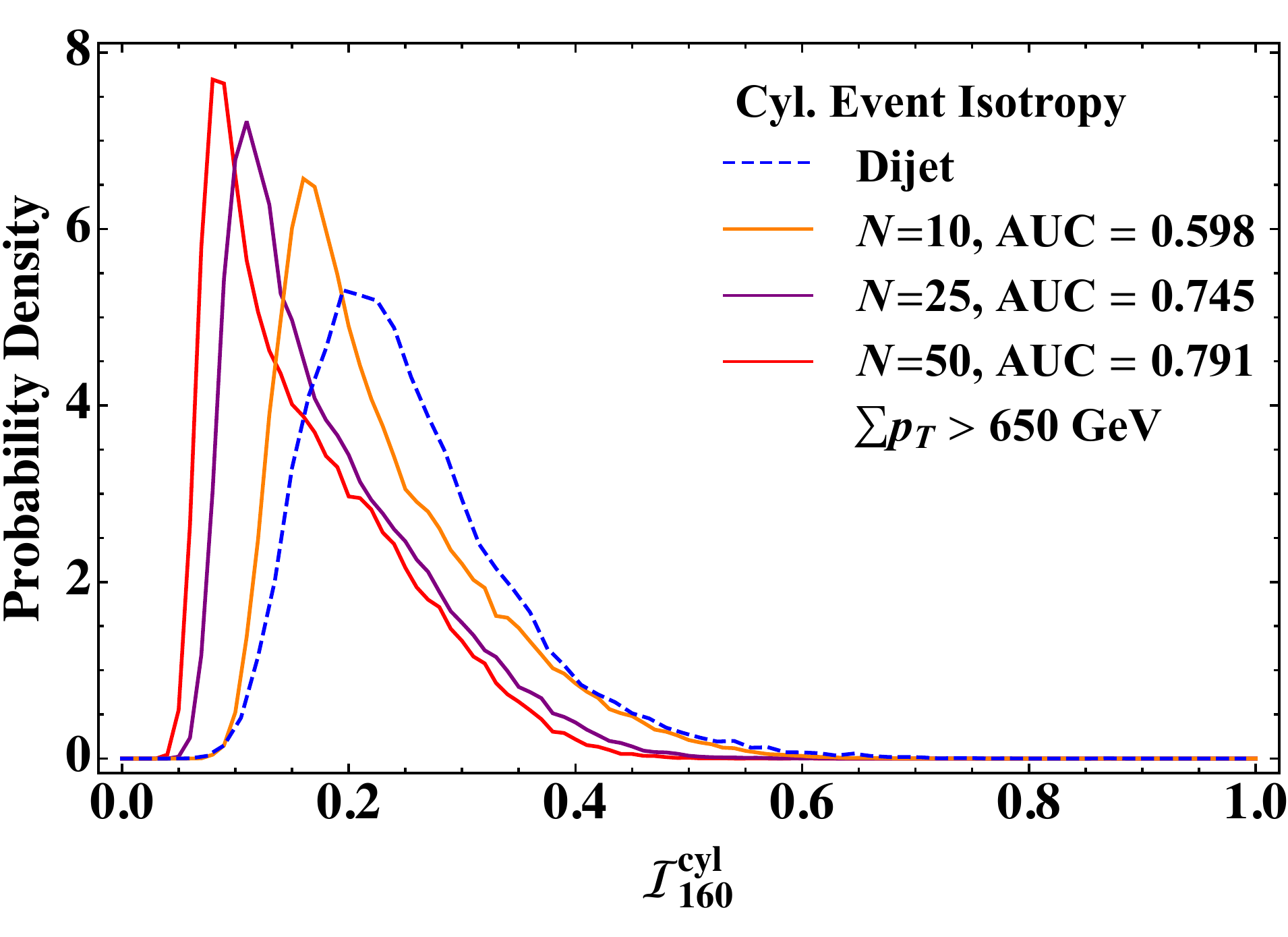}
       \label{fig:ppRamboSpec_cyl}
     }
     \hfill
     \subfloat[]{
       \includegraphics[width=0.45\textwidth]{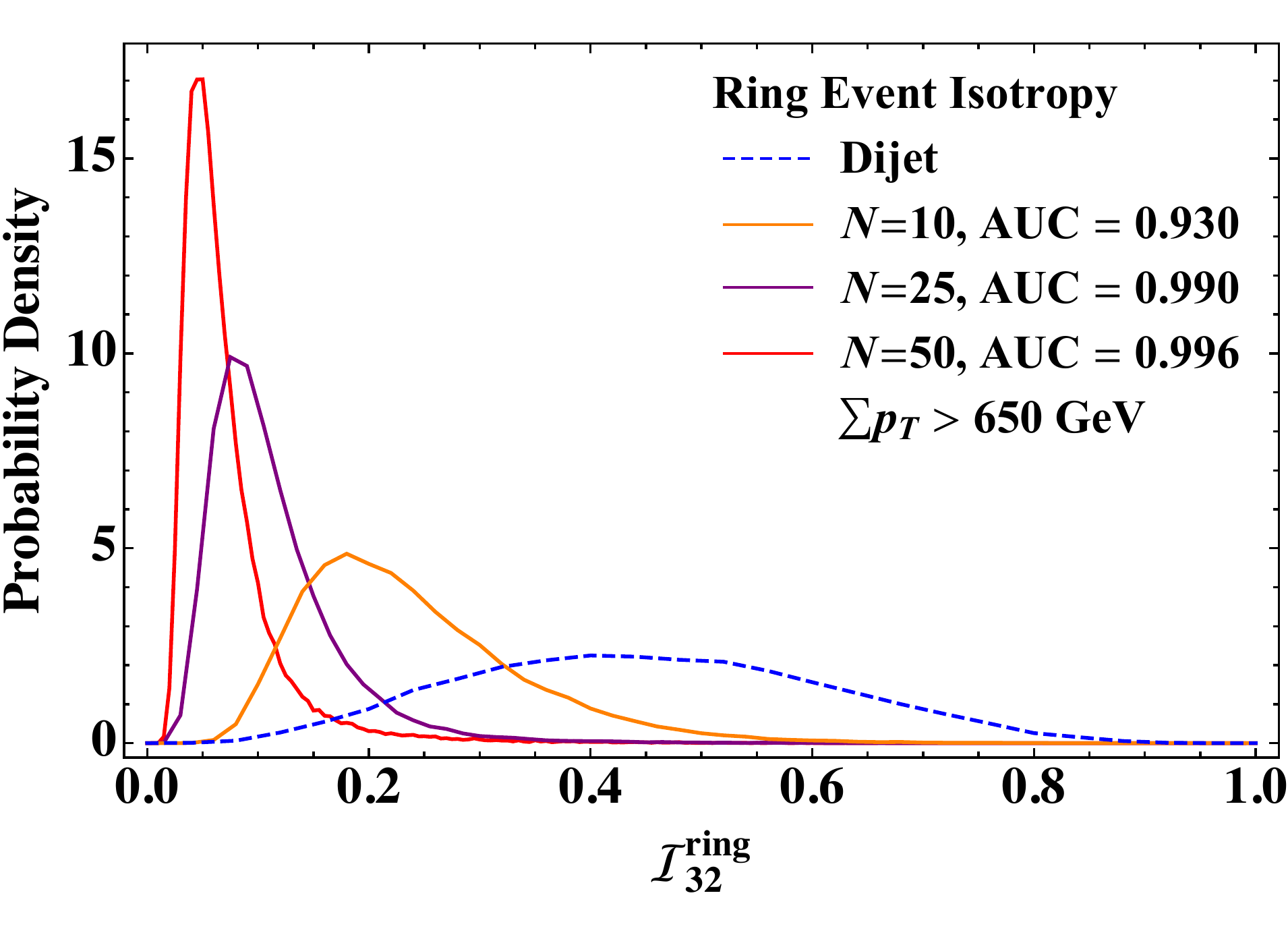}
              \label{fig:ppRamboSpec_ring}
     }
     \hfill
     \subfloat[]{
       \includegraphics[width=0.45\textwidth]{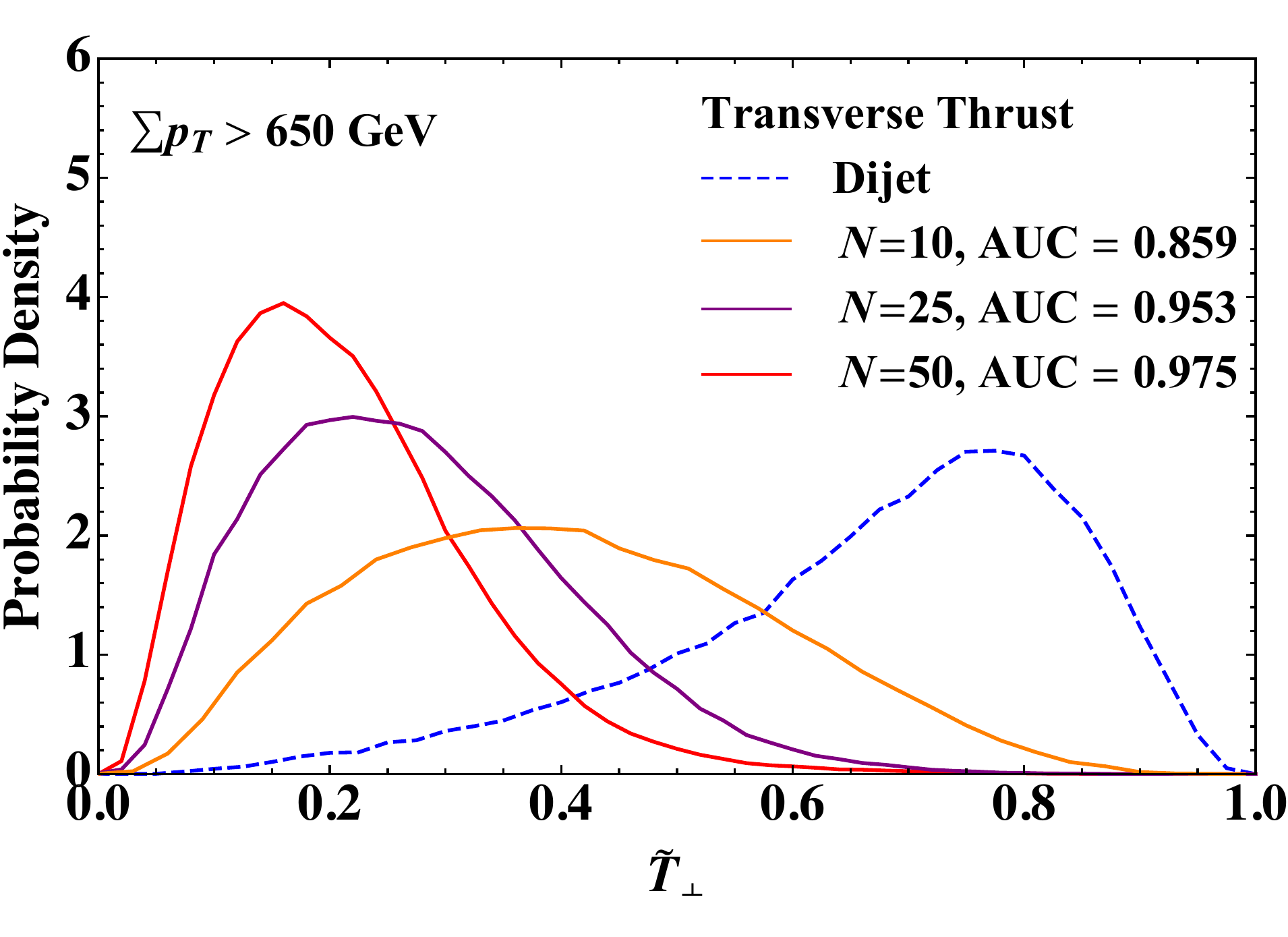}
     }
\caption{Comparing the QCD dijet background to uniform $N$-body samples with $N = \{10,25,50\}$, using the same observables as \Fig{fig:ppSpec} in $pp$ collisions at $\sqrt{s} = 14$ TeV.
The AUC values in the legend correspond to the separation power with respect to the dijet sample.
}
\label{fig:ppRamboSpec}
\end{figure}
\begin{figure}[t!]
\centering
\subfloat[]{
       \includegraphics[width=0.45\textwidth]{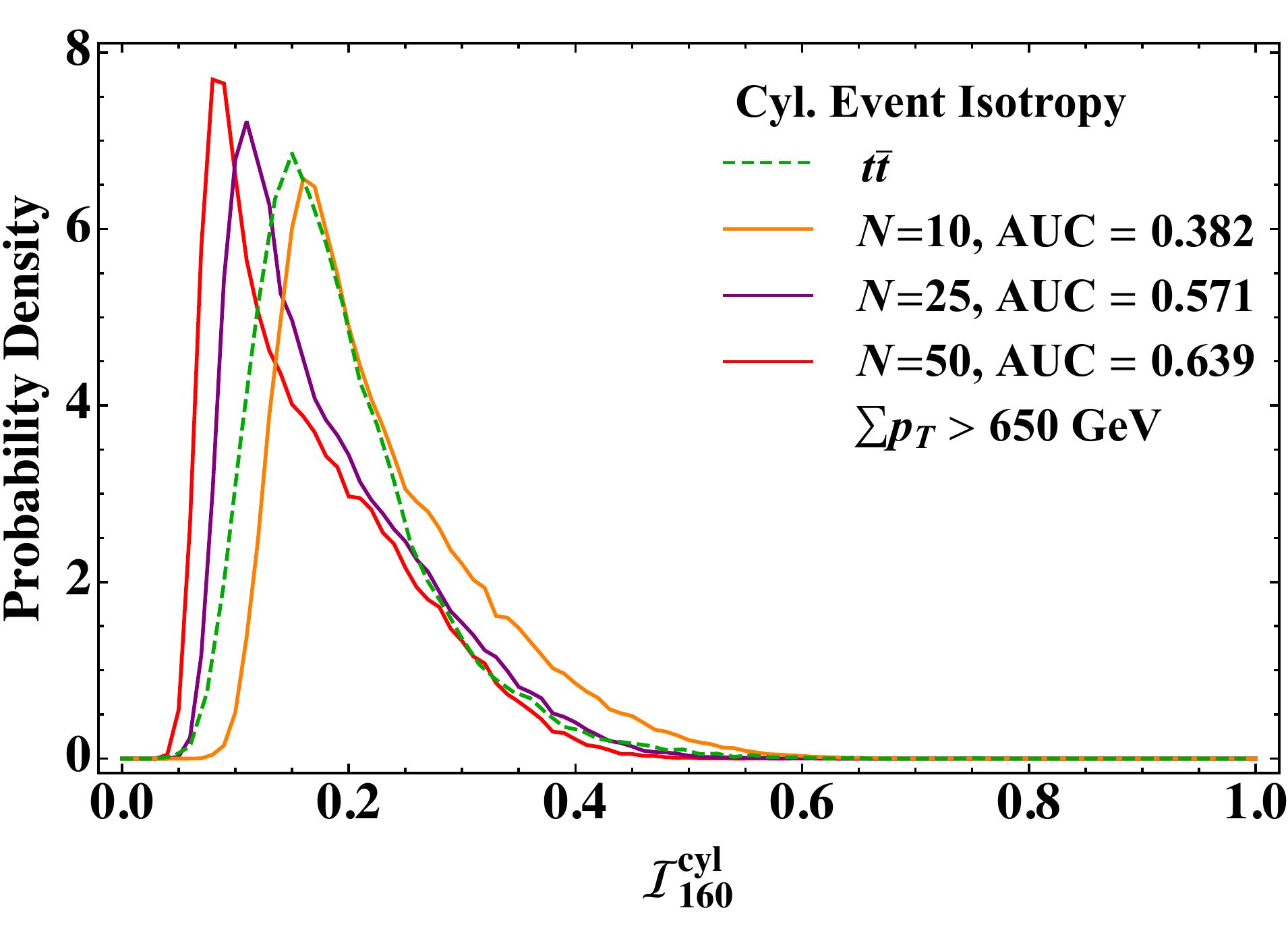}
       \label{fig:ppRamboSpecTT_cyl}
     }
     \hfill
     \subfloat[]{
       \includegraphics[width=0.45\textwidth]{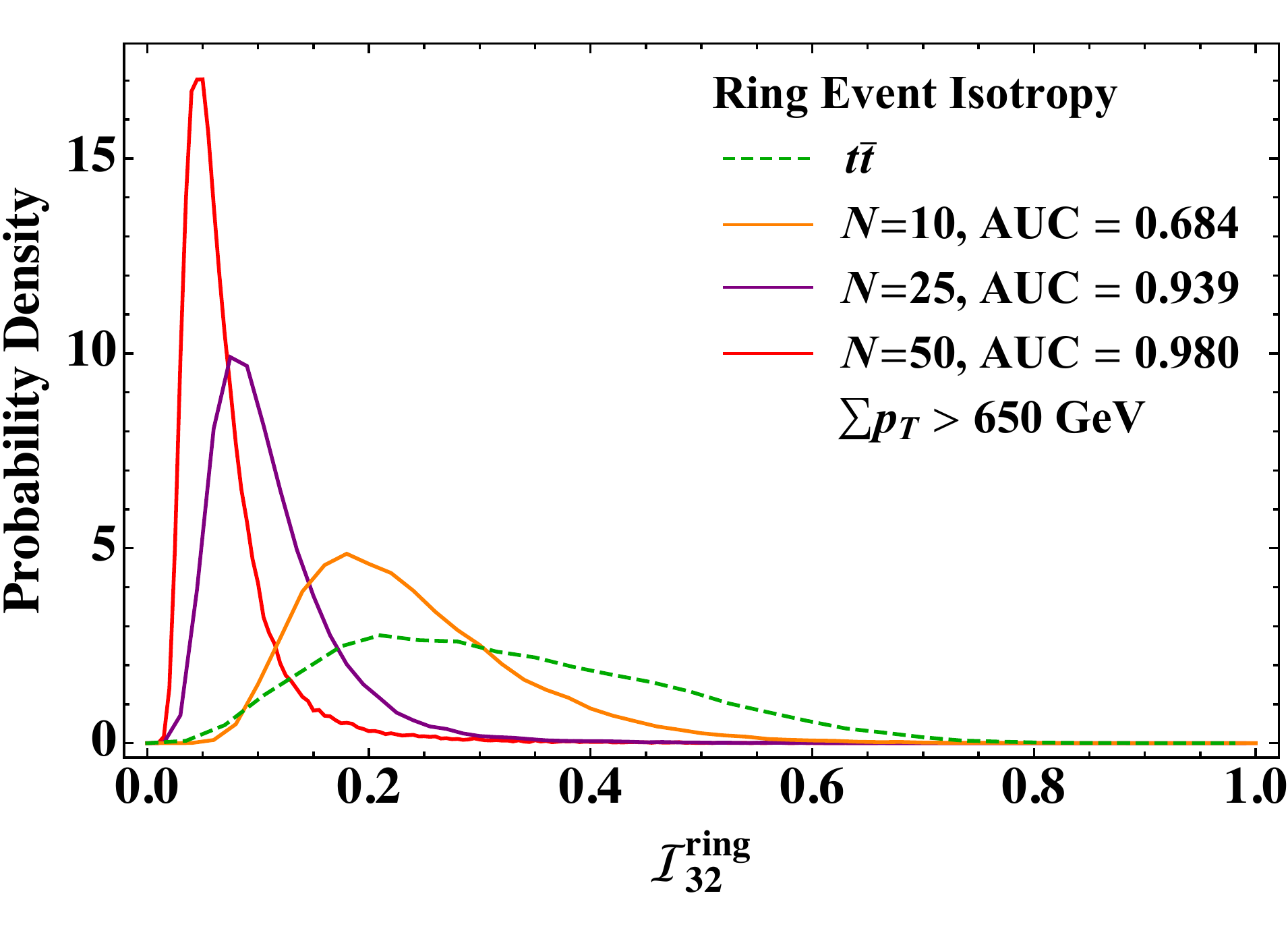}
     }
     \hfill
     \subfloat[]{
       \includegraphics[width=0.45\textwidth]{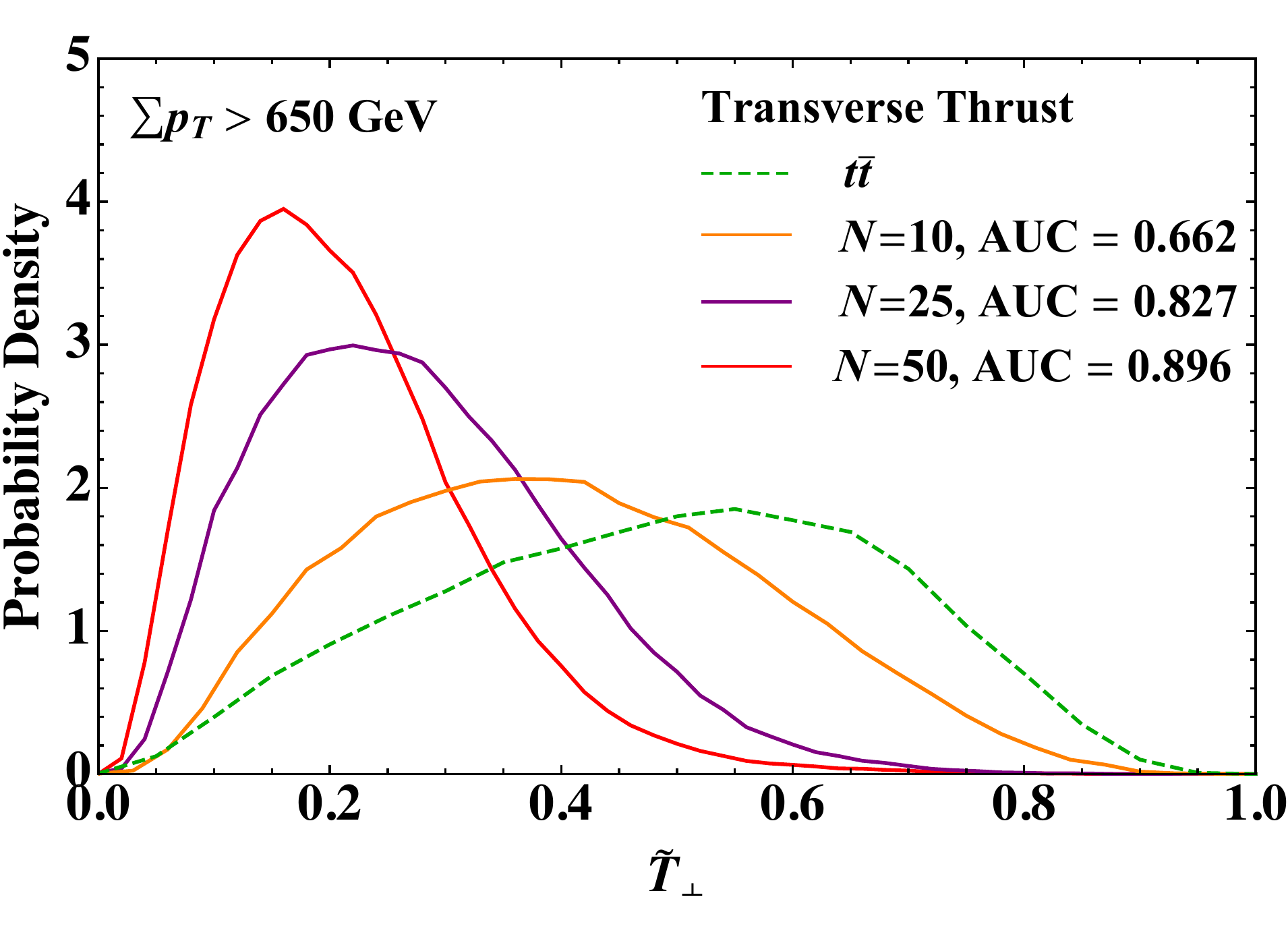}
     }
\caption{The same as \Fig{fig:ppRamboSpec}, but now compared to the $t\bar{t}$ background.}
\label{fig:ppRamboSpecTT}
\end{figure}

\begin{figure}[t!]
\centering
\subfloat[]{
       \includegraphics[width=0.45\textwidth]{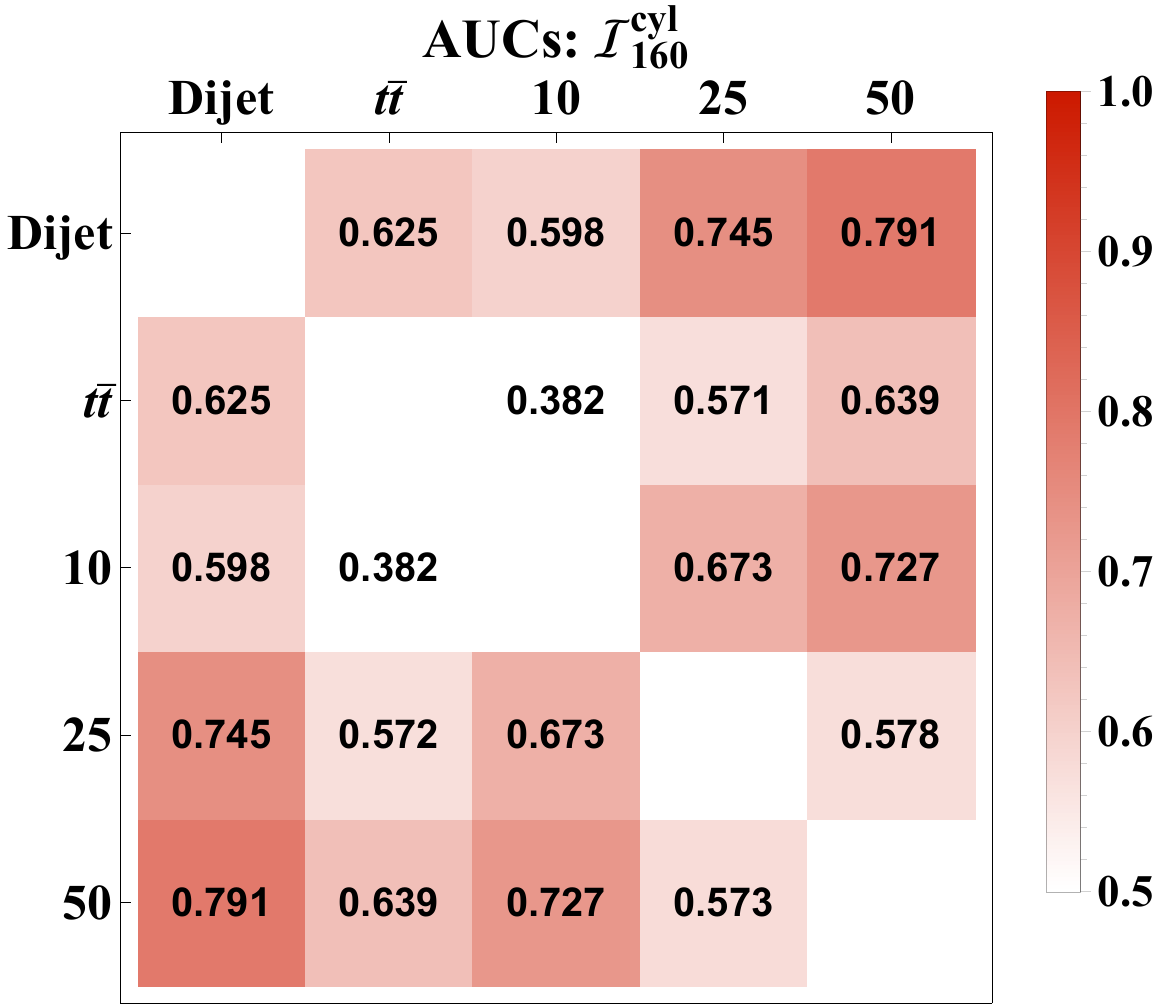}
     }
     \hfill
     \subfloat[]{
       \includegraphics[width=0.45\textwidth]{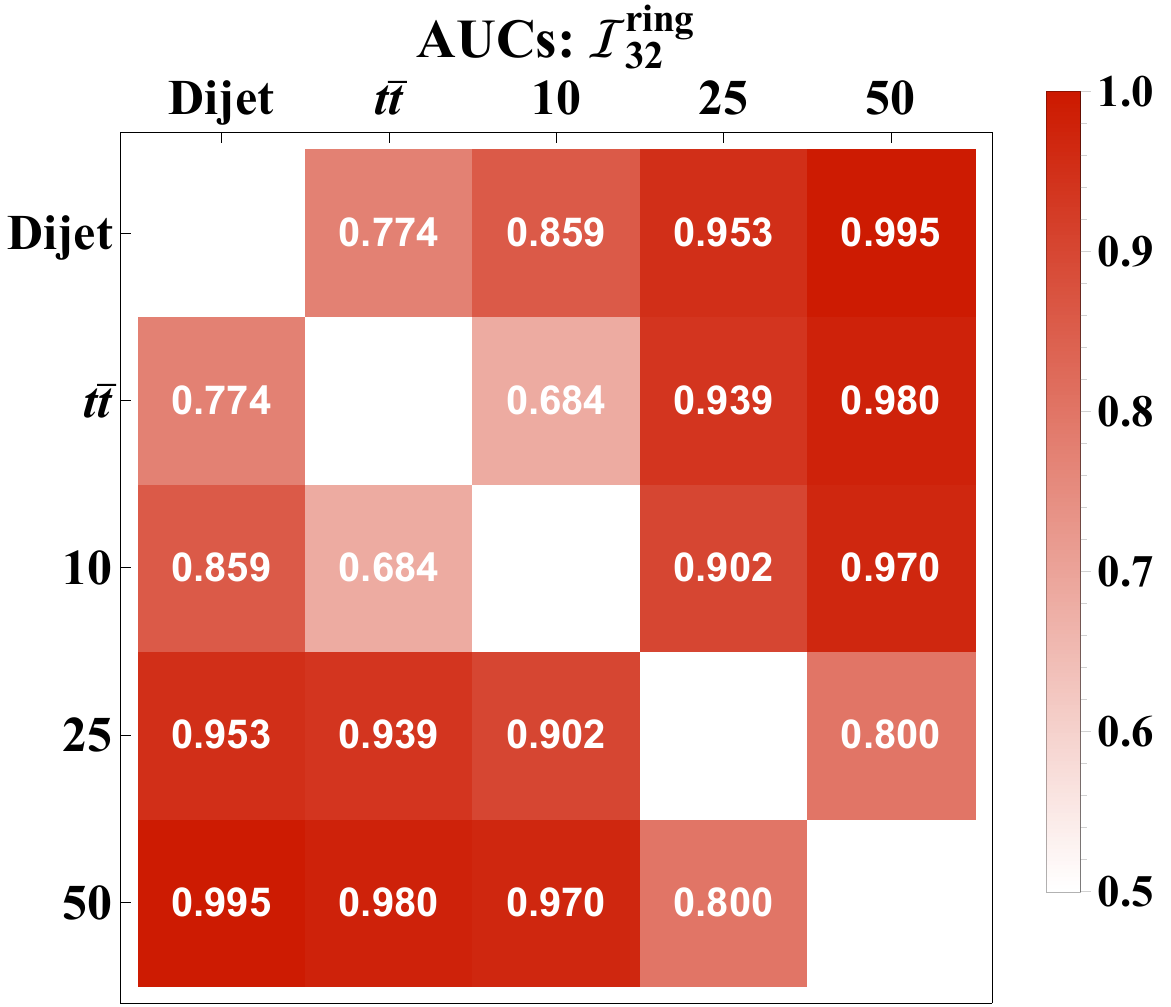}
     }
     \hfill
     \subfloat[]{
       \includegraphics[width=0.45\textwidth]{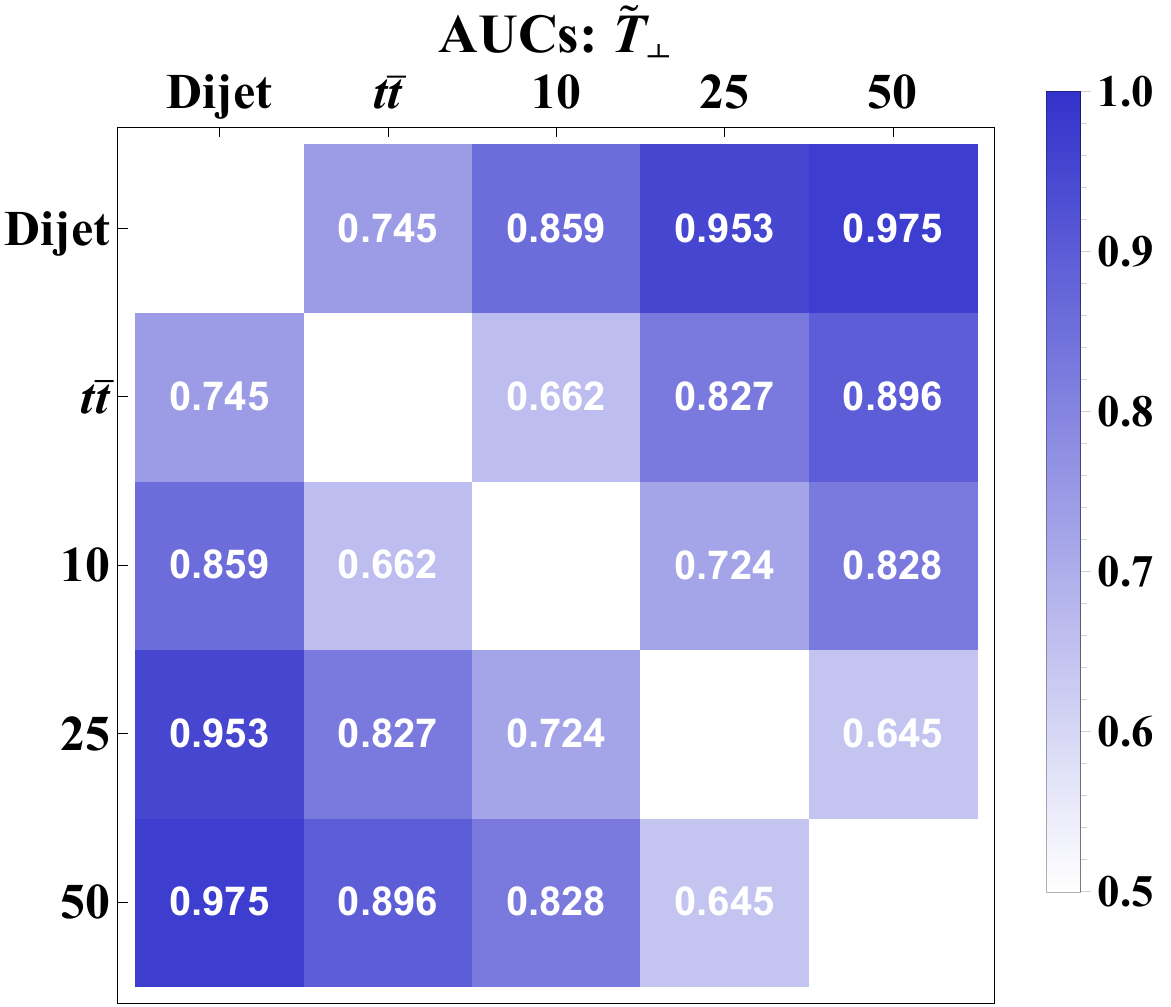}
     }
\caption{
The discrimination power between each pair of samples plotted in \Figs{fig:ppRamboSpec}{fig:ppRamboSpecTT}, as summarized by the AUC for (a) cylindrical event isotropy with $n = 160$, (b) ring-like event isotropy with $n = 32$, and (c) transverse thrust.
Darker shades indicate more efficient discrimination.
Cylindrical event isotropy is not as powerful of a discriminant as the other two observables which project out longitudinal information.
Ring-like event isotropy is the most powerful discriminant for $N\gg1$.
}
\label{fig:ppAUCtables}
\end{figure}

Distributions for $N$-body and QCD dijet samples with $10^5$ events are shown in \Fig{fig:ppRamboSpec}.
As in the dijet versus $t \bar{t}$ study, the two observables that project out the rapidity information---transverse thrust and ring-like event isotropy---are much more effective discriminants, with the best performance achieved by ring-like event isotropy when $N\gg1$.
Comparing the $N$-body samples to the $t\bar{t}$ sample in \Fig{fig:ppRamboSpecTT}, there is significant overlap between $N=10$ and $t\bar{t}$ for all event shape observables, analogous to what was seen in the $e^+e^-$ study from \Sec{subsec:uniformNbody_ee}. 
We also consider the separation power of these observables when comparing uniform $N$-body samples with different values of $N$.
The AUC values are summarized in \Fig{fig:ppAUCtables}, which demonstrates that ring-like event isotropy is consistently more effective than transverse thrust for this task.

\begin{figure}[p]
\centering
 \subfloat[]{
\includegraphics[width=0.45\textwidth]{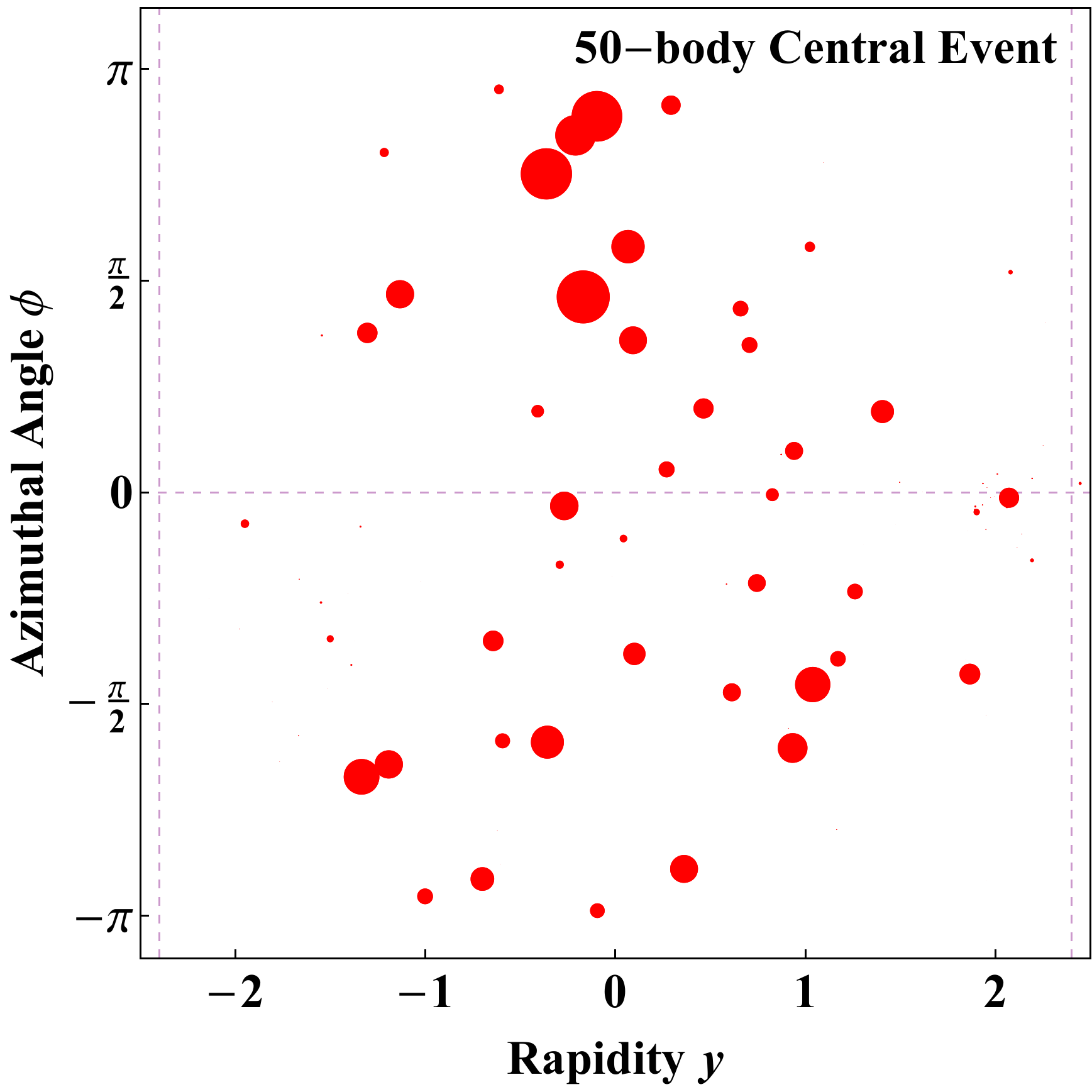}
}
  \hfill
\subfloat[]{
       \includegraphics[width=0.45\textwidth]{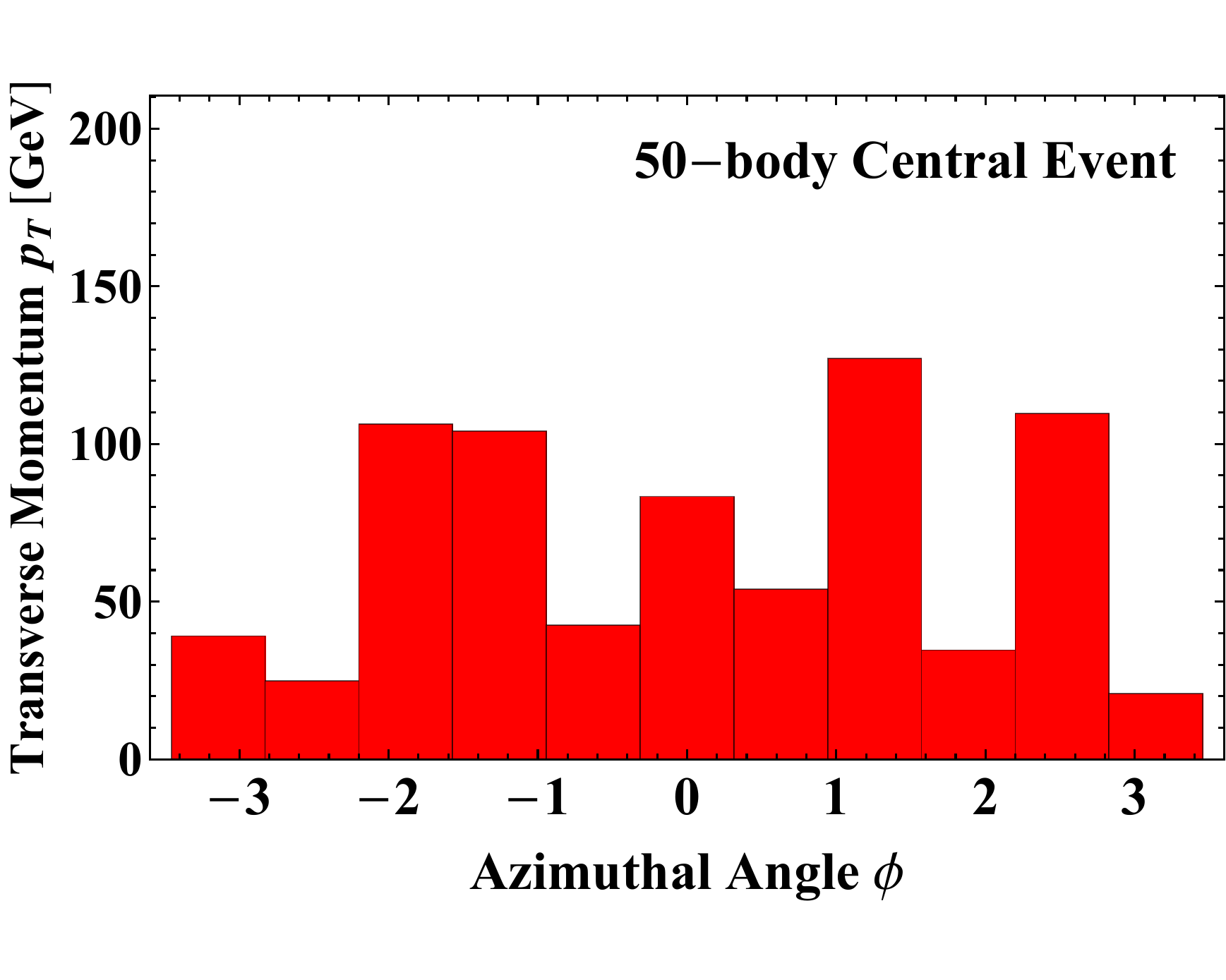}
     }
 \hfill
 \subfloat[]{
\includegraphics[width=0.45\textwidth]{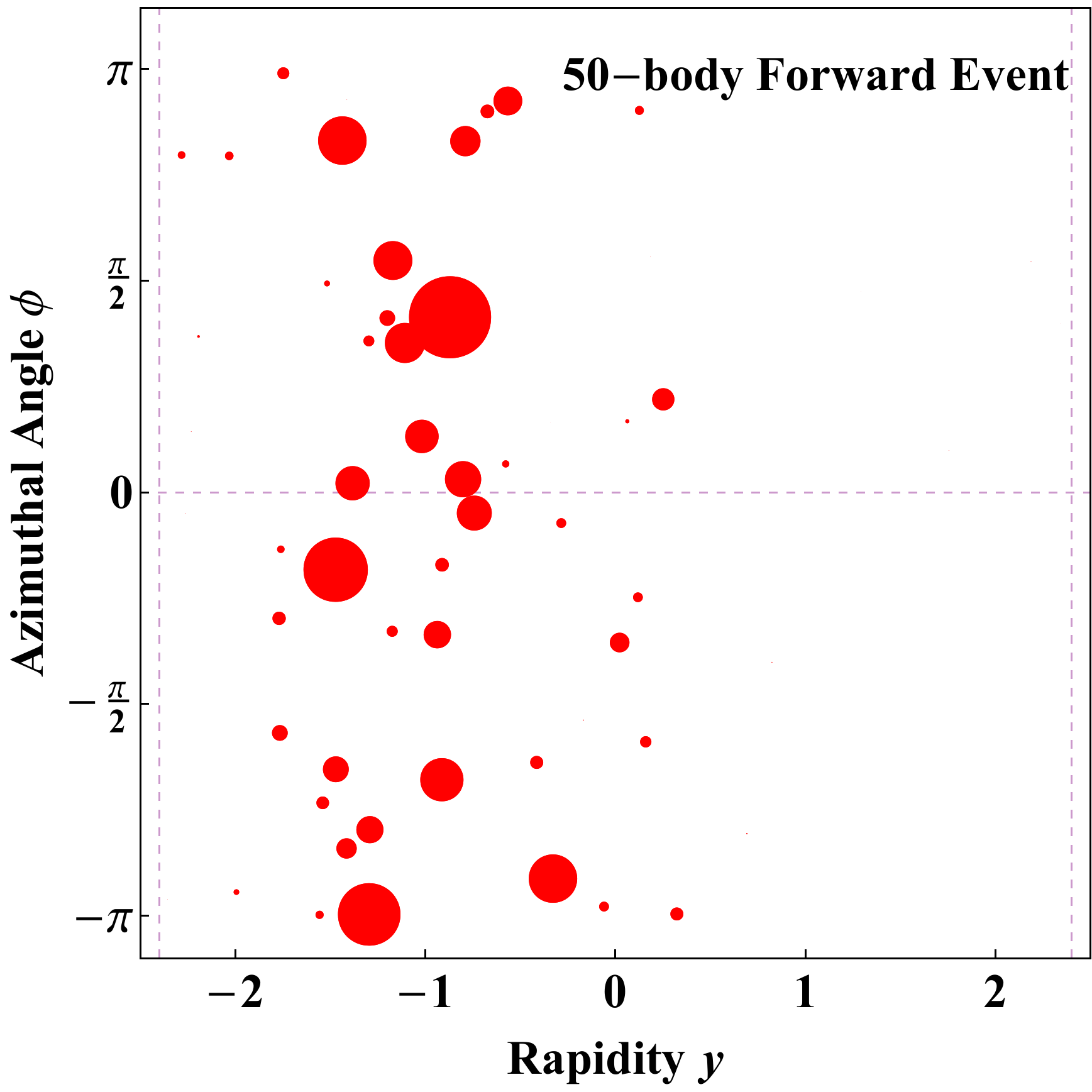}
}
  \hfill
\subfloat[]{
       \includegraphics[width=0.45\textwidth]{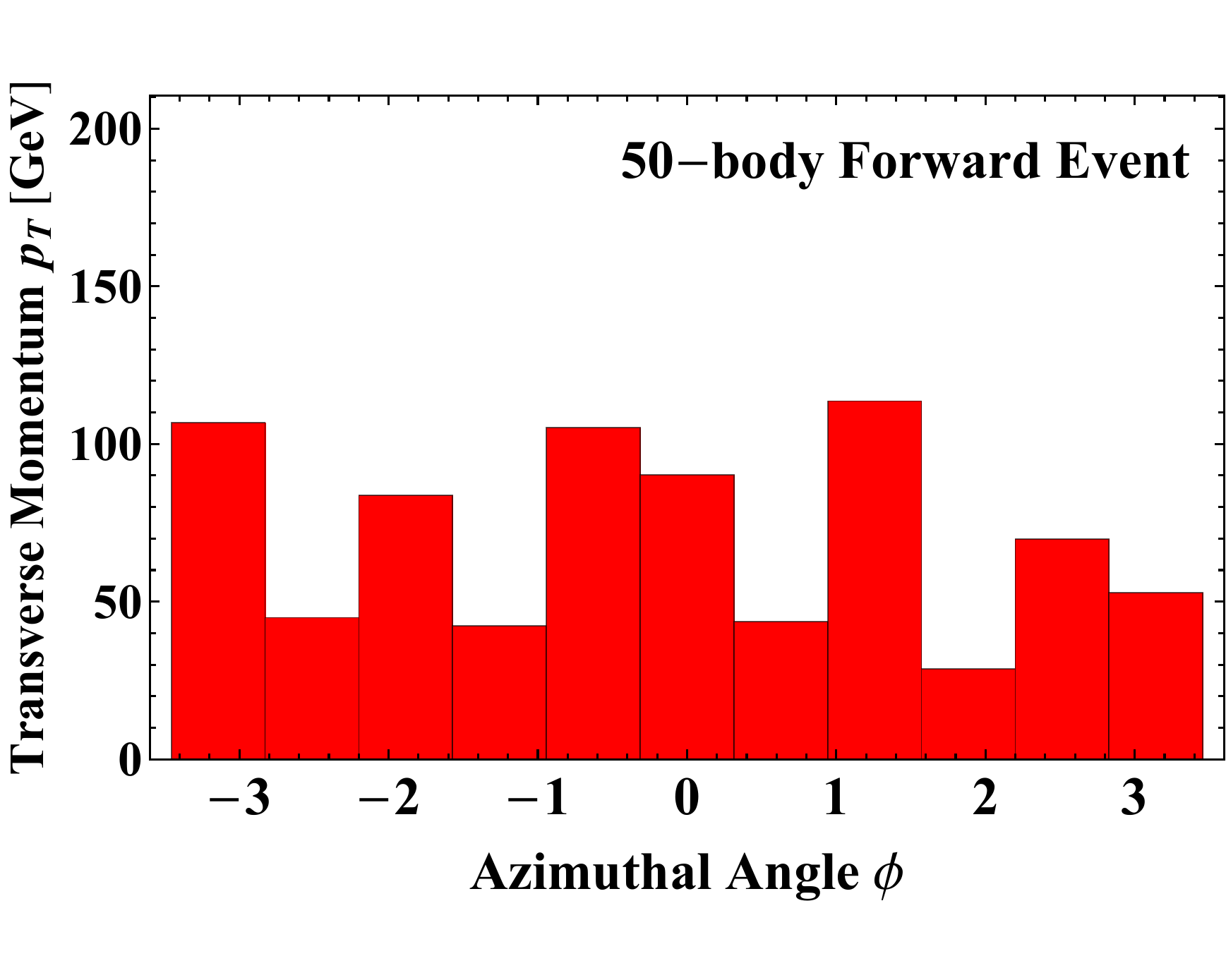}
     }

\caption{Two uniform 50-body phase space events from $pp$ collisions, in (left column) the rapidity-azimuth plane where the size of the marker corresponds to the $p_T$ of the particle and (right column) projected to the azimuthal direction.
The distributions in azimuth are reasonably isotropic in both events.
The first event (top row) is centered in rapidity, while the second event (bottom row) is longitudinally boosted with the particles clustered to one side of the cylinder.}
\label{fig:projVis}
\end{figure}

\begin{figure}[p]
\centering
 \subfloat[]{
\includegraphics[width=0.45\textwidth]{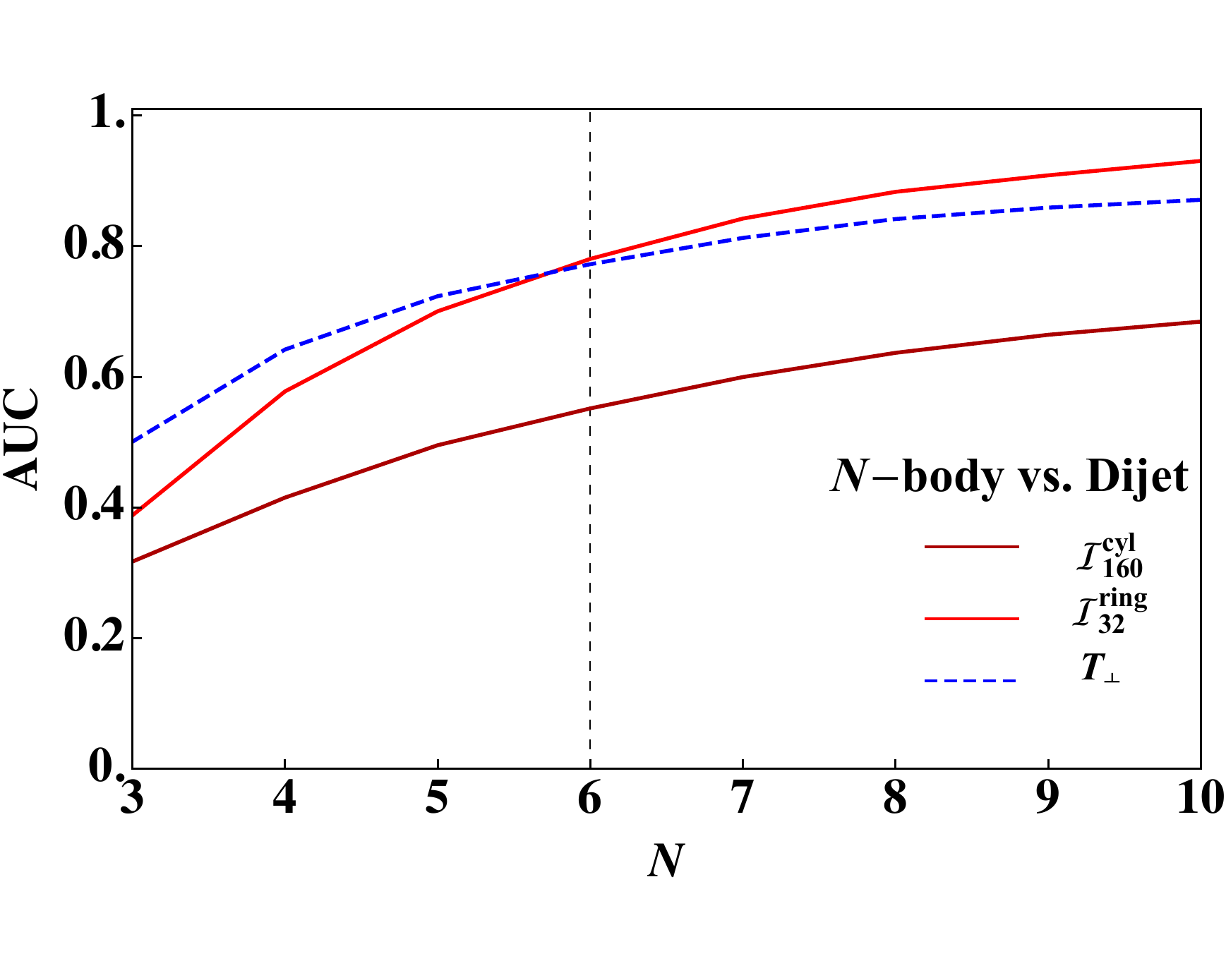}   
}
\hfill
 \subfloat[]{
\includegraphics[width=0.45\textwidth]{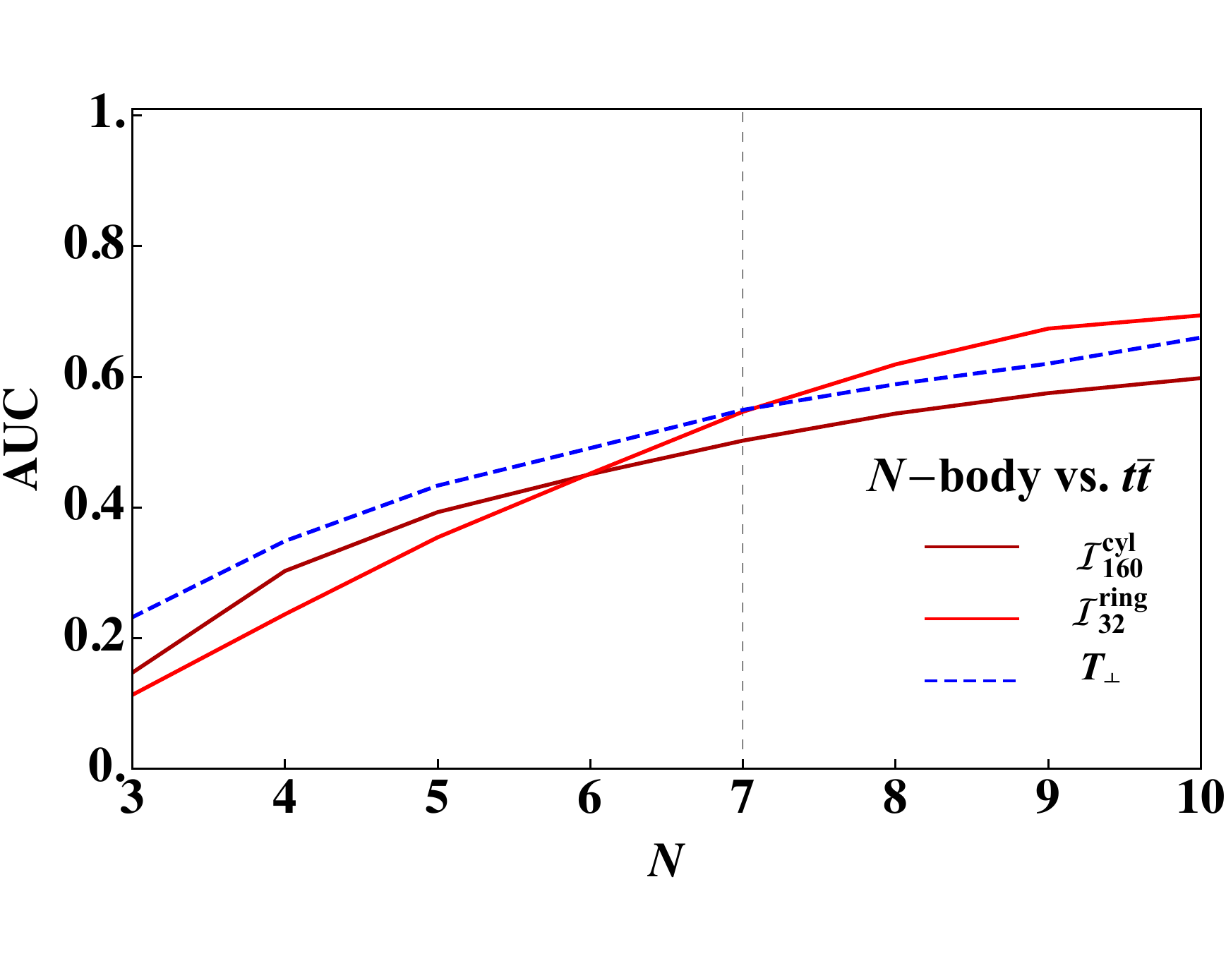}   
}
\caption{The AUC values as a function of $N$ for discriminating uniform $N$-body decays of a 1 TeV resonance from (a) QCD dijet events and (b) $t\bar{t}$ events.
Shown are cylindrical event isotropy, ring-like event isotropy, and transverse thrust, as computed in $pp$ collisions at $\sqrt{s} = 14$ TeV.
In general, transverse thrust is a better discriminant for low multiplicity events, but ring-like event isotropy is a better discriminant beyond $N = 6$ ($N = 7$) for the dijet ($t\bar{t}$) background. 
Values of $\text{AUC} <0.5$ occur when the SM sample is more isotropic than the $N$-body phase space sample according to the event shape observable.
}
\label{fig:AUCofRamboPP}
\end{figure}

\begin{figure}[p]
\centering
     \subfloat[]{
       \includegraphics[width=0.45\textwidth]{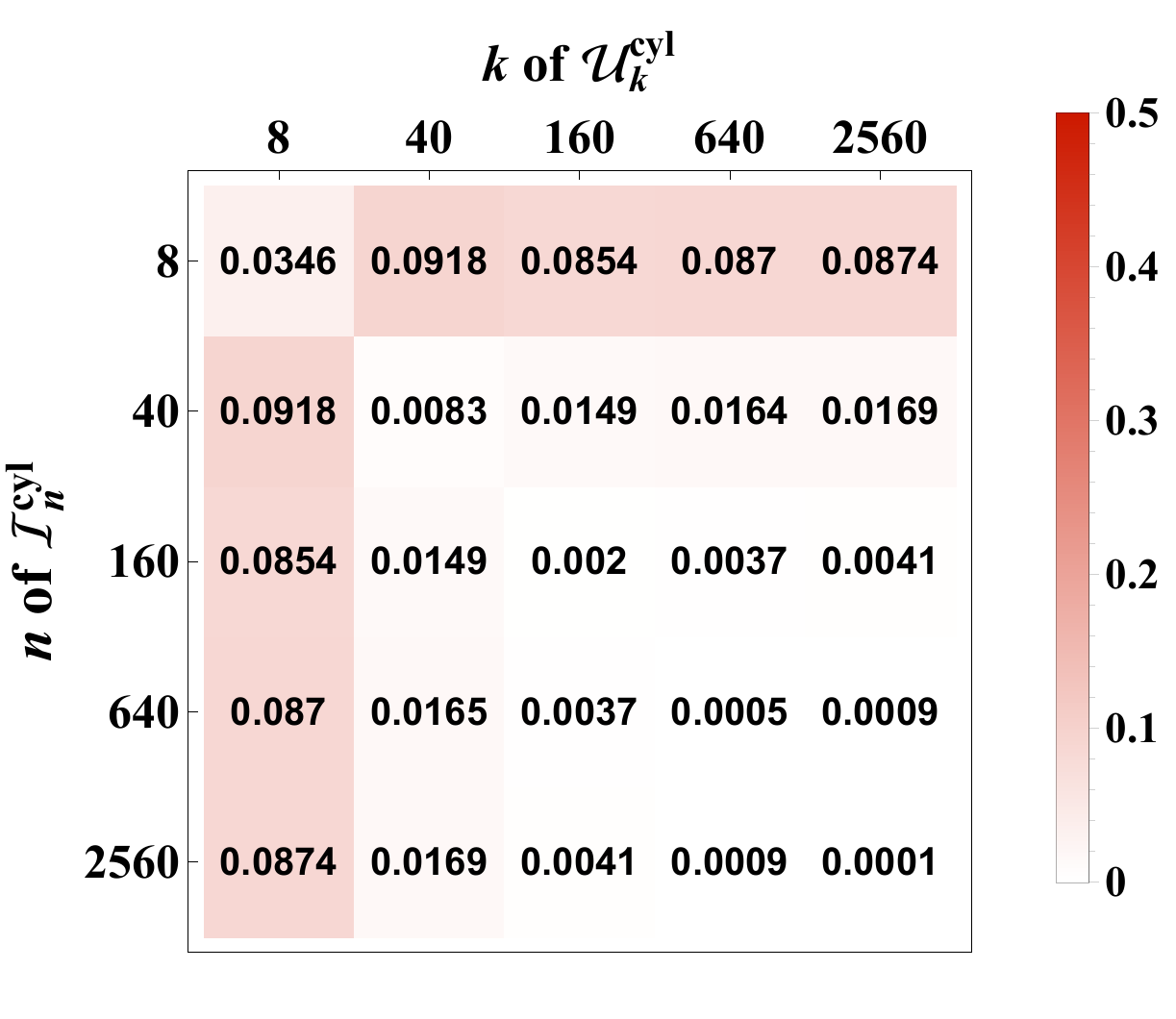}
     }
     \hfill
\subfloat[]{
       \includegraphics[width=0.45\textwidth]{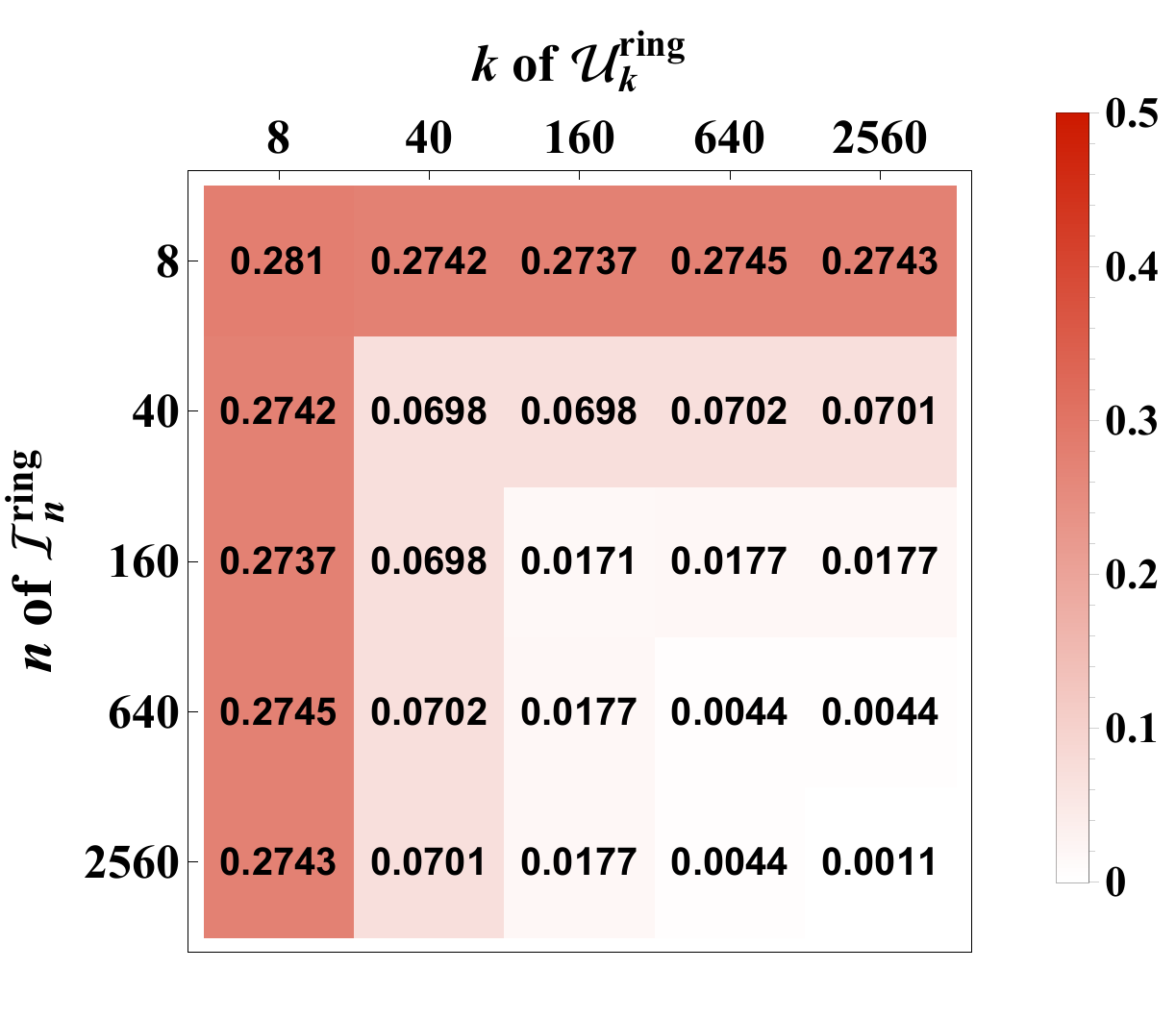}
     }
     \hfill
\caption{The event isotropy for $k$-body quasi-isotropic samples, averaged over random orientations.  Shown are (a) $\iso{cyl}{n}$ computed on cylindrical events and (b) $\iso{ring}{n}$ computed on ring-like events.
The distance from the coarsest cylinder to the finest cylinder ($\iso{cyl}{2560}(\mathcal{U}^\text{cyl}_{8})$ =  $\iso{cyl}{8}(\mathcal{U}^\text{cyl}_{2560}) \simeq 0.09$) is useful benchmark for when an event should be considered ``cylindrically isotropic.''
For the ring-like case, we recommend $\iso{ring}{64}(\mathcal{U}^\text{ring}_{8})$ =  $\iso{ring}{8}(\mathcal{U}^\text{ring}_{64}) \simeq 0.07$ as a more appropriate isotropy benchmark.
}
\label{fig:ppisoAUCtables}
\end{figure}

In general, cylindrical event isotropy exhibits poor discrimination power between the $N$-body samples.
While the peaks of the $N$-body samples are distinct in \Fig{fig:ppRamboSpec_cyl}, the separation power is degraded by the large tails from varying longitudinal boosts of the hard process along the beamline.
While the $N$-body event isotropy distributions approach zero when using the ring-like geometry, there is a persistent shift away from zero for the cylindrical geometry. 
This is because $\iso{cyl}{}$ can only be zero when the momentum distribution is uniform across both the rapidity and azimuthal directions.
Since we cannot impose longitudinal momentum conservation within the detector, events will often be centered around a nonzero value of $y$ and smeared asymmetrically.
An example of this is shown in \Fig{fig:projVis}, which compares two events that are isotropic in azimuth but have different rapidity distributions due to this longitudinal boost.
We advise that the ring-like geometry be used for a robust search for isotropic new physics signatures at hadron colliders, and that the cylindrical geometry only be used when motivated by the parameters of the model.%
\footnote{We also considered a variant of cylindrical event isotropy where the entire event is first boosted to $y = 0$ prior to computing the isotropy.
This variant exhibited roughly the same (poor) performance as the original unboosted version.}

In \Fig{fig:AUCofRamboPP}, we plot the AUC between the SM backgrounds and the uniform $N$-body phase space samples as a function of $N$.
Note that AUC $< 0.5$ means that the $N$-body sample is \emph{less} isotropic than the SM events.
For $N \lesssim 6$, the cylindrical event isotropy finds both the QCD dijet and $t\bar{t}$ samples to be more isotropic than the $N$-body sample. 
This can be seen from \Fig{fig:ppVis}, where the QCD dijets often have at least three prongs of energy with some non-negligible smearing and the $t\bar{t}$ events have around six hard prongs. 
Transverse thrust is the best discriminant for $N=\{3,4,5,6\}$ against both SM backgrounds.
Near $N=7$, however, ring-like event isotropy overtakes the performance of transverse thrust, and it continues to have better separation power for all higher values of $N$.
We observe that the value of $N$ at which (ring-like) event isotropy outperforms (transverse) thrust is similar between $e^+e^-$ and $pp$ collisions.

\subsection{Toy Model: Quasi-Isotropic Distributions}
\label{subsec:pp_kbody}

As a benchmark to understand when a $pp$-collider event should be considered isotropic, we calculate the event isotropy for $k$-body discretized uniform events. 
Cylindrically symmetric events are not very well motivated as a toy model for an isotropic process at $pp$ colliders since it requires events to be centered in $y$, which is nontrivially model-dependent. 
Once projected to the azimuthal direction, though, ring-like symmetry should be a generic new physics signature.

After averaging over $10^4$ random orientations, the event isotropies for the appropriate geometry are given in \Fig{fig:ppisoAUCtables}. 
We checked that these values only have weak dependence on the rapidity acceptance limits, though the computation time of $\iso{cyl}{}$ suffers when considering an extended rapidity region due to increasing numbers of particles in the comparison event.
As expected, the largest values of event isotropy are seen for the coarsest tilings.
For the cylindrical case, this value is $\iso{cyl}{n}(\mathcal{U}^\text{cyl}_{12}) \simeq 0.09$, which aligns with the peak of the $N = 50$ sample in \Fig{fig:ppRamboSpec_cyl} and is therefore a reasonable benchmark for when an event should be regarded as ``cylindrically isotropic.''
For the ring-like case, $k = 4$ is probably insufficiently isotropic to be useful as a benchmark.
Instead, we recommend $\iso{cyl}{n}(\mathcal{U}^\text{ring}_{8}) \simeq 0.07$ as corresponding to ``ring isotropic,'' which again aligns with the $N = 50$ peak in \Fig{fig:ppRamboSpec_ring}.


\section{Soft Unclustered Energy Patterns at the LHC}
\label{sec:SUEP}

In this section, we move our discussion from the use of event isotropy for discriminating between event categories to its use in characterizing new physics.
A novel signal topology was introduced in \Ref{Strassler:2008bv} and dubbed ``soft unclustered energy patterns'' (SUEP) by the authors of \Ref{Knapen:2016hky}.%
\footnote{SUEPs were previously called ``soft bombs'' but were rebranded to not tempt any negative press on CERN.}
This topology is motivated by strongly-coupled hidden valley models that have a weak coupling to the SM via a heavy scalar mediator. 
We assume that the hidden sector is quasi-conformal such that strong dynamics persist over the energy range of showering. 
If we consider the mass of the lightest meson in the hidden sector to be much smaller than the mass of the scalar mediator, then the final state will be high multiplicity with many soft particles. 
Here, we explore the structure of SUEP-like events by considering both the charged particle multiplicity and the event isotropy of the final state to understand how the new physics signature changes as a function of the model parameters. 

\subsection{Model Details and Analysis Strategy}

We simulate events of this model with the SUEP Generator \cite{suep_gen2020}, an add-on to \texttt{Pythia 8.243}.%
\footnote{We thank Simon Knapen for useful discussions, instruction, and access to the SUEP Generator.} 
The generator first simulates production of a massive scalar with mass $m_S$. 
The mediator is forced to decay with a 100\% branching ratio into the hidden sector and cascades to a final state composed of the lightest scalar meson $\phi_{\rm hid}$ with mass $m$. 
The momenta of the hidden mesons is modeled as an isotropic relativistic Maxwell-Boltzmann distribution at a temperature $T$.
These mesons decay promptly via the effective operator $\frac{\alpha_{D}}{\Lambda} \phi_{\rm hid} \bar{u} u$, then undergo standard hadronization in \texttt{Pythia}.
For simplicity, we only consider coupling to $u\bar{u}$ although future studies could examine additional couplings to gluons or other quarks.
The full SUEP event contains not only the SM final state of the hidden sector decay after hadronization, but also ISR and underlying event associated with the $pp$ scattering process.

For our study, the relevant free parameters of the model are $m_S$, $m$, and $T$.
In the original benchmark studies in \Ref{Knapen:2016hky}, several values of $m_S$ were considered, and the parameters of the model were chosen to produce very soft particles in the hidden valley:  $m = 1$ GeV, $T = 0.5$ GeV. 
Here, we focus on the rare Higgs decay scenario $m_S = m_h$ and allow all SM production modes for the Higgs boson, which now plays the role of the mediator. 
The specifics of the model set the values of the coupling $\alpha_D$ and energy scale $\Lambda \sim T, m$ of the theory. 
These parameters determine the lifetime and therefore the displacement of the $\phi_{\rm hid}$ meson decays. 
While displaced vertex signatures have been considered for SUEP signal detection, to do so is nontrivial \cite{Alimena:2019zri}, so displacement will not be a component of the present study.

We run the SUEP Generator with $pp$ collisions at $\sqrt{s} = 14$ TeV for $10^5$ events.
We consider different values of $T$ and $m$ around the GeV scale, but restrict our attention to $T \sim m$ as predicted by gauge/gravity duality~\cite{Hatta:2008qx}.   
As appropriate for $m_S = 125$ GeV, we apply kinematic cuts of $\sum_i {p_T}_i > 100$ GeV on all samples.
An analysis strategy is proposed in \Ref{Knapen:2016hky} for both existing ATLAS L1 $E_T^\text{miss}$ triggers as well as for a new high-level trigger specific to the SUEP model.
Here, we present a simplified analysis of truth-level information, neither simulating the ATLAS detector nor accounting for pileup.

The goal of this study is to understand how event isotropy can be used to characterize new physics event radiation patterns, without considering cuts needed to suppress SM backgrounds.
As shown in \Ref{Knapen:2016hky}, charged particle multiplicity is already an effective way to suppress backgrounds for SUEP events.
While event isotropy offers complementary information to charged particle multiplicity, in preliminary studies we did not find much of a gain from using event isotropy to suppress SM backgrounds. 
One reason is that event isotropy is an IRC-safe observable, so it only probes the degree of isotropy of an event and not the number of soft emissions.
The analyses in \Secs{subsec:uniformNbody_ee}{subsec:uniformNbody_pp} show that event isotropy is an effective probe of $\mathcal{O}(10)$-body phase space, but for multiplicities of $\mathcal{O}(100)$ or more as in SUEP events, simply counting the number of final-state particles is more effective as a discriminant against SM backgrounds.
On the other hand, we will find that event isotropy offers additional information beyond charged particle multiplicity in the context of signal characterization.

Since we do not attempt to simulate the full detector, we apply the fiducial and kinematic cuts of the ATLAS inner tracker \cite{Aaboud:2016itf}.
We assume that all charged SM particles with $p_T > 100$ MeV within the region $|\eta| < 2.4$ are reconstructed.
When computing the ring-like event isotropy, we include all particles (charged and neutral) that satisfy these fiducial cuts.%
\footnote{Note that this implies a slight mismatch for massive particles between pseudorapidity (used to define the fiducial region) and rapidity (used to compute the EMD ground measure).}
In practice, a charged particle with $p_T$ near 100 MeV does not leave hits in all the tracker layers, so a larger $p_T$ threshold might be necessary for higher fidelity track reconstruction.

One key difference from \Ref{Knapen:2016hky} is that in our study, we do not require the SUEP to be recoiling against another object in the event.
A SUEP event often produces a \textit{belt of fire} ring of high track multiplicity near the primary interaction point.
For such a signal to meet the triggering conditions of the L1 ATLAS trigger, the SUEP has to recoil against an ISR jet or associated vector boson or have a substantial amount of missing transverse energy $E_T^\text{miss}$.
In the case of SUEP production from rare Higgs decays, one would likely make use of triggers designed for vector-boson-plus-Higgs production~\cite{ATLAS-CONF-2020-007}.
A boost other than along the beam axis degrades the axial symmetry of the signature, and therefore leads to larger values of the ring-like event isotropy (i.e.~less isotropic).
For this exploratory study, we focus on the case where the SUEP is produced with small transverse recoil, leaving a study of more sophisticated quasi-isotropic observables for boosted signals to future work.
As one possible way to deal with signals with substantial $E_T^\text{\rm miss}$, one could first boost the recoiling SUEP by the missing transverse momentum vector to recover the quasi-uniformity of the radiation pattern before computing the event isotropy.

\subsection{Results}

We now compare the distributions of charged track multiplicity and event isotropy in SUEP models with different parameters.
Fixing $m = 1$ GeV, we consider $T = \{0.5, 1, 3, 5\}$ GeV, leaving additional values of $m$ to \App{app:suep}. 
As the model begins to break down when $T \gg m$, we do not consider values of $T \sim O(10)$ GeV.
When $T \lesssim m$, the hidden mesons are typically non-relativistic, leading to higher multiplicity events.
When $T > m$, the hidden mesons are typically relativistic, leading to lower multiplicity events.

\begin{figure}[t!]
\centering
\subfloat[]{
       \includegraphics[width=0.485\textwidth]{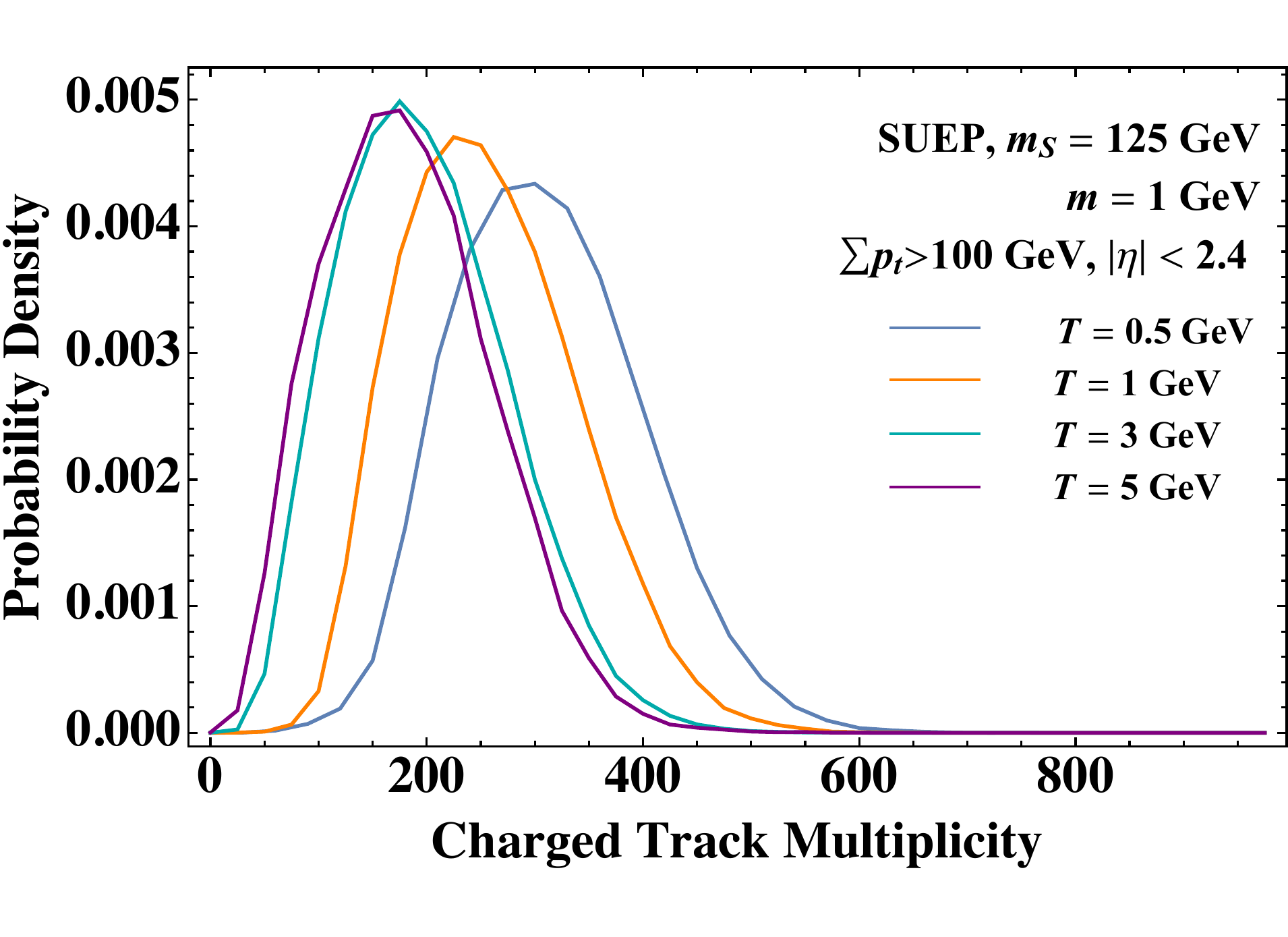}
       \label{fig:smallMassSUEP_mult}
     }
     \hfill
\subfloat[]{
       \includegraphics[width=0.46\textwidth]{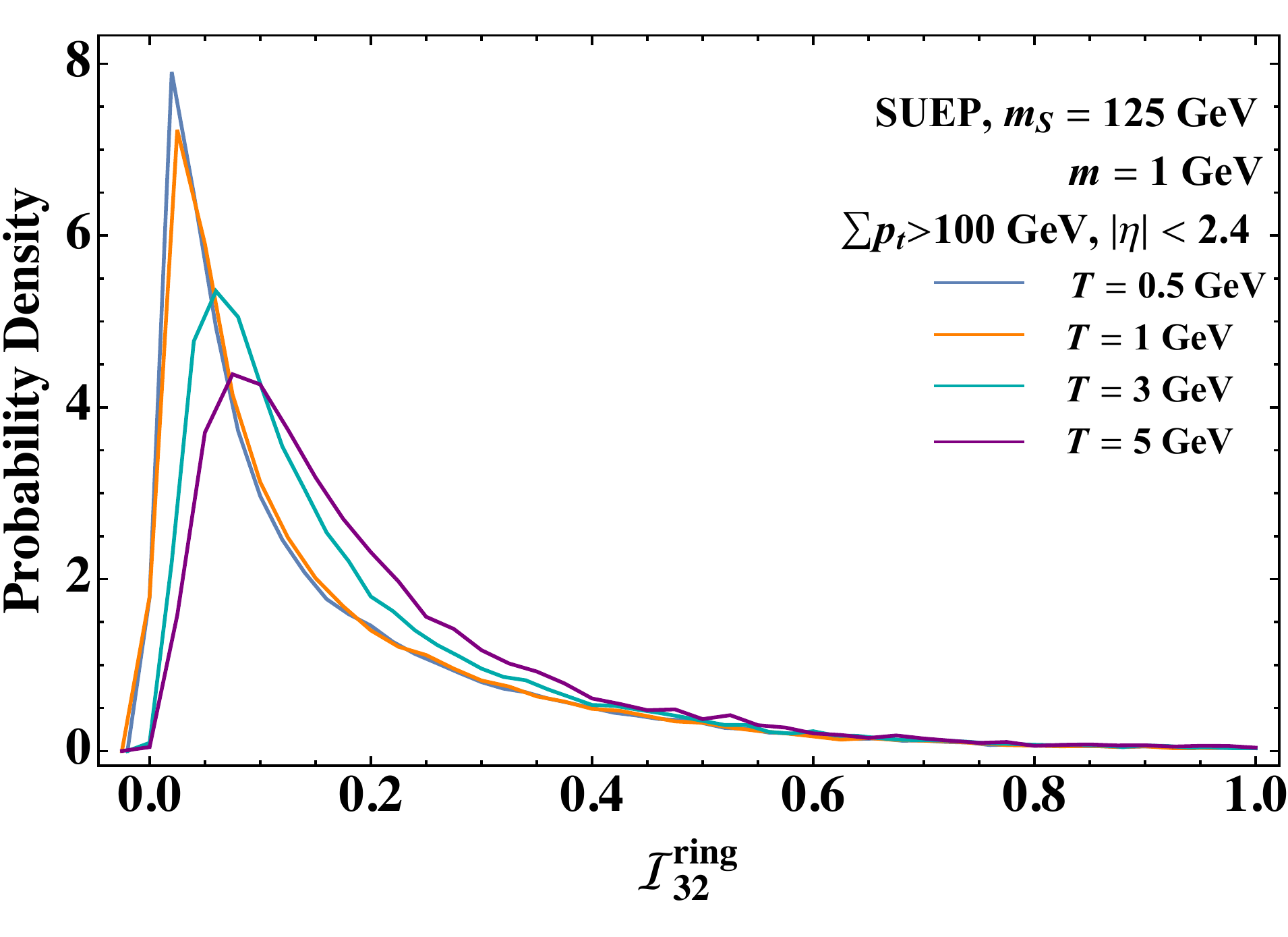}
       \label{fig:smallMassSUEP_iso}
     }
\caption{The distributions of (a) charged track multiplicity and (b) ring-like event isotropy for SUEP samples of mass $m = 1$ GeV and temperatures $T = \{0.5, 1, 3, 5\}$ GeV. 
We see that the distributions of track multiplicity for $T = \{3,5\}$ GeV are nearly overlapping, whereas the curves with similar event isotropy are $T = \{0.5, 1\}$ GeV.
For additional values of $m$, see \App{app:suep}.}
\label{fig:smallMassSUEP}
\end{figure}

We can identify several characteristic features in different regimes of parameter space. 
First, consider the one-dimensional distributions of charged track multiplicity and event isotropy, shown in \Fig{fig:smallMassSUEP}.
The charged track multiplicity distributions for $T = \{3, 5\}$ GeV are nearly identical, but they are noticeably different from $T = \{0.5, 1\}$ GeV.
The reverse is true for the event isotropy distributions, where the  distributions for $T = \{0.5, 1\}$ GeV are very similar but distinct from $T = \{3, 5\}$ GeV.
This can be understood as follows.
When $T \lesssim m$, the hidden mesons have low velocities and the events are more isotropic, so the main role of $T$ is to control the number of $\phi_{\rm hid}$ particles produced.
When $T > m$, the hidden mesons have high velocities, so the events have more of a clustered structure, and $T$ controls the degree of clustering without affecting much the number of $\phi_{\rm hid}$ particles.
In this way, charged track multiplicity and event isotropy offer complementary probes of the SUEP behavior.

\begin{figure}[t!]
\centering
\subfloat[]{
       \includegraphics[width=0.45\textwidth]{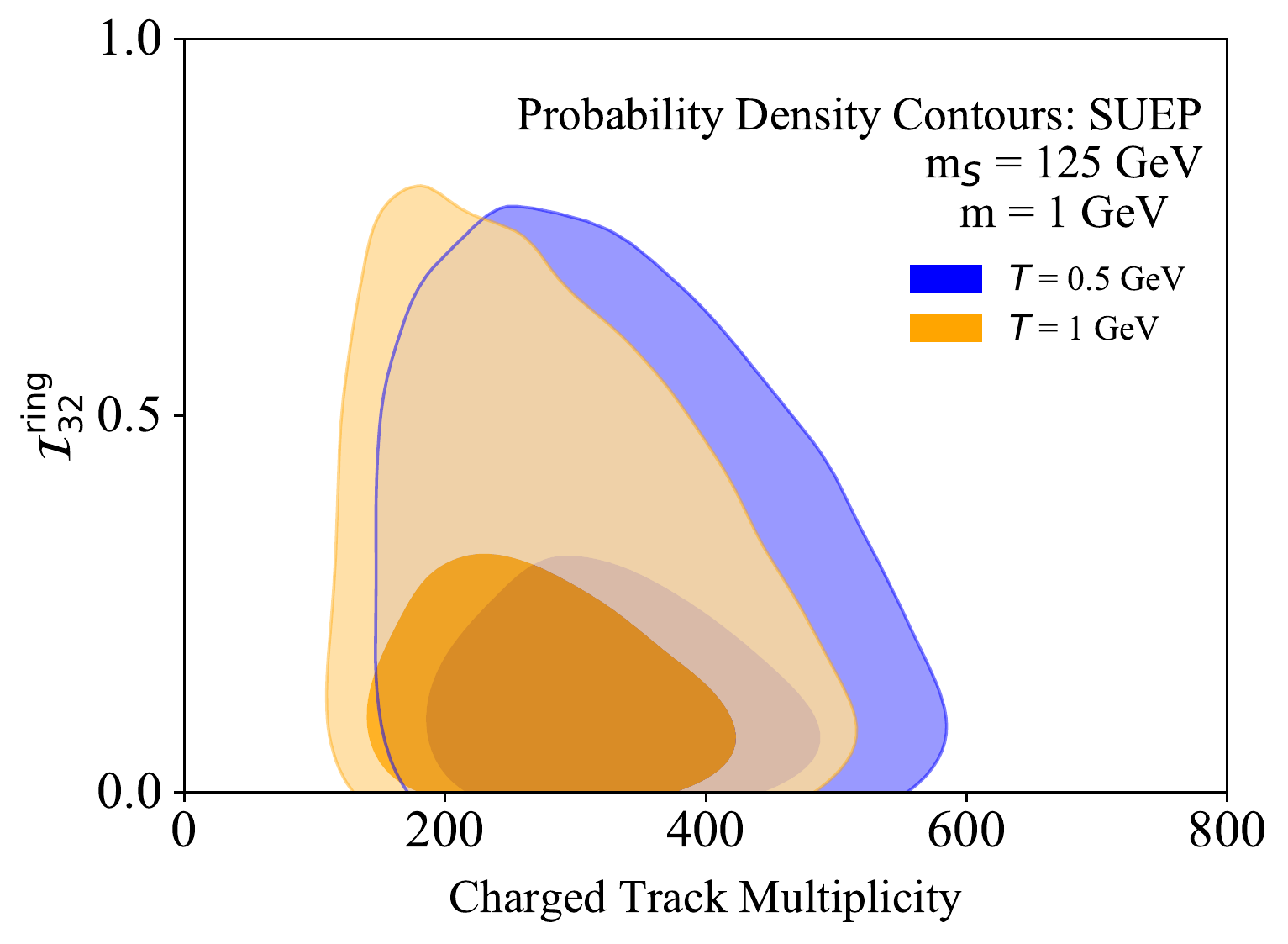}
     }
     \hfill
\subfloat[]{
       \includegraphics[width=0.45\textwidth]{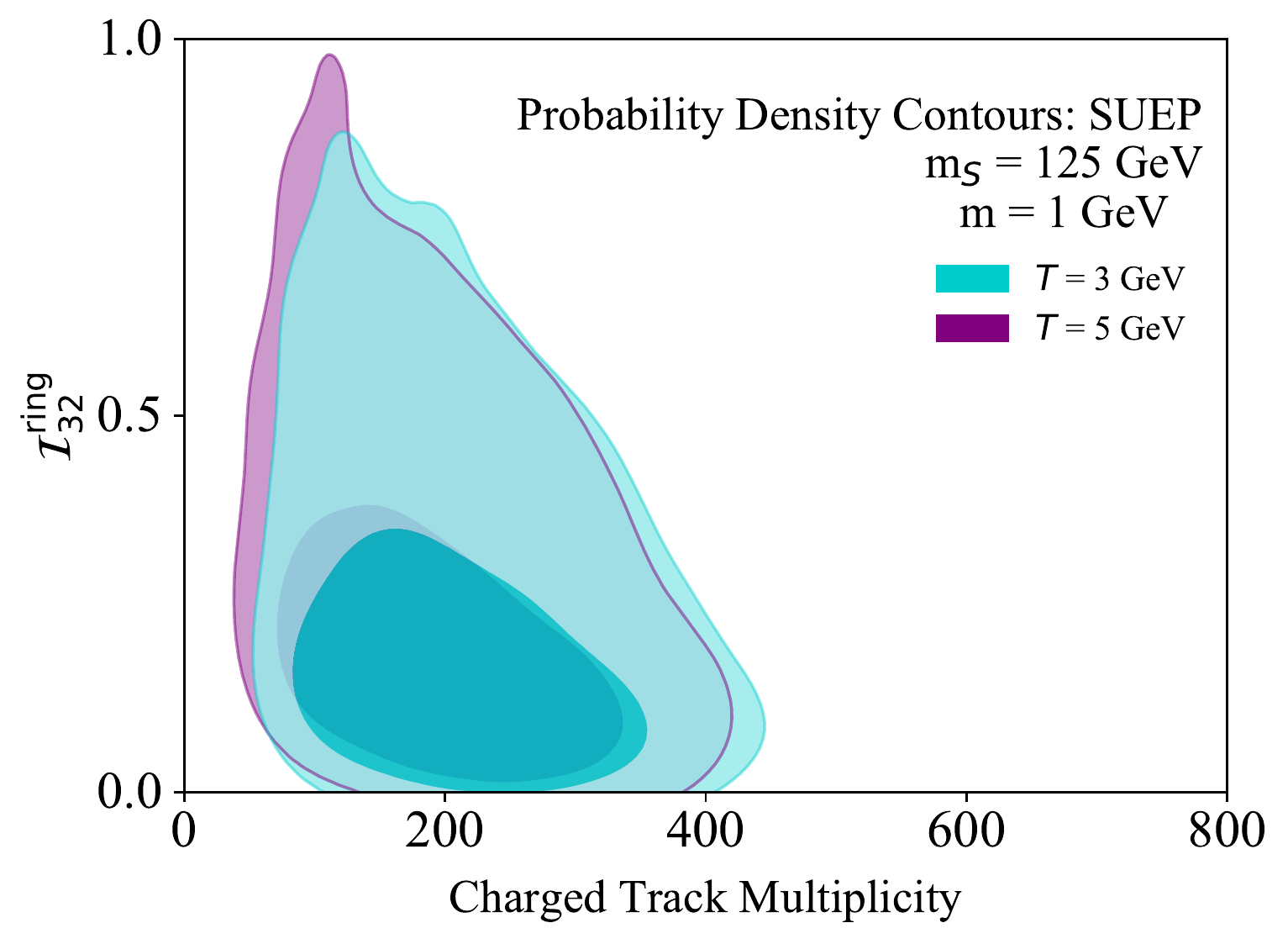}
     }
\caption{Probability density contours in track multiplicity and event isotropy space for SUEP scenarios with $m$ = 1 GeV and (a) $T = \{0.5, 1\}$ GeV and (b) $T =  \{3, 5\}$ GeV.
The darker regions contain 68\% of the events, while the lighter regions contain 95\% of the events.}
\label{fig:smallMassSUEP2d}
\end{figure}

We can gain further insights by plotting the two-dimensional distributions of these observables.
As shown in \Fig{fig:smallMassSUEP2d}, the charged track multiplicity and event isotropy are only somewhat correlated.
In the SUEP scenario, the quasi-isotropy is due in part to the high multiplicity of the final state.
But for a fixed degree of isotropy, there is a relatively wide range of multiplicity values.
For the case of $T = \{3, 5\}$ GeV, there is some anti-correlation between multiplicity and isotropy (i.e.\ higher multiplicity implies more isotropic).
But as mentioned above, as the temperature $T$ increases, the hidden valley particles are produced with larger boost factors, which tends to degrade the event isotropy.
We conclude that the multiplicity of the event is only one aspect that determines the radiation pattern, and event isotropy offers complementary information about the overall event kinematics.

\begin{figure}[t!]
\centering
\subfloat[]{
       \includegraphics[width=0.47\textwidth]{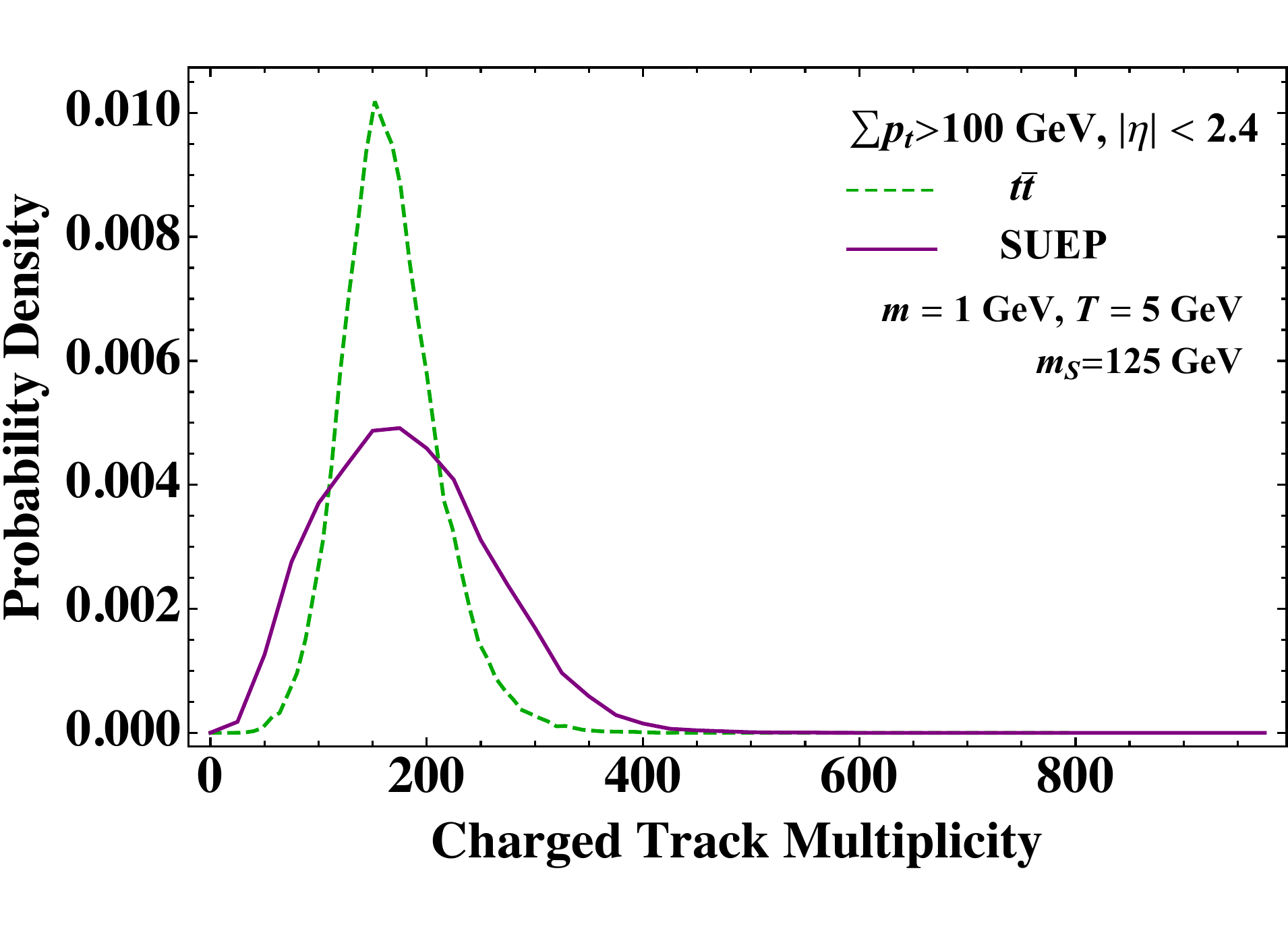}
     }
     \hfill
\subfloat[]{
       \includegraphics[width=0.45\textwidth]{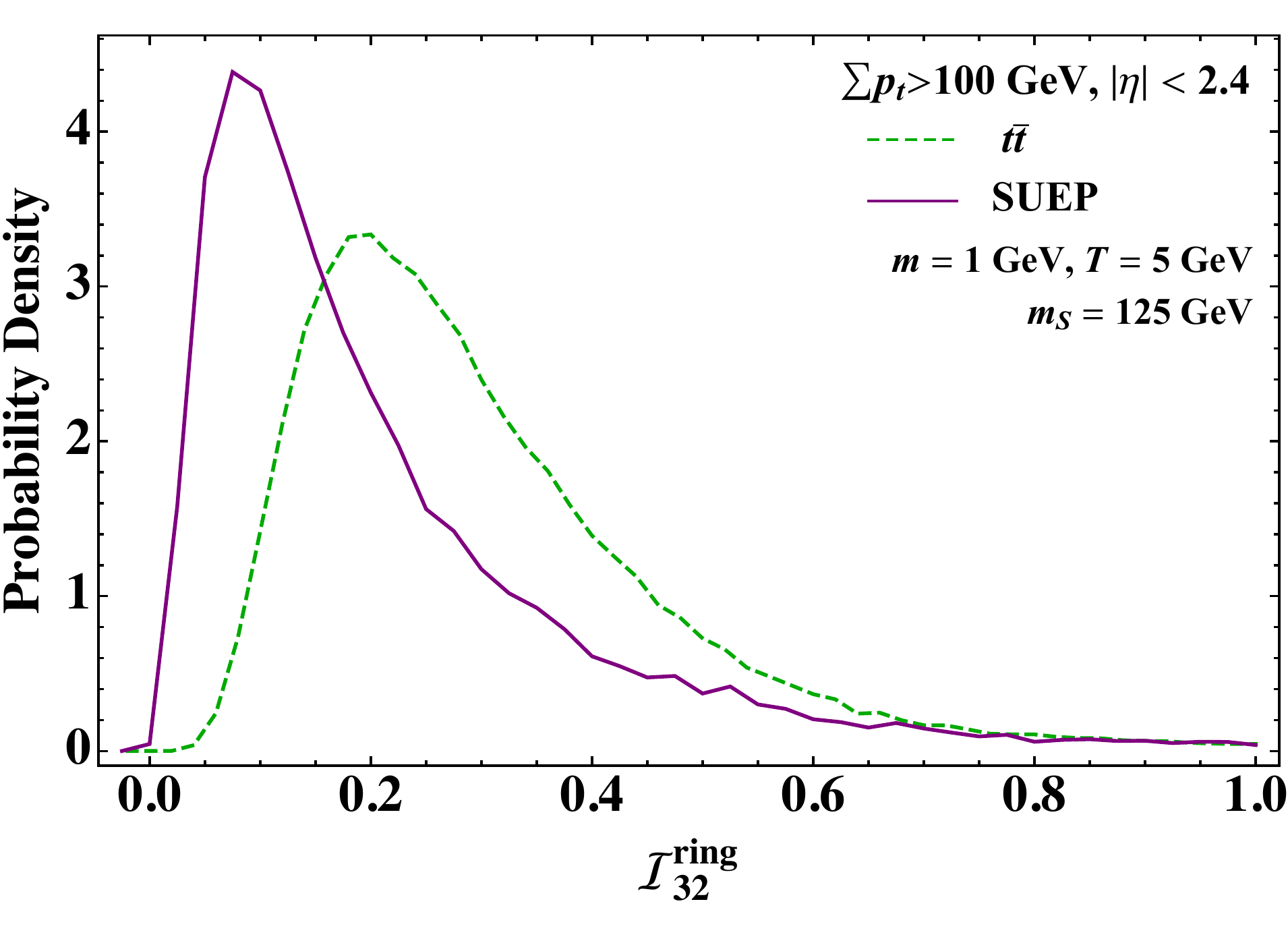}
     }
\caption{The same observables as \Fig{fig:smallMassSUEP} but comparing the $t\bar{t}$ sample (green) to the $m=1$ GeV, $T = 5$ GeV SUEP sample (purple). 
The distributions of charged track multiplicity cover a similar range for both samples, but the peaks in ring-like event isotropy are better separated.}
\label{fig:ttbarSuep}
\end{figure}

\begin{figure}[t!]
\centering
\includegraphics[width=0.5\textwidth]{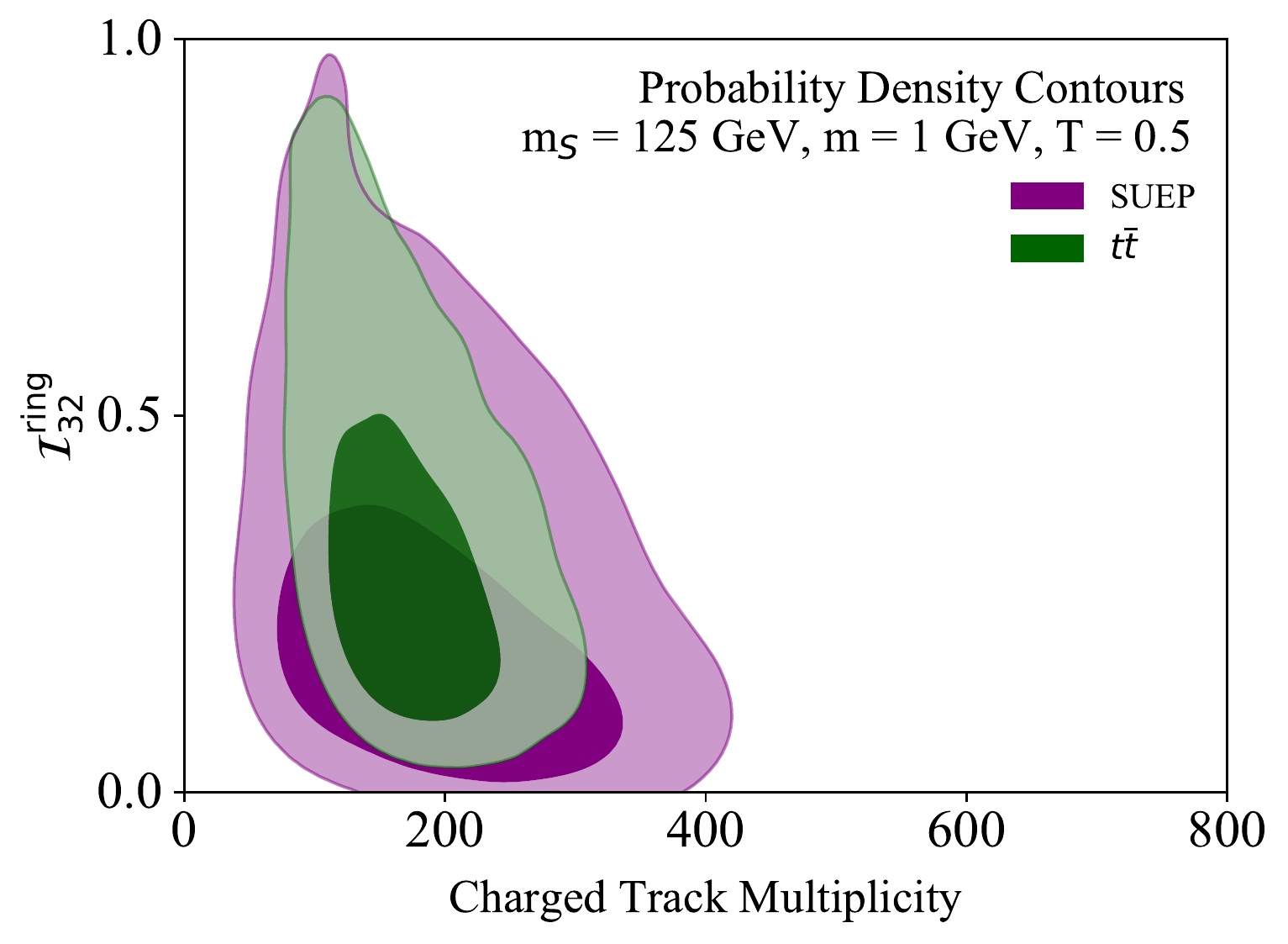}
\caption{The same observables as \Fig{fig:smallMassSUEP2d}, but comparing the $t\bar{t}$ sample (green) to the $m=1$ GeV, $T = 5$ GeV SUEP sample (purple).}
\label{fig:ttbarSuep2d}
\end{figure}

To exemplify the sensitivity of event isotropy to different event topologies with similar multiplicity distributions, we compare a SUEP scenario to SM $t\bar{t}$ production. 
In \Fig{fig:ttbarSuep}, the multiplicity distributions of the $t\bar{t}$ sample and the $m = 1$ GeV, $T=5$ GeV SUEP sample peak at nearly the same value of multiplicity and have support over a similar region. 
However, the radiation pattern of the two types of events are intrinsically different. 
Since $t\bar{t}$ events typically have six hard prongs of energy smeared into jets, high-multiplicity events will not necessarily look isotropic. 
Indeed, it is clear from \Fig{fig:ttbarSuep} that the $t\bar{t}$ events have larger event isotropy (i.e.\ are less isotropic) than the SUEP events, even for events of the same multiplicity.

Comparing the two-dimensional distributions in \Fig{fig:ttbarSuep2d}, we see that SUEP events and $t\bar{t}$ events dominate distinct regions of phase space.
This emphasizes that for an event to have small event isotropy, it must have both high multiplicity and isotropically distributed particles.
Event isotropy is sensitive to these combined features, and we expect it will be useful measure for characterizing new physics signatures, especially in combination with other observables.

\section{Conclusions and Future Studies}
\label{sec:conclusions}

In this paper, we introduced a new event shape observable designed to robustly identify isotropic radiation patterns in collider events. 
For the task of distinguishing multi-body final states from SM backgrounds, we found that event isotropy performed as well as or better than previous event shapes, such as thrust and sphericity at $e^+e^-$ colliders and transverse thrust at $pp$ colliders.
Thrust quantifies how far an event is from a back-to-back two particle configuration, whereas event isotropy quantifies the distance from isotropic configurations.
Event isotropy therefore excels in a regime where previous observables had little dynamic range:  distinguishing various high-multiplicity signatures from each other.

In general, event isotropy outperforms thrust for separating SM jetty backgrounds from quasi-isotropic events when the multiplicity of hard prongs in the event is $\mathcal{O}(10)$.
Various new physics scenarios can generate these kind of high-multiplicity signals, so we expect event isotropy to play a useful role both in reducing SM backgrounds and, if a signal is found, in characterizing the new physics behavior.
In our toy model of the uniform $N$-body phase space, event isotropy could differentiate the $N= \{10, 25, 50\}$ samples from QCD backgrounds as well as from each other.
This suggests that event isotropy will be a powerful observable for identifying inherently isotropic signals at the LHC or a future collider even with large multi-jet or fat jet backgrounds. 
Of course, in a realistic analysis, additional cuts on features such as missing $p_T$, jet multiplicity, or object classifiers would be applied, so future studies are needed to understand the correlation of event isotropy with other discriminants.

To explore the potential role of event isotropy in characterizing new physics signals, we used event isotropy to study the topology of SUEP events.
High-multiplicity final states are a generic feature of theories with large 't Hooft couplings and arise ubiquitously in hidden valley scenarios~\cite{Strassler:2006im}.
The original literature on SUEPs searches recommended track multiplicity as a primary discriminant~\cite{Knapen:2016hky}, and our studies support this conclusion.
That said, high multiplicity by itself is not sufficient to produce an isotropic event, so event isotropy provides complementary information to track multiplicity to help disentangle SUEP model parameters. 
While more model-specific modifications to this search would need to be implemented for a full analysis, event isotropy provides a new handle for characterizing rare collider signatures.

One limitation of event isotropy in its current form is that it only tests for event configurations that are uniform in the lab frame of the collider.
In scenarios where a resonance decays to a quasi-isotropic distribution of particles, that resonances could acquire a significant Lorentz boost, for example by recoiling against another hard object in the event.
This issue also arose in our study of cylindrical event isotropy, where longitudinal boosts along the beam line due to parton distribution functions led to worse discrimination performance than ring-like event isotropy.
Just as thrust involves optimizing over the thrust axis, one could imagine an extension of event isotropy where one optimizes over the choice of isotropy frame.
Alternatively, one could follow the logic from the jet substructure literature~\cite{Larkoski:2017jix,Asquith:2018igt,Marzani:2019hun} and first identify a candidate fat jet region before computing its isotropy.
More generally, one could test for various event patterns using the analogous EMD logic to \Eq{eq:evIsoDef} and compute the transportation cost to a different target event (or set of events) of interest~\cite{Komiske:2020qhg}.
For example, another event topology of generic new physics distinct from standard model predictions is radiation confined to a plane, as is predicted in vanishing dimension models \cite{Anchordoqui:2010hi,Anchordoqui:2010er}.

As the energy scale of future colliders continues to increase, the final-state particle multiplicity can also grow.
Event isotropy can therefore be used to understand how inherently isotropic a new physics signature is as a function of multiplicity~\cite{cesarotti:2020mm}. 
Another key challenge for future colliders as well as for the high-luminosity LHC is backgrounds from pileup.
Pileup yields uniform event contamination, and could therefore mask genuinely uniform new physics signatures.
On the other hand, pileup mitigation can be phrased as an optimal transport problem~\cite{Komiske:2020qhg}, so there may be avenues to compute event isotropy in a robust and computationally efficiency way, even in the presence of pileup.
Similarly, because both thrust and event isotropy can be defined as optimal transport problems, one can analytically study their susceptibility to exactly uniform pileup contamination.
For example, one can show that if uniform pileup constitutes a fraction $x < 1$ of the total transverse energy of an event, then both transverse thrust and ring-like event isotropy shift linearly in $x$ towards more isotropic values, leaving the discrimination power largely unchanged.
As noted in \Ref{Knapen:2016hky}, triggering on isotropic events can be challenging, but a fast approximation to ring-like event isotropy~\cite{Rabin:2011jd} could potentially be incorporated into trigger-level analyses.
In addition, future studies would need to assess the achievable experimental resolution on event isotropy, to validate that the discrimination power is robust to energy and angular smearing effects. 

Finally, isotropy would interesting to consider as a jet substructure observable, both for new physics searches and for studying QCD.
Just as event isotropy was a robust way to identify $N$-body configurations at high $N$, jet isotropy could be a robust probe of high-multiplicity jet configurations.
Following \Ref{Aguilar-Saavedra:2017rzt}, one could use event isotropy as an anti-QCD jet tagger, either by itself or in combination with other discriminants, since boosted new physics resonances are expected to yield more uniform radiation patterns than quark or gluon jets.
For discriminating quark jets from gluon jets, most of the literature has focused on either multiplicity-like or prong-like observables; see e.g.~\cite{Gallicchio:2012ez,Gras:2017jty,Frye:2017yrw,Larkoski:2019nwj}.
Jet isotropy is a fundamentally different type of observable, and it could provide a complementary perspective on the quark/gluon discrimination challenge, especially if one could compute jet isotropy distributions from first principles.
%
%
We hope that event isotropy motivates precision calculations of EMD-based observables as well as inspires other new ways to characterize the shape of jets and collider events.

\acknowledgments{We thank Simon Knapen, Patrick Komiske, Eric Metodiev, Matt Reece, and Matt Strassler for helpful conversations.
CC is supported in part by an NSF Graduate Research Fellowship Grant DGE1745303 and by the U.S. Department of Energy (DOE) under Grant No.\ DE-SC0013607.
JT is supported by the Office of High Energy Physics of the DOE under Grant Nos.\ DE-SC0012567 and DE-SC0019128 and by the Simons Foundation through a Simons Fellowship in Theoretical Physics.
}
\appendix

\section{Approximations of Event Isotropy}
\label{app:eiApp}

In this appendix, we derive approximate expressions for event isotropy in the case of $k$-body symmetric event configurations, as studied in \Secs{subsec:ee_kbody}{subsec:pp_kbody}.
In the main text, we computed the event isotropy as the EMD between an event and the $n$-particle quasi-isotropic event $\mathcal{U}_n$. 
Here, we estimate the event isotropy using the idealized uniform event $\mathcal{U}_\infty$.

\begin{figure}[t!]
\centering
\includegraphics[width=0.4\textwidth]{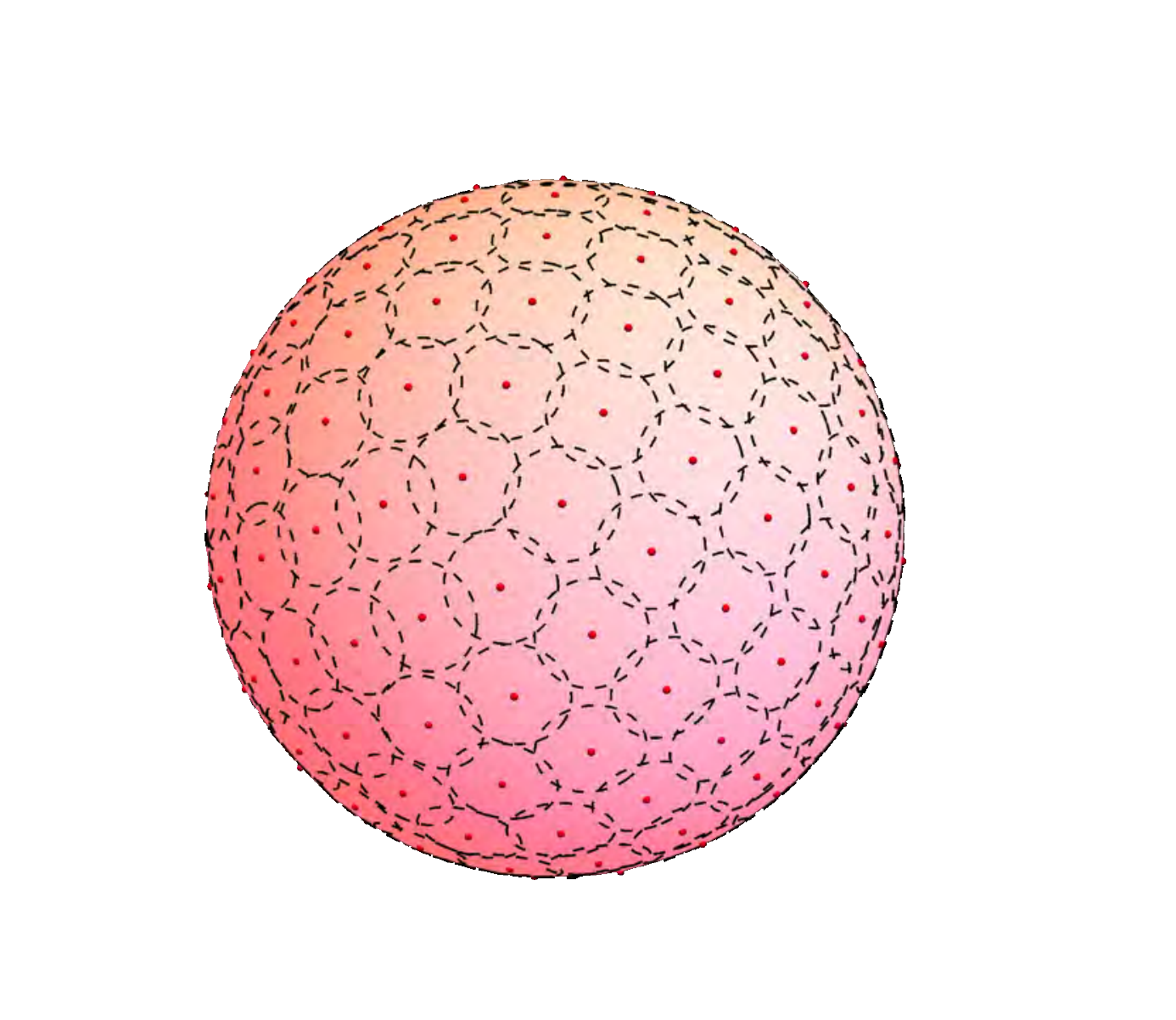}
\caption{The approximate disk tiling of a sphere for $k=192$ particles.}
\label{fig:sphAnalytic}
\end{figure}

To approximate the event isotropy of a $k$-body symmetric distribution, we partition the uniform geometry into $k$ tiles of equal size centered around the $k$ particles, as shown in \Fig{fig:sphAnalytic} for the spherical case.
We then compute the transportation cost to move the energy from each uniform tile to the corresponding particle at its center.
Each particle has weight $w_i = \frac{1}{k}$ and there are $k$ such particles, such that the final answer can be computed by moving unit energy within a single tile. 
For large enough $k$, the scaling of the event isotropy with $k$ will be independent of the exact geometry of the tiling, though the overall prefactor depends on the details of the tiling.
Our analytic expressions should be a reasonably good approximation to the numerically computed values with finite $n$, assuming the hierarchy $1 \ll k \ll n$.

\begin{figure}[t!]
\centering
\subfloat[]{
\label{fig:analytic_sph}
\includegraphics[width=0.45\textwidth]{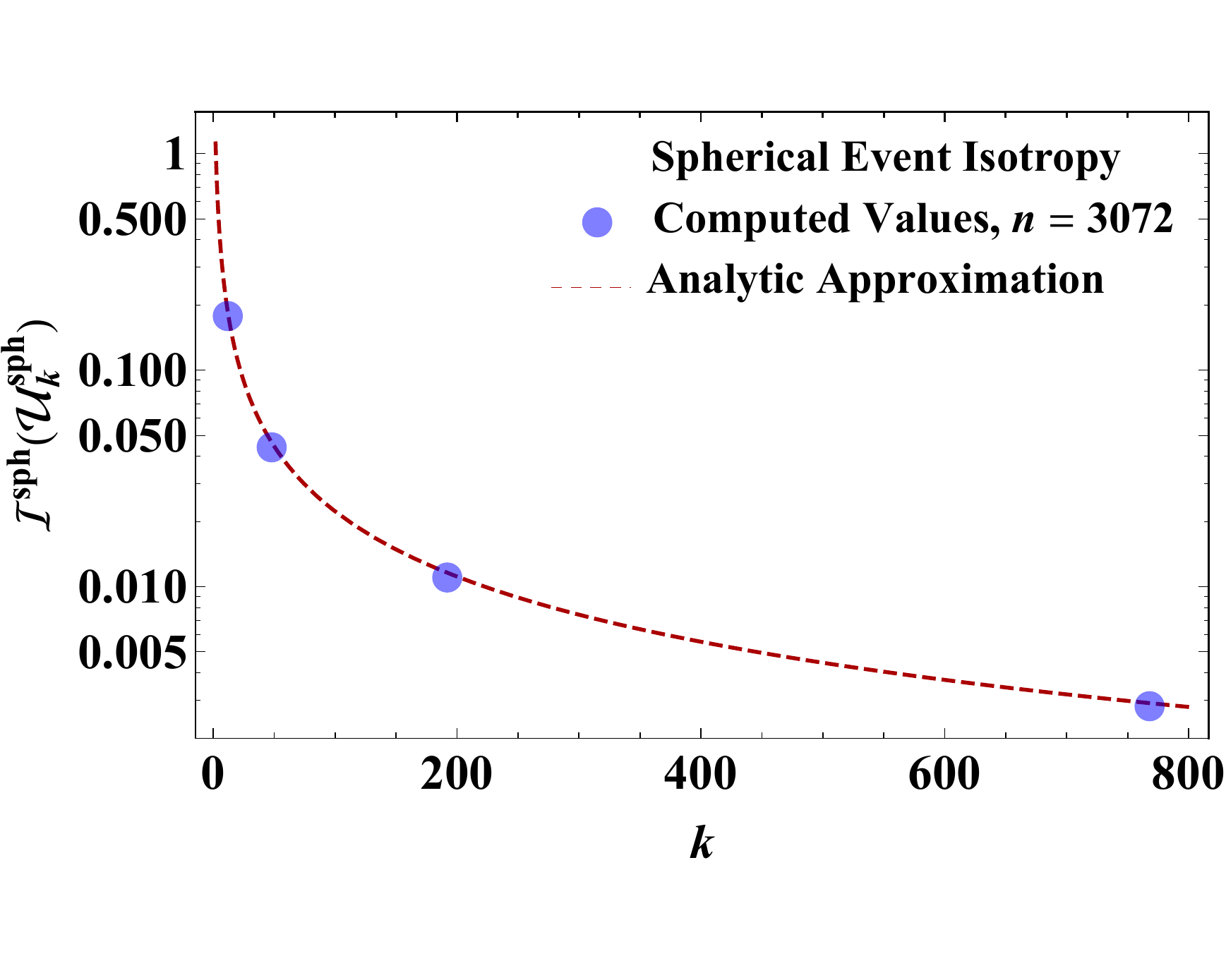}
}
\hfill
\subfloat[]{
\label{fig:analytic_cyl}
\includegraphics[width=0.45\textwidth]{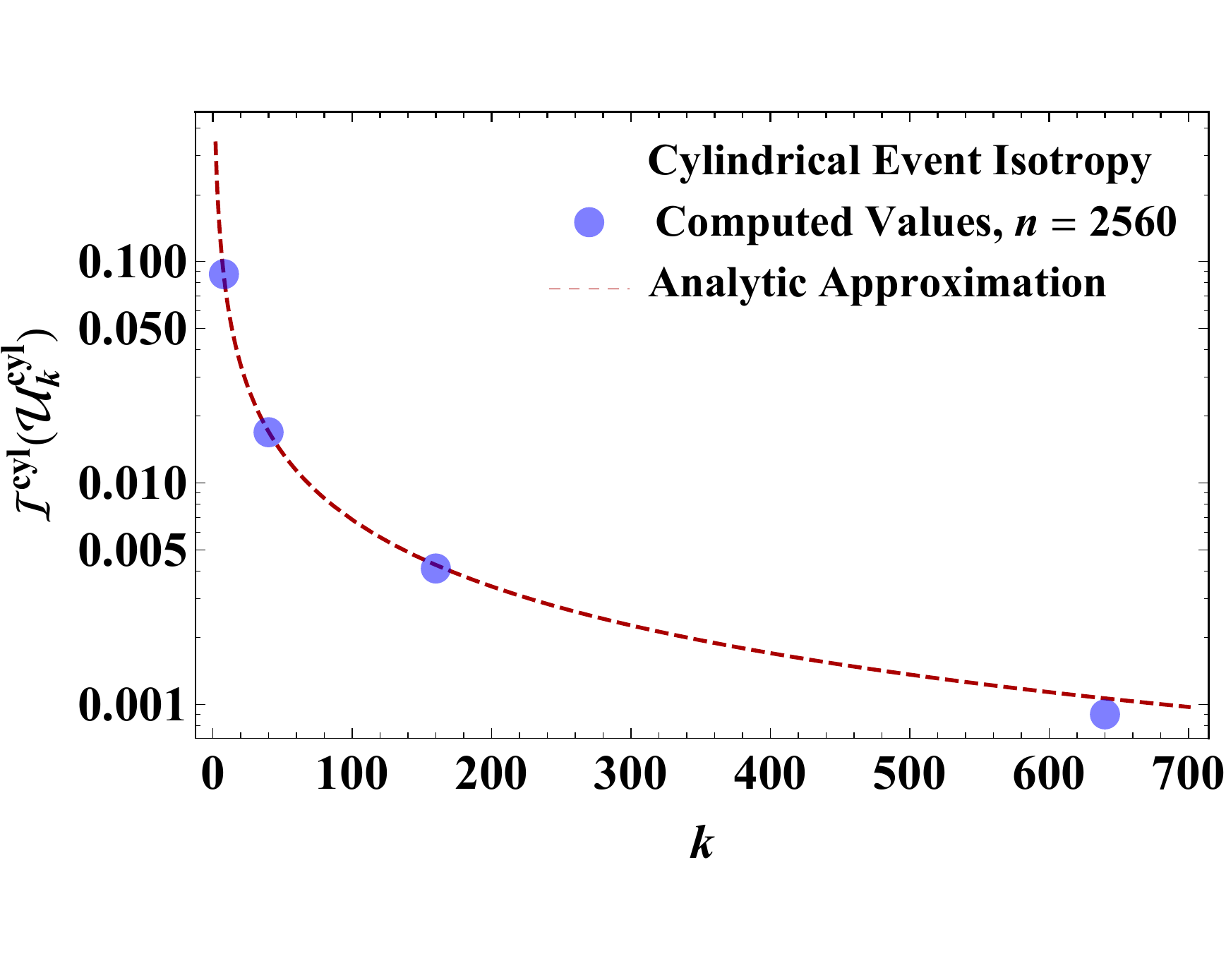}
}
\hfill
\subfloat[]{
\label{fig:analytic_ring}
\includegraphics[width=0.45\textwidth]{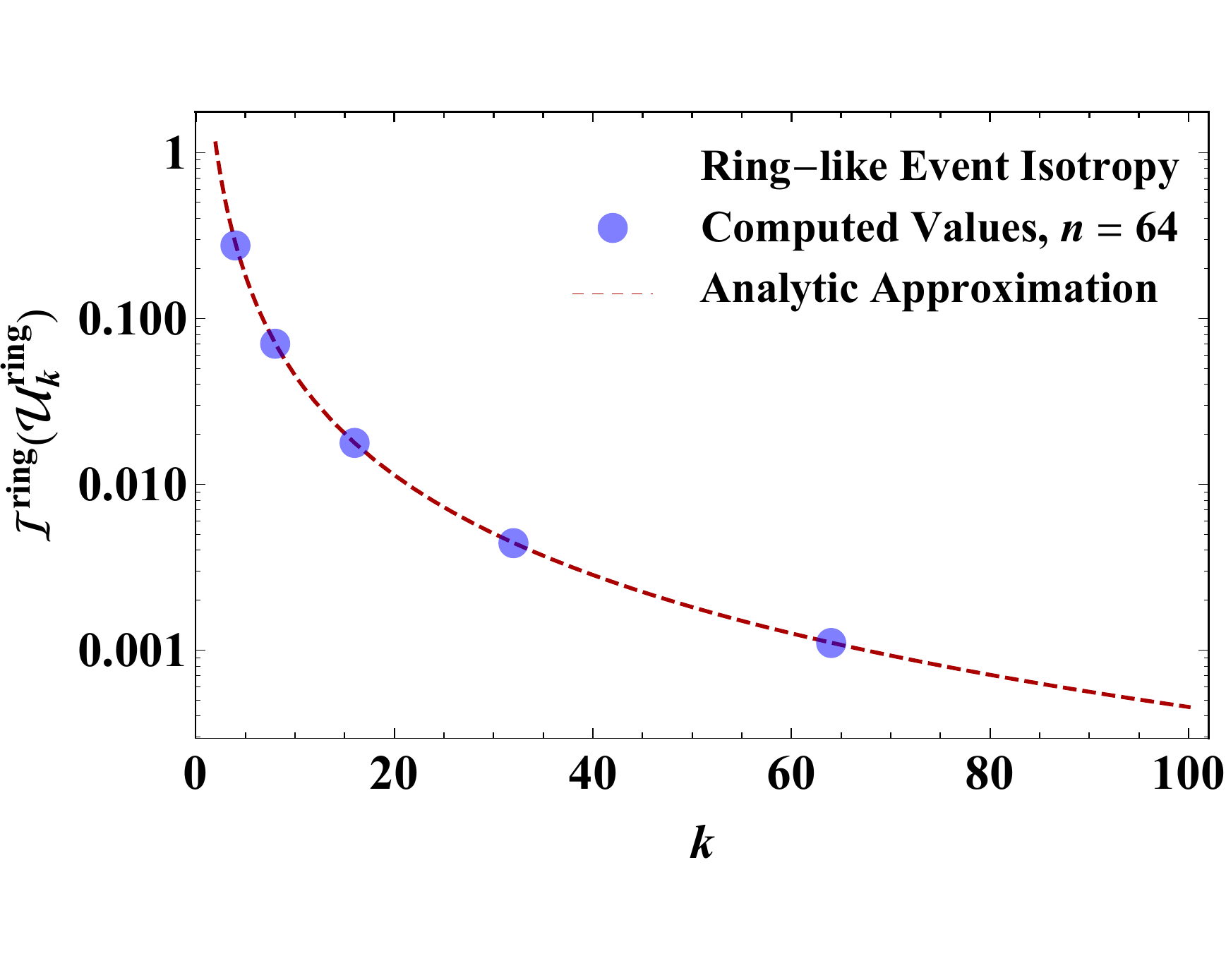}
}
\caption{The event isotropy of quasi-uniform $k$-body configurations for the (a) spherical, (b) cylindrical, and (c) ring-like geometries.
We find good agreement between the analytic approximation of event isotropy $\iso{}{\infty}$ (red line) and the numerically computed values of event isotropy (blue markers) using (a) $\iso{sph}{3072}$, (b) $\iso{cyl}{2560}$, and (c) $\iso{ring}{64}$.}
\label{fig:analytic}
\end{figure}

\begin{itemize}

\item \textbf{Spherical.}  To compute the event isotropy of a $k$-body sphere, we partition the uniform sphere of area $A_{\rm tot} = 4\pi$ into $k$ equal area disks of area $A_k$ and radius $r_k$:
\begin{equation}
A_k = \frac{A_{\rm tot}}{k}, \qquad r_k = \sqrt{\frac{A_k}{\pi}}.
\end{equation}
In the limit of small opening angle $\theta$, the spherical distance measure takes the form
\begin{equation}
\label{eq:smallangledij}
d_{ij} \simeq a \, \theta_{ij}^\beta,
\end{equation}
where $a = 1$ and $\beta = 2$ for the default measure in \Eq{eq:spherical_metric}.
The transportation cost can be computed from a single disk:
\begin{equation}
\iso{sph}{\infty}(\mathcal{U}^\text{sph}_k) \approx \frac{1}{A_k} \int_0^{r_k} 2 \pi \theta \, \text{d} \theta \left(a \, \theta^\beta \right) = \frac{2 a r_k^\beta}{2+\beta} \Rightarrow \frac{2}{k},
\label{eq:estimateKbody_sph}
\end{equation}
where in the last step, we substituted in the values of $r_k$, $a$, and $\beta$ for the default measure.
We compare this analytic estimate to the numerical $n = 3072$ result in \Fig{fig:analytic_sph}, finding the characteristic $\iso{sph}{\infty} \sim 1/k$ scaling as predicted by \Eq{eq:estimateKbody_sph}.

\item \textbf{Cylindrical.}
The cylindrical case is similar to the spherical case, though we can get an improved estimate because the tiling is known exactly.
For particles uniformly distributed in the $y-\phi$ plane, the relevant total area is that of a cylinder of circumference $2\pi$ and length $2y_\text{max}$, yielding $A_{\rm tot} = 4\pi y_\text{max}$.
Assuming a square lattice arrangement, the cylinder can be tiled with $k$ equal area squares of area $A_k$ and half-width $r_k$:
\begin{equation}
A_k = \frac{A_{\rm tot}}{k}, \qquad r_k = \sqrt{\frac{A_k}{4}}.
\end{equation}
For the default cylindrical measure in \Eq{eq:cylindrical_metric}, the small opening angle behavior is the same as \Eq{eq:smallangledij}, but with:
\begin{equation}
a = \frac{12}{\pi^2 + 16 y_{\rm max}^2}, \qquad \beta = 2.
\end{equation} 
Repeating \Eq{eq:estimateKbody_sph} but for a square tiling and specializing to the $\beta = 2$ case, we have
\begin{equation}
\iso{cyl}{\infty}(\mathcal{U}^\text{sph}_k) \approx \frac{1}{A_k} \int_{-r_k}^{r_k} \text{d} y \int_{-r_k}^{r_k} \text{d} \phi \, a \, (y^2 + \phi^2)  = \frac{2 a r_k^2}{3} \Rightarrow \frac{1}{k} \frac{8 \pi y_\text{max}}{\pi^2 + 16 y_\text{max}^2}.
\label{eq:estimateKbody_cyl}
\end{equation}
Like the spherical case, we find the scaling $\iso{cyl}{\infty} \sim 1/k$, which agrees with the numerical $n=2560$ result plotted in \Fig{fig:analytic_cyl}.
When expressed in terms of $r_k$, \Eq{eq:estimateKbody_cyl} is a factor of $4/3$ larger than \Eq{eq:estimateKbody_sph} with $\beta = 2$, due to the fact that there is a larger transportation cost associated with the corners of the squares tiles.
It is straightforward to check that this factor is $\{10/9, 4/3, 2\}$ for a \{triangular, square, hexagonal\} lattice.

\item \textbf{Ring-like.}  For the ring-like geometry with $k$ equally spaced particles around a total length $L_{\rm tot} = 2\pi$, the geometry can be broken up into $k$ segments of half-length $r_k$:
\begin{equation}
r_k = \frac{L_{\rm tot}}{2k}.
\end{equation}
The default ring-like measure in \Eq{eq:ring_metric} has the same small opening angle behavior as \Eq{eq:smallangledij}, with:
\begin{equation}
a = \frac{\pi}{2\pi -4}, \qquad \beta = 2.
\end{equation} 
The event isotropy is approximately:
\begin{equation}
\iso{ring}{\infty}( \mathcal{U}^\text{ring}_k ) \approx \frac{1}{2 r_k} \int_{-r_k}^{r_k} \text{d} \phi \left(a \, \phi^\beta \right) = \frac{a r_k^\beta}{1+\beta} \Rightarrow \frac{1}{k^2} \frac{\pi^3}{6 \pi -12}.
\label{eq:ringApprox}
\end{equation}
The ring-like event isotropy scales like $\iso{ring}{\infty} \sim 1/k^2$, which is expected since a ring is a lower dimensional geometry than a cylinder or sphere.
This $1/k^2$ scaling is shown numerically for $n=64$ in \Fig{fig:analytic_ring}.

\end{itemize}


\section{Events with Dijet Plus Spherical Structure}
\label{app:dijetSphere}

\begin{figure}[t!]
\centering
\includegraphics[width=0.5\textwidth]{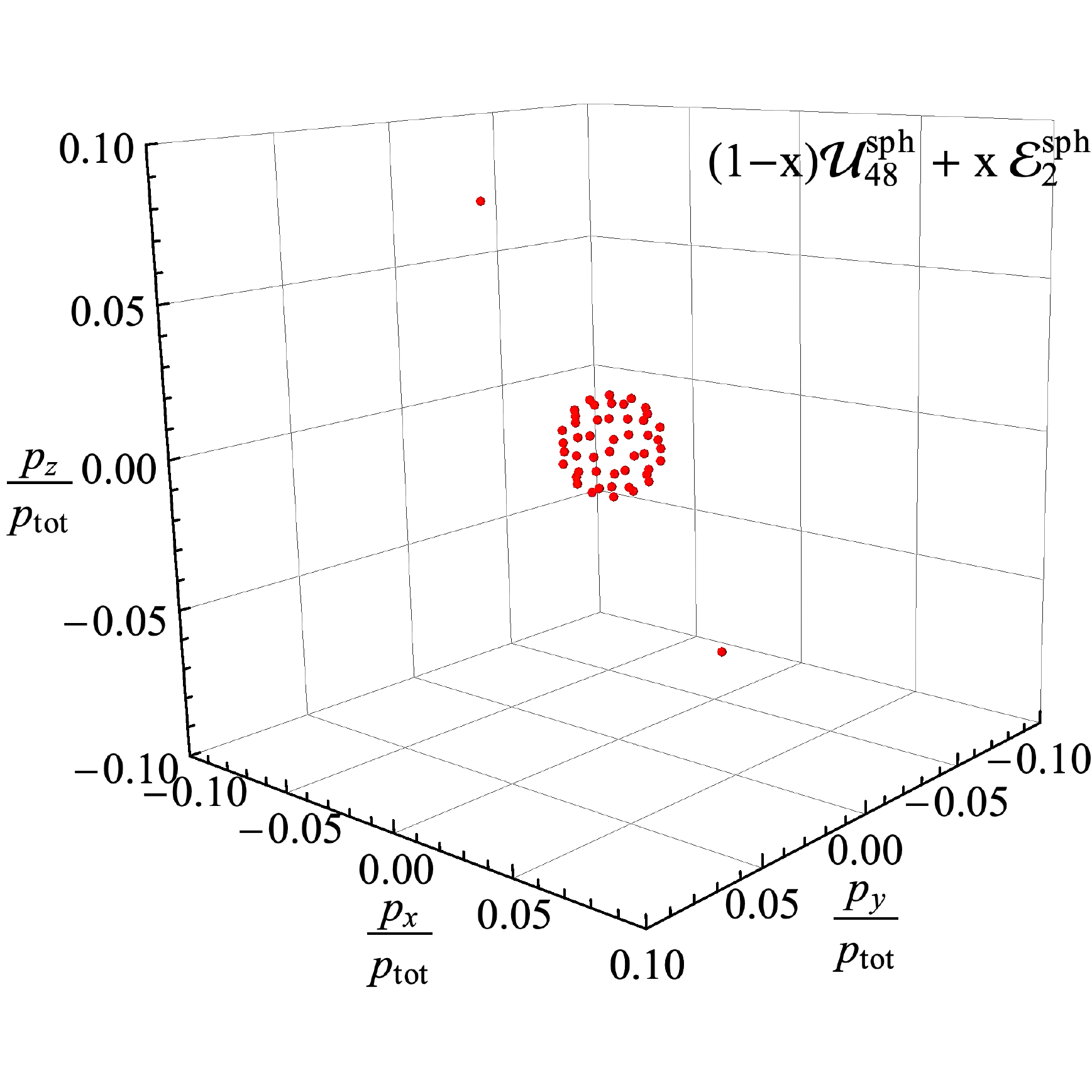}
\caption{Two back-to-back particles $\mathcal{E}^\text{sph}_2$ superimposed on a quasi-isotropic event $\mathcal{U}^\text{sph}_{48}$.
The total momentum of the two particles is 20\% of the total event momentum ($x=0.2$).}
\label{fig:sphDijetVis}
\end{figure}

In this appendix, we study hybrid event configurations that have both jet-like and isotropic features.
Such patterns could arise in the associated production of two new physics resonances, where one decays into a conformal hidden sector while the other decays immediately back into SM jets. 
As a toy model, the radiation pattern we investigate is two back-to-back particles superimposed on a quasi-isotropic event, as shown in \Fig{fig:sphDijetVis}.
We normalize the energy of the event to be 1, and we scan over different percent allocations of the energy to the jetty configuration versus the spherical configuration.

In addition to understanding the behavior of event isotropy for such hybrid configurations, we are also interested in illustrating the effect of choosing different ground measures to compute event isotropy.
In the main text, we focused on a ground measure that satisfies \Eq{eq:pwasserstein} with $\beta = 2$:
\begin{equation}
d_{ij}^{\beta=2} = 2 \, (1 - \cos \theta_{ij} ).
\label{eq:2was}
\end{equation} 
This choice was motivated by the direct comparison to thrust in \Eq{eq:thrust_as_EMD} and because the quadratic dependence on opening angle makes event isotropy more robust to small changes in the event configuration.
Here, we compare to an alternative measure that satisfies the ordinary triangle inequality in \Eq{eq:beta1met} (i.e.\ \Eq{eq:pwasserstein} with $\beta = 1$):
\begin{equation}
d_{ij}^{\beta=1} = \frac{3}{2} \sqrt{ 1 - \cos \theta_{ij} },
\label{eq:met}
\end{equation} 
where the factor of $\frac{3}{2}$ ensures that the event isotropy of two back-to-back particles $(\mathcal{E}^\text{sph}_2)$ is 1, in analogy to \Eq{eq:computeMaxIsoSpherical}.
For a generic $\beta$, the normalized distance would be $\big(1 + \frac{\beta}{2}\big) (1 - \cos \theta_{ij})^{\beta/2}$.
As discussed in \Ref{Komiske:2020qhg}, the $\beta$ parameter in the EMD ground measure plays a similar role to the $\beta$ parameter for the jet angularities~\cite{Ellis:2010rwa,Larkoski:2014pca} and $N$-subjettiness~\cite{Thaler:2010tr,Thaler:2011gf}, with $\beta = 2$ exhibiting more mass-like behavior and $\beta = 1$ exhibiting more $k_t$-like behavior.
To see more examples of event isotropy being calculated with this $\beta=1$ measure, see \Ref{cesarotti:2020mm}.

\begin{figure}[t]
\centering
\subfloat[]{
\label{fig:dijetSphereFig_beta1}
\includegraphics[width=0.45\textwidth]{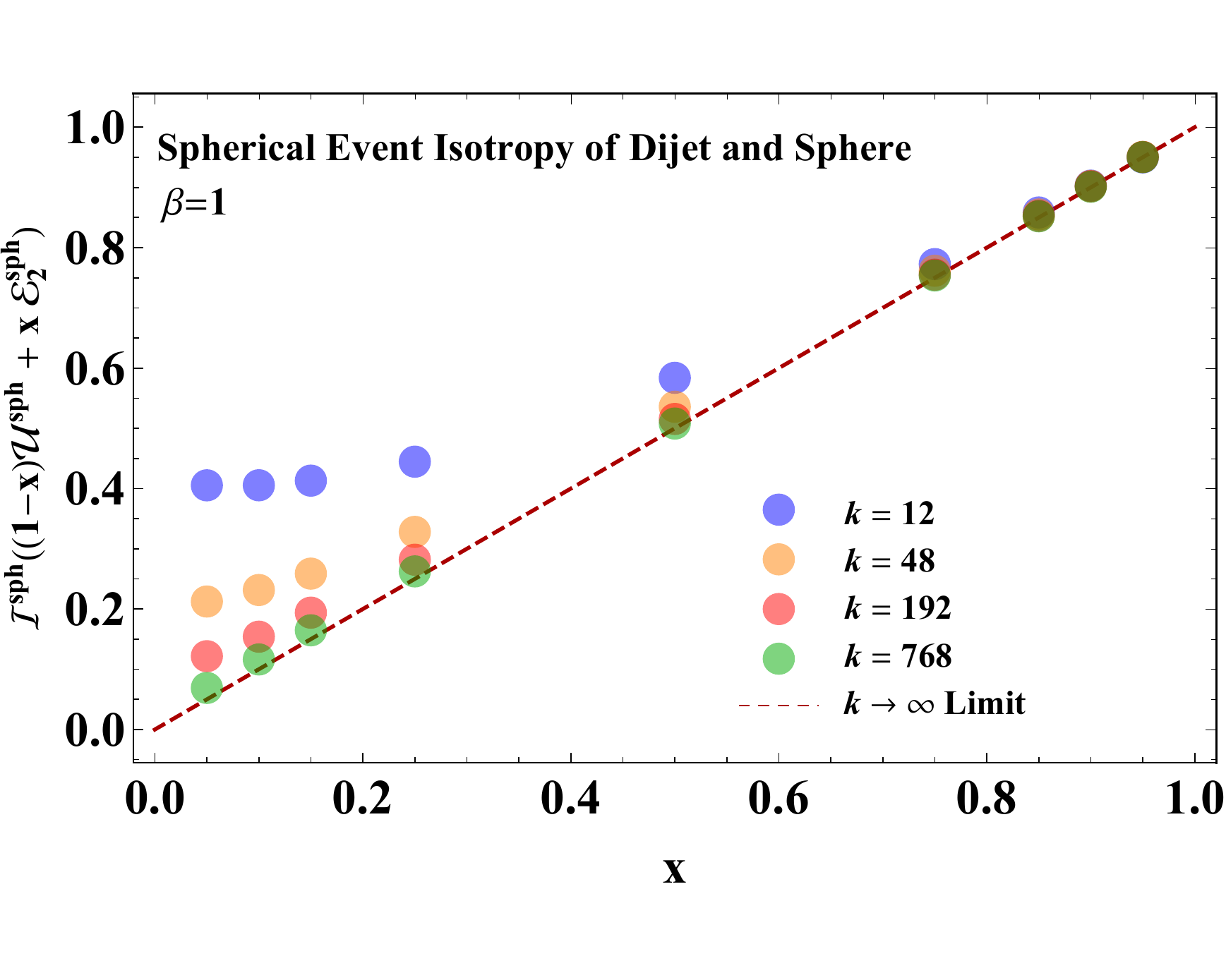}
}
\hfill
\subfloat[]{
\label{fig:dijetSphereFig_beta2}
\includegraphics[width=0.45\textwidth]{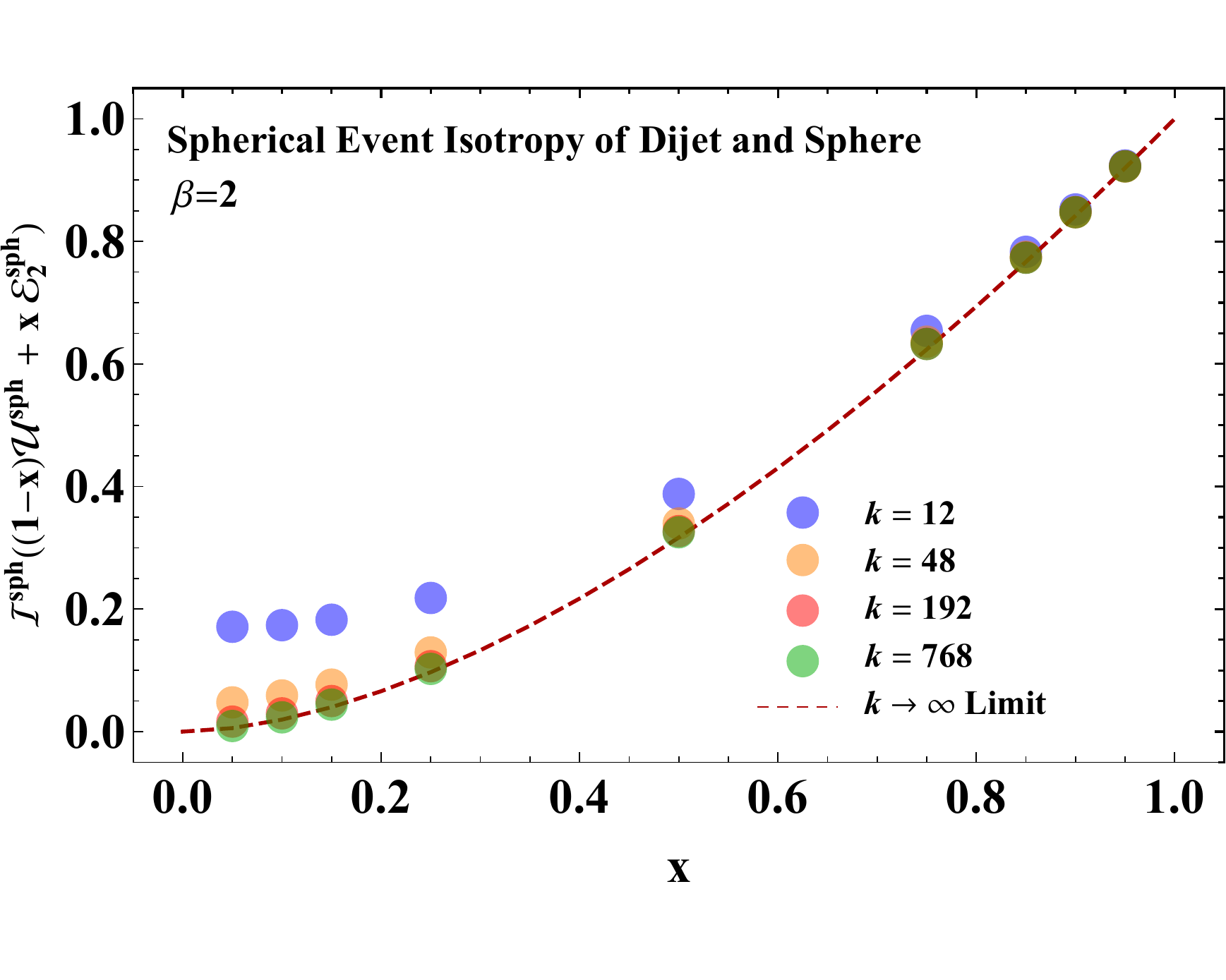}
}
\caption{Spherical event isotropy computed with (a) $d_{ij}^{\beta=1}$ and (b) $d_{ij}^{\beta=2}$ for back-to-back two-particle configurations superimposed on $k$-particle spheres. 
The numerical values are computed using $\iso{sph}{768}$ after averaging over $10^2$ random orientations, and we sweep over the fraction $x$ of the total energy allocated to the two-particle configuration. 
When using the $\beta = 1$ measure, the event isotropy is linearly sensitive to $x$ at large $k$, as computed in \Eq{eq:analyticSDijet}. 
For the $\beta =2$ measure, the large $k$ limit has a more complicated behavior, as computed in \Eq{eq:beta2analyticSDijet}.}
\label{fig:dijetSphereFig}
\end{figure}

Because $d_{ij}^{\beta=1}$ satisfies the usual triangle inequality, the shortest distance any net energy can travel from one configuration to another is directly from the excess to the deficient region. 
This has a simple interpretation for the event isotropy of the two-particle-plus-sphere configuration:  the energy of each hard particle is redistributed uniformly among the soft particles in its respective hemisphere. 
For a configuration where a fraction $x$ of the total energy is in the two-particle configuration $\mathcal{E}_2$ and a fraction $1-x$ is in a uniform $k$-body sphere $\mathcal{U}^\text{sph}_k$, the event isotropy is approximately
\begin{equation}
\beta =1: \quad \iso{sph}{n}\left[ (1-x)\mathcal{U}^\text{sph}_{k} + x \, \mathcal{E}^\text{sph}_2 \right] \approx (1-x) \, \iso{sph}{n}( \mathcal{U}^\text{sph}_{k}) + x \, \iso{sph}{n}(\mathcal{E}_2),
\label{eq:spherAndDijet}
\end{equation}
where this expression becomes an equality for $k,n \to \infty$.
For large enough $n$, $\iso{sph}{n}(\mathcal{E}_2) \approx 1$.
To approximate the event isotropy of $\mathcal{U}^\text{sph}_{k}$, we repeat the calculation of \Eq{eq:estimateKbody_sph} but with $a = \sqrt{9/8}$ and $\beta = 1$, as appropriate for the small angle limit of \Eq{eq:met}:
\begin{equation}
\beta =1: \quad \iso{sph}{\infty}\left( \mathcal{U}^\text{sph}_{k} \right) \approx \sqrt{\frac{2}{k}}.
\end{equation}
Thus, in the large $n$ limit, we can approximate \Eq{eq:spherAndDijet} for the $\beta =1$ measure as 
\begin{equation}
\beta =1: \quad \iso{sph}{n}\left[ (1-x)\mathcal{U}^\text{sph}_{k} + x \, \mathcal{E}^\text{sph}_2 \right] \approx \sqrt{\frac{2}{k}}(1-x) + x.
\label{eq:analyticSDijet}
\end{equation}
In \Fig{fig:dijetSphereFig_beta1}, we numerically compute $\iso{sph}{768}$ for two-particle-plus-sphere events as a function of $x$ and $k$, finding linear behavior in $x$ that agrees with \Eq{eq:analyticSDijet} in the large $k$ limit.

Our default angular measure $d_{ij}^{\beta=2}$ does not satisfy the usual triangle inequality, resulting in a more complicated energy flow when computing event isotropy.
In particular, the flow of energy out of the two-particle component is non-trivial, and the spherical component no longer stays static.
In the $n,k \to \infty$ limit, it is possible to compute the event isotropy analytically.
Letting $z = 2(1-\cos \theta)$ and considering just one hemisphere, the uniform configuration $\mathcal{U}^\text{sph}_\infty$ has a flat distribution in $z$ and the two-particle event $\mathcal{E}_2$ is localized at $z = 0$.
The optimal transportation path is obtained by sorting the energy deposits in increasing $z$ and transporting each quantile in one distribution to the corresponding quantile in the other.
Mathematically, this corresponds to:
\begin{equation}
\label{eq:beta2analyticSDijet}
\beta =2: \quad \iso{sph}{\infty}\left[ (1-x)\mathcal{U}^\text{sph}_{\infty} + x \, \mathcal{E}^\text{sph}_2 \right] = 2 \left(\int_{0}^x \text{d}z \, d(0,z) + \int_{x}^{1} \text{d}z \,d\Big(\frac{z-x}{1-x}, z \Big) \right),
\end{equation}
where the overall factor of 2 accounts for the two hemispheres, and the $\beta = 2$ transportation distance in these $z$ coordinates is
\begin{equation}
d(z_1, z_2) = z_1 + z_2 - \frac{z_1 z_2}{2} - \frac{1}{2} \sqrt{z_1 z_2 (z_1 - 4)(z_2 - 4)}.
\end{equation}
While it is possible to find a closed form expression for \Eq{eq:beta2analyticSDijet}, it is not particularly illuminating.
In \Fig{fig:dijetSphereFig_beta2}, we show the numerical values for $\iso{sph}{768}$ as a function of $x$ and $k$, which indeed asymptote to \Eq{eq:beta2analyticSDijet} in the large $k$ limit.

\section{Additional SUEP Benchmark Scenarios}
\label{app:suep}

In \Sec{sec:SUEP}, we discussed the use of event isotropy to characterize quasi-isotropic new physics signals.
We showed results from a SUEP model where the lightest meson mass was $m = 1$ GeV and plotted several values of the effective temperature $T \sim m$.
In this appendix, we explore other values of $m$ and $T$ to identify general trends of the event isotropy and charged track multiplicity in SUEP scenarios.

\begin{figure}[t]
\centering
\subfloat[]{
       \includegraphics[width=0.47\textwidth]{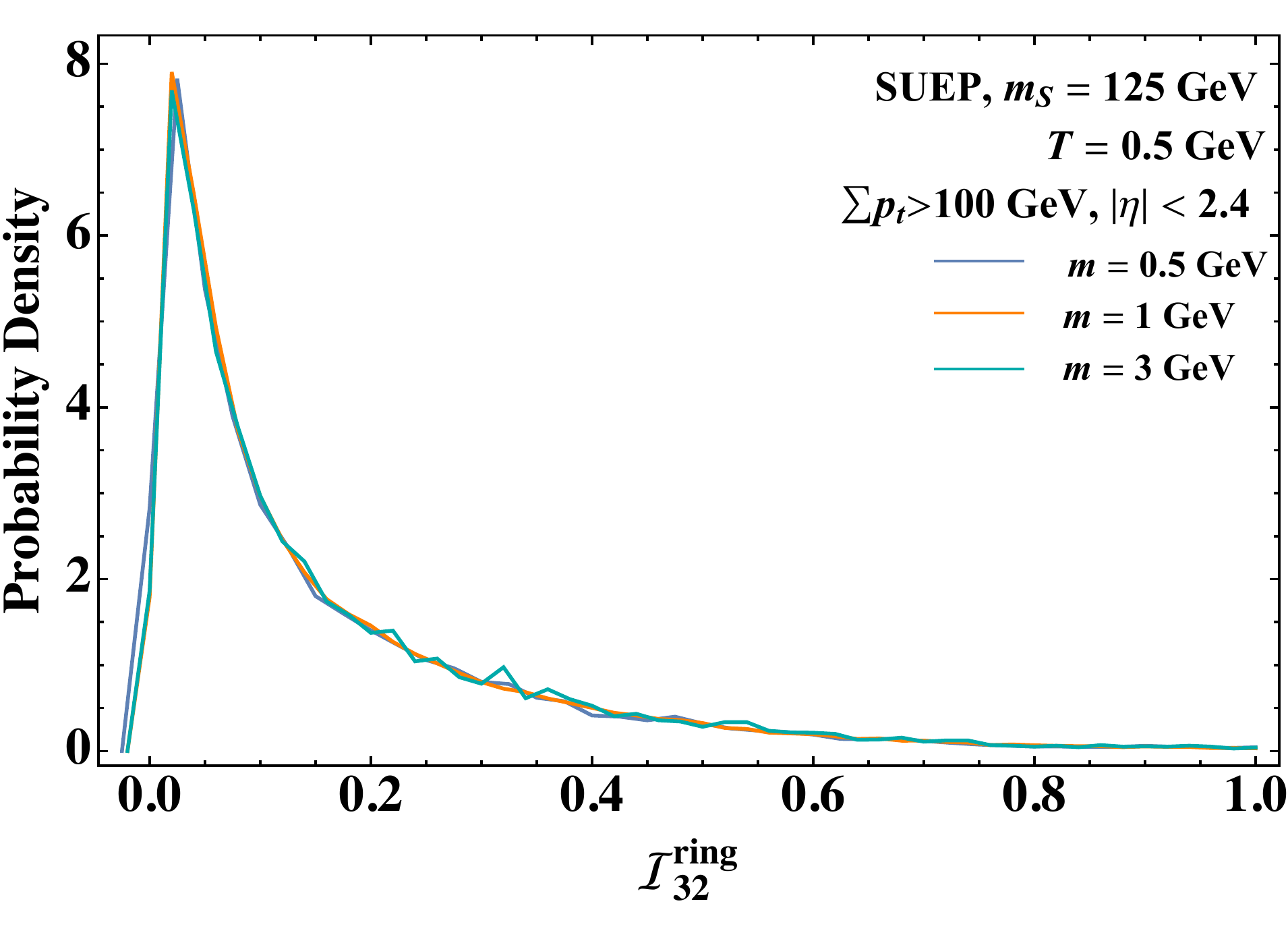}
     }
     \hfill
\subfloat[]{
       \includegraphics[width=0.47\textwidth]{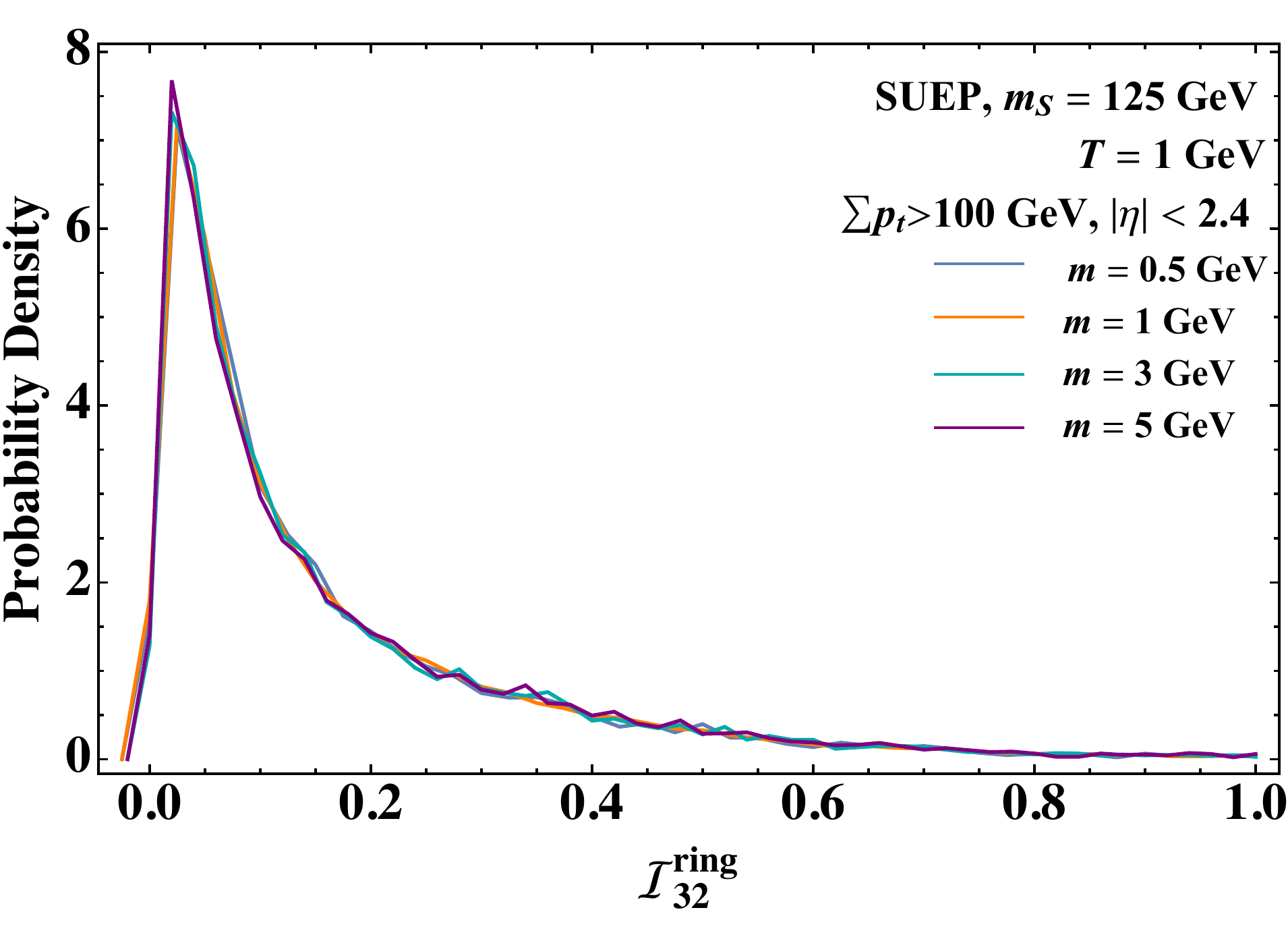}
     }
          \hfill
 \subfloat[]{
            \includegraphics[width=0.47\textwidth]{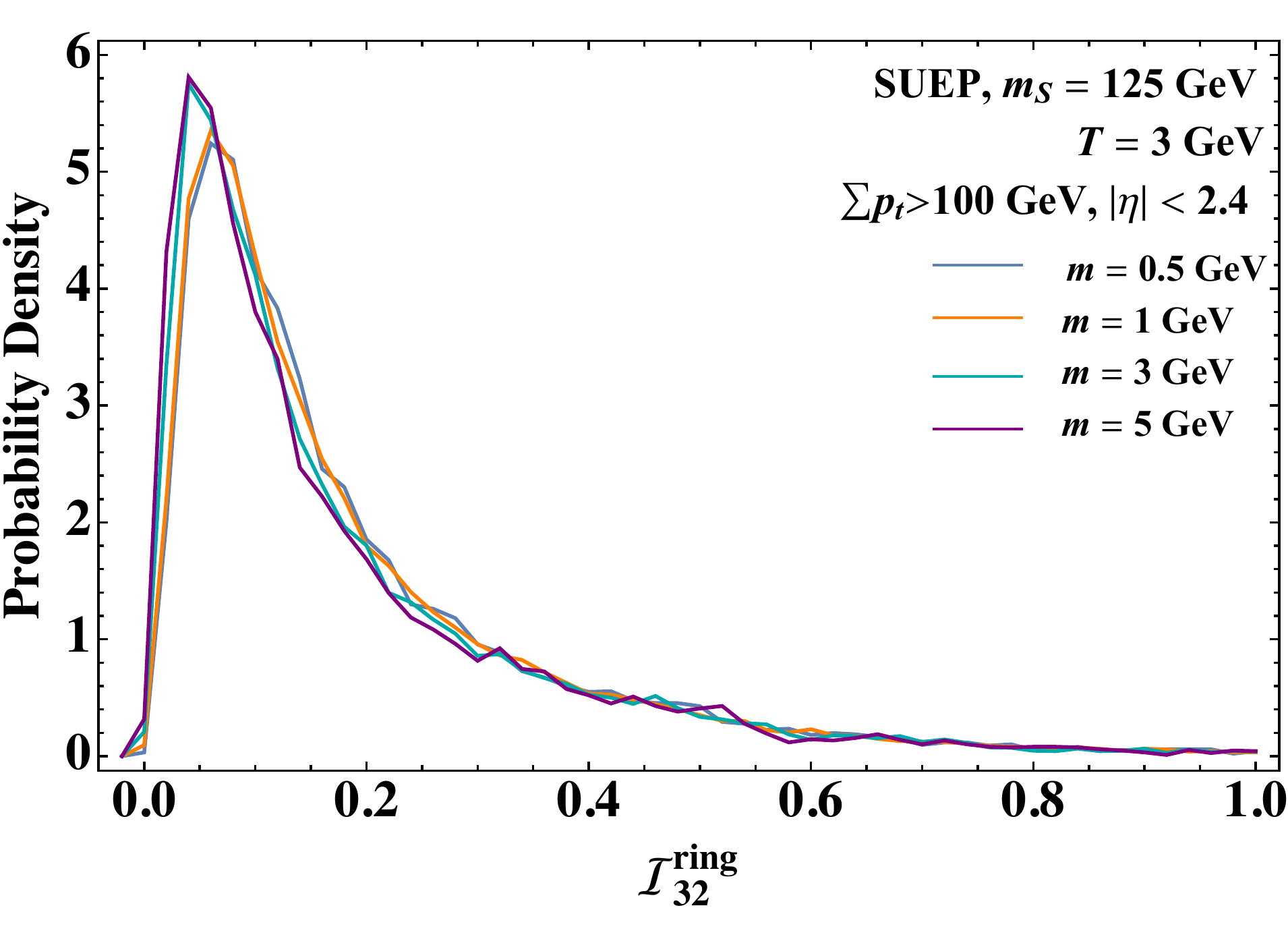}
     }
     \hfill
     \subfloat[]{
            \includegraphics[width=0.47\textwidth]{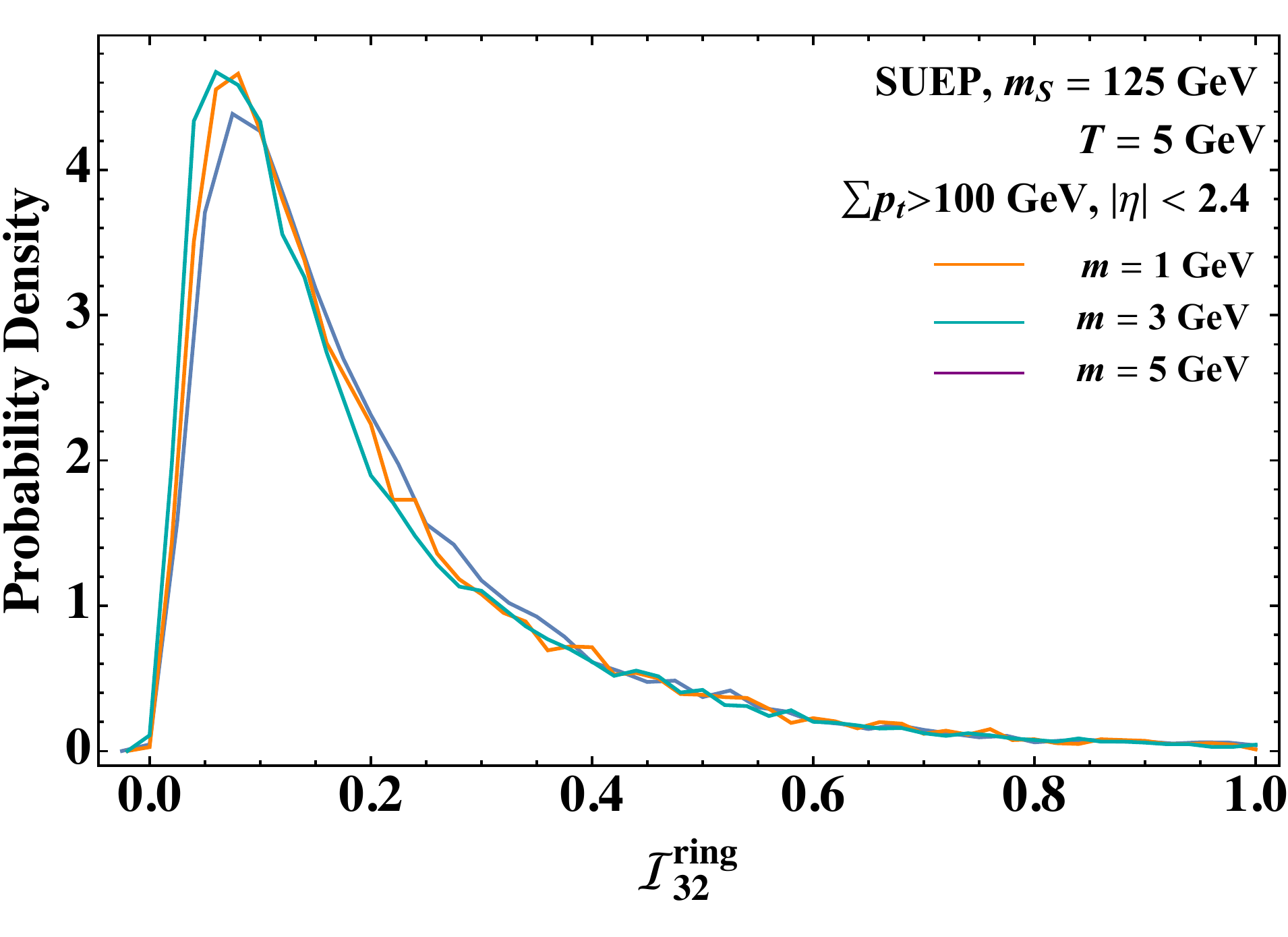}
     }
     \hfill
\caption{Distributions of ring-like event isotropy in the SUEP samples for (a) $T = 0.5$ GeV, (b) $T=1$ GeV, (c) $T=3$ GeV, and (d) $T=5$ GeV.
The different curves on each plot correspond to different values of $m$ satisfying $m\sim T$.
For this observable, the dominant behavior is determined by the temperature $T$ with only weak dependence on the lightest meson mass $m$.}
\label{fig:appsuepNoT_remd}
\end{figure} 

Fixing the mediator mass at $m_S = 125$ GeV, we consider hidden meson masses of $m = \{0.5, 1, 3, 5 \}$ GeV and temperatures of $T = \{0.5, 1, 3, 5\}$ GeV.
We generate $5\times 10^4$ events per sample, but only show scenarios where $T \sim m$, since this is a condition for the SUEP Generator to yield physically sensible results.
Because the SUEP Generator populates phase space as an approximate relativistic Maxwell-Boltzmann distribution, larger values of $m  \sim T$ are inaccessible without raising the mediator mass $m_S$.

Distributions for the ring-like event isotropy are shown in \Fig{fig:appsuepNoT_remd}.
For a fixed value of $T$, the distributions have only mild dependence on the value of $m$.
This make sense because event isotropy is an IRC-safe observable, so it is dominantly sensitive to the overall kinematics of the event (as determined by $T$) and not to the detailed partitioning of the energy from meson decays (as determined by $m$).

\begin{figure}[t]
\centering
\subfloat[]{
       \includegraphics[width=0.47\textwidth]{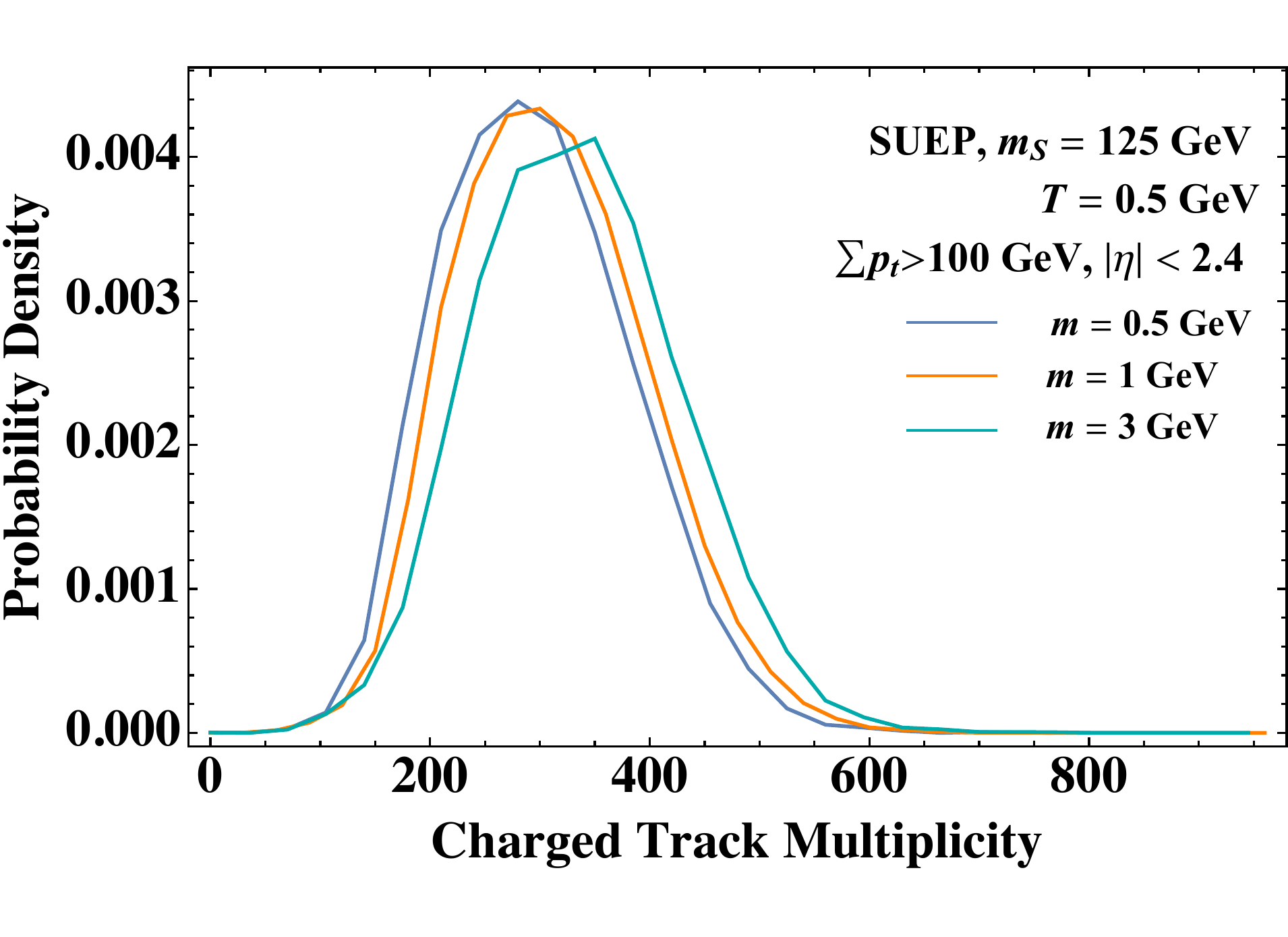}
     }
     \hfill
\subfloat[]{
       \includegraphics[width=0.47\textwidth]{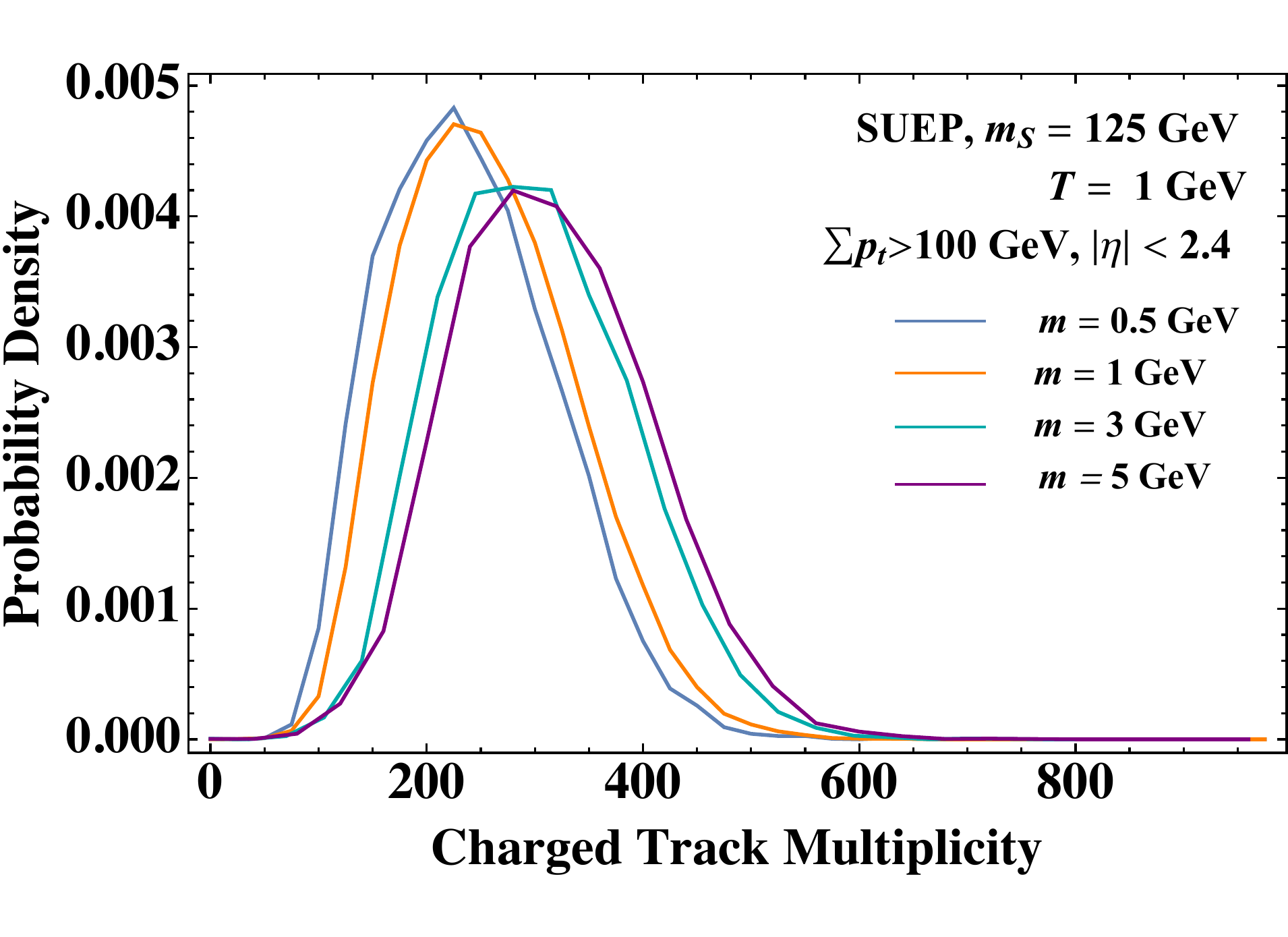}
     }
          \hfill
 \subfloat[]{
            \includegraphics[width=0.47\textwidth]{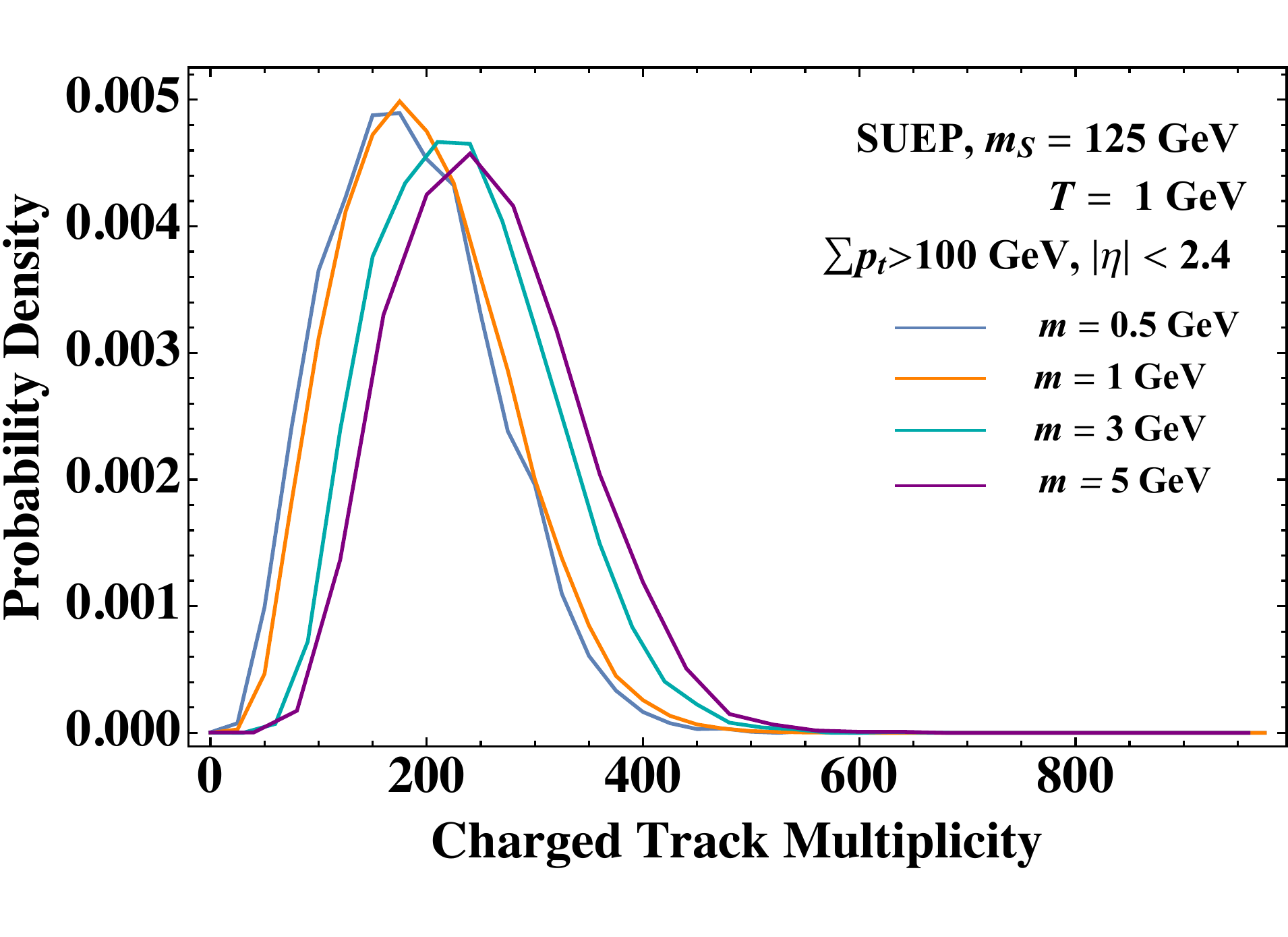}
     }
     \hfill
     \subfloat[]{
            \includegraphics[width=0.47\textwidth]{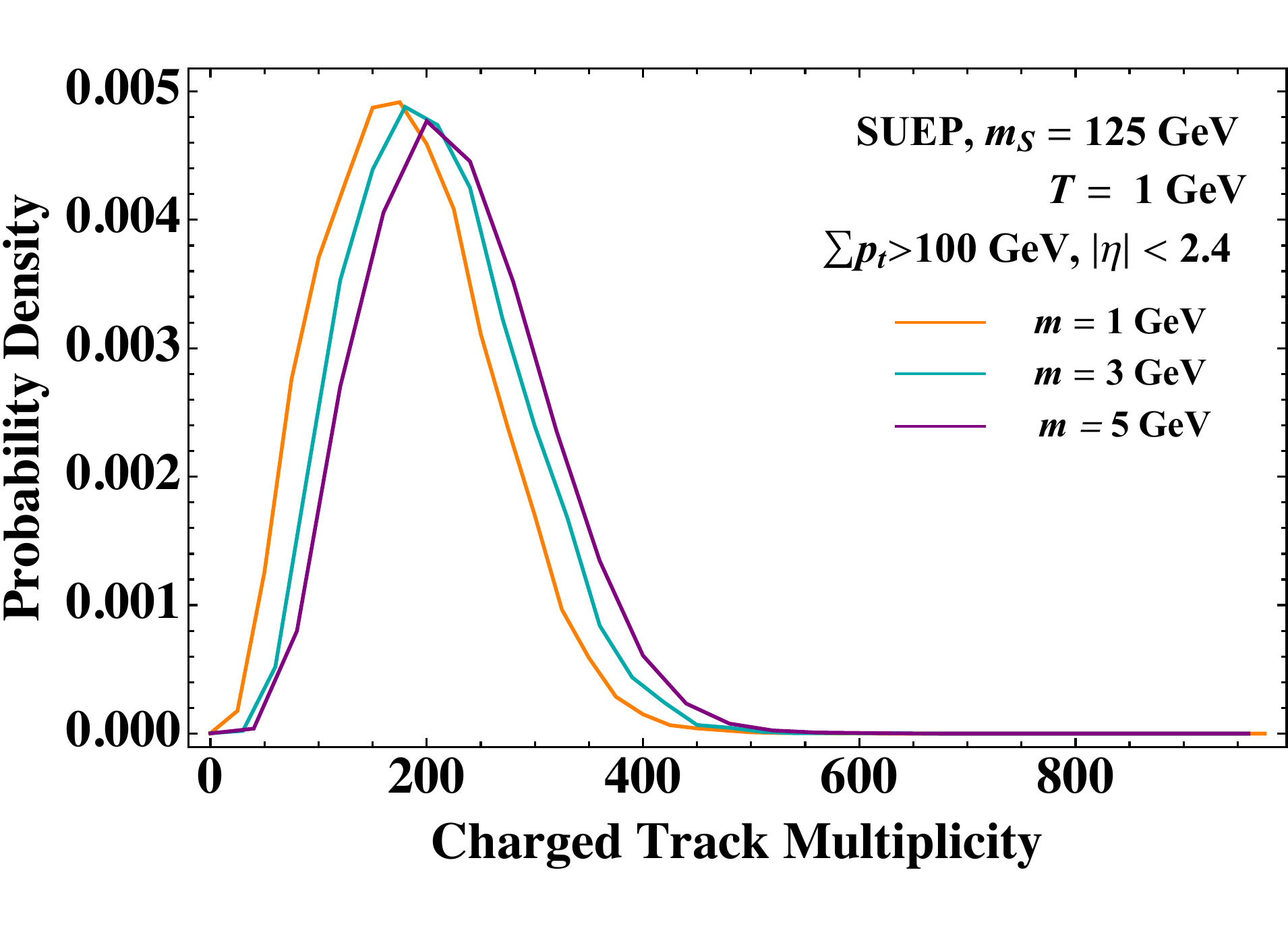}
     }
     \hfill
\caption{
The same as \Fig{fig:appsuepNoT_mult} but for the charged track multiplicity.
For fixed $T$, larger meson masses $m$ yield higher multiplicity events.}
\label{fig:appsuepNoT_mult}
\end{figure} 

\begin{figure}[t]
\centering
\includegraphics[width=0.47\textwidth]{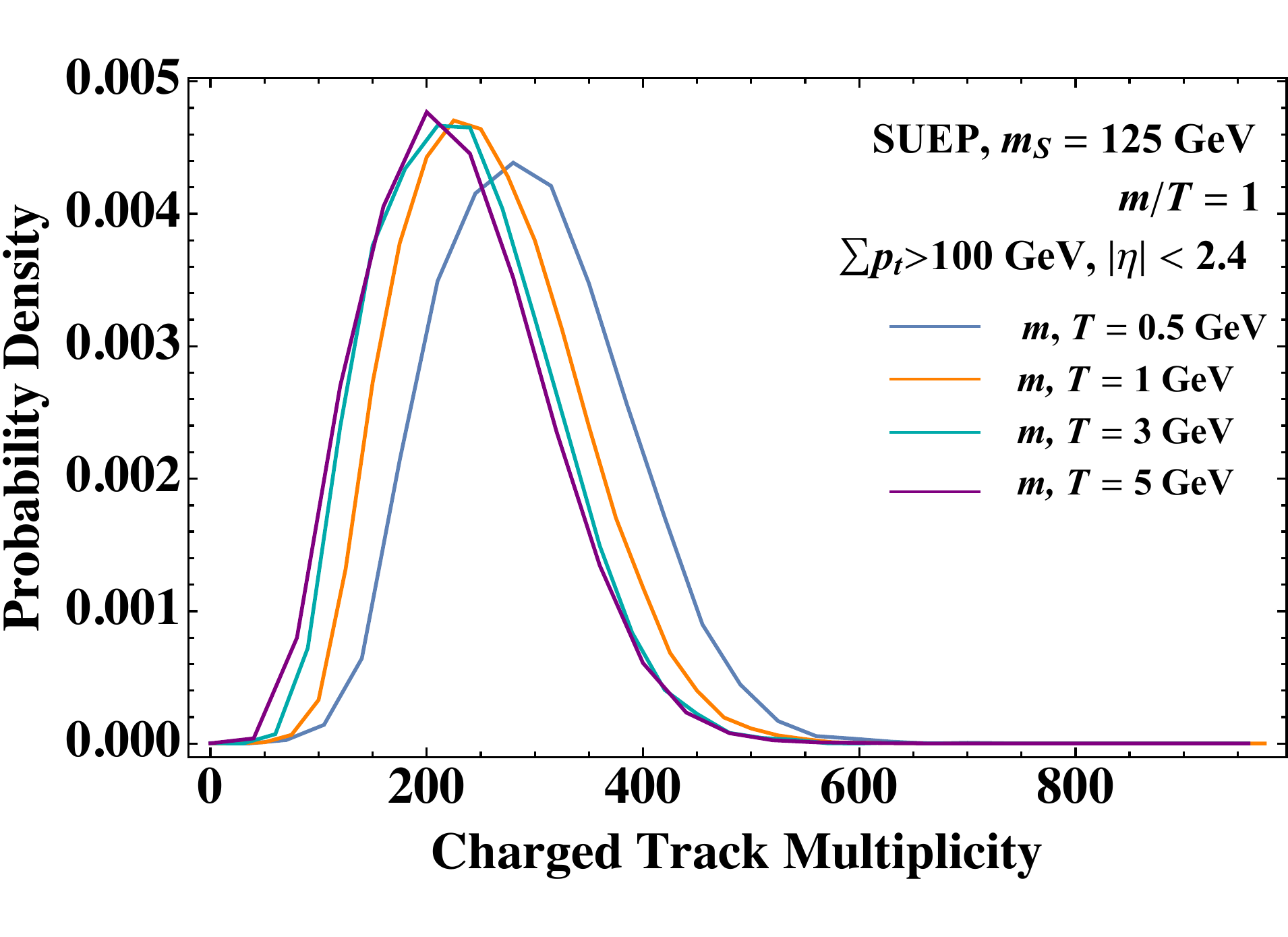}
\caption{ Distributions of the charged track multiplicity when $m/T =1$.
For $m = T$ values away from the QCD confinement scale, the distributions are very similar.}
\label{fig:mt1}
\end{figure} 

By contrast, the charged track multiplicity distributions in \Fig{fig:appsuepNoT_mult} show much larger dependence on the meson mass $m$.
For fixed $T$, the multiplicity increases with larger $m$, as expected since the decay of heavier mesons yield a higher multiplicity of SM final states.
Roughly speaking, the parameter controlling the multiplicity distributions is $m/T$.
This is shown for $m/T = 1$ in \Fig{fig:mt1}, where the multiplicity distributions are nearly overlapping for $m \geq 1$ GeV.
The increased multiplicity at $m = T = 0.5$ GeV arises from an interplay between the hidden sector shower and QCD confinement.

\bibliographystyle{utphys}
\bibliography{emd}

\end{document}